\documentclass[a4paper,11pt]{article}
\usepackage{jheppub} 
\usepackage{lineno}

\arxivnumber{2410.16226} 

\title{\boldmath IR finite correlation functions in de Sitter space, a smooth massless limit, and an autonomous equation}

\author[a,b]{Alexander Kamenshchik} \author[a,b]{and Polina Petriakova} 
\affiliation[a]{Dipartimento di Fisica e Astronomia, Universit\`{a} di Bologna, \\
via Irnerio 46, 40126 Bologna, Italy}
\affiliation[b]{I.N.F.N., Sezione di Bologna, I.S. FLAG, \\
viale B. Pichat 6/2, 40127 Bologna, Italy}

\emailAdd{kamenshchik@bo.infn.it} 
\emailAdd{polina.petriakova@bo.infn.it}

\abstract{We explore two-point and four-point correlation functions of a massive scalar field on the flat de Sitter background in the long-wavelength approximation. By employing the Yang{--}Feldman-type equation, we compute the two-point correlation function up to $\lambda^3$ order and the four-point correlation function up to $\lambda^2$ order. In contrast to the standard theory of a massive scalar field based on the de Sitter-invariant vacuum, we develop a vacuum-independent reasoning that may not possess de Sitter invariance but results in a smooth massless limit of the correlation function's infrared part. Our elaboration allows us to calculate correlation functions of a free massive scalar field and to proceed with quantum corrections, relying only on the known infrared part of the two-point correlation function of a free massless one. Remarkably, the two-point correlation function of a free massive scalar field coincides with that of the Ornstein{--}Uhlenbeck stochastic process and has a clear physical interpretation. We compare our results with those obtained using the Schwinger{--}Keldysh diagrammatic technique, Starobinsky's stochastic approach, and the Hartree{--}Fock approximation. At last, we construct a renormalization group-inspired autonomous equation for the two-point correlation function. Integrating~its approximate version, one obtains the non-analytic expression with respect to a self-interaction coupling constant $\lambda$. That solution reproduces the correct perturbative series up to the two-loop level. In the late-time limit, it almost coincides with the result of Starobinsky's stochastic approach over the whole interval of a new dimensionless parameter $0 \leq \tfrac{\pi^2 m^4}{3\lambda H^4} < \infty$.}

\dedicated{To the memory of Alexei~A.~Starobinsky}

\begin{document}
\maketitle
\flushbottom

\section{Introduction}
Quantum field theory in curved spacetime is of interest from both the purely theoretical and phenomenological points of view. Its connection with the cosmology of the very early universe becomes especially important due to the explosive growth of observational data. Scalar fields might play a central role during the inflationary (quasi-)de Sitter stage of the evolution of the universe, whereas their quantum fluctuations might be responsible for the observable large-scale structure formation. Therefore, even a relatively simple example of the quantum scalar field on the de Sitter background is already very instructive.


When someone sets the curved background, the choice of the vacuum becomes a non-trivial task; see, e.g., a book~\cite{Birrell:1982ix} and references therein. Due to the maximal symmetry of de Sitter space, one would attempt to define the quantum field's vacuum state as an invariant one under the full symmetry group. In the mainstream approach, people choose in the de Sitter set up the so-called Bunch{--}Davies vacuum~\cite{Bunch:1978yq}, where the short-wavelength modes behave as the corresponding modes in Minkowski space. For a massive scalar field, there exists a one-parameter family of de Sitter-invariant vacuum states~\cite{1985PhRvD..32.3136A}, while for a massless one it does not exist~\cite{1987PhRvD..35.3771A}. Consequently, an abyss appears between the results of computations for massive and massless scalar fields in de Sitter space: there is no smooth massless limit. 

In contrast to the standard theory of a massive scalar field based on the de Sitter-invariant vacuum, we develop a vacuum-independent reasoning that may not possess de Sitter invariance but results in a smooth massless limit of the correlation function's infrared part. We consider a rather particular theory of a minimally coupled massive scalar field living on the flat de Sitter background
\begin{equation}\label{S0}
\mathcal{L}_m = \sqrt{-g} \, \left( \,\, \frac12 \, \phi^{\, , \mu}\phi_{, \mu} - \frac12 m^2 \phi^2 - \frac{\lambda}{4} \, \phi^4 \right), \qquad ds^2=dt^2- \e^{2Ht}d \x^2 
\end{equation}  
and retain only the long-wavelength modes: the simplest ones technically and potentially the most relevant for upcoming observational tests. Our elaboration can be considered a theory of a massive scalar field with the vacuum "inherited" from a massless one. That should not be a concern, since the most symmetric vacuum is not always the physical one. 
We employ the Yang{--}Feldman formalism~\cite{Yang:1950vi} that recursively defines the interacting field as a formal power series in the coupling constant through the free field. Such a formalism in de Sitter space appears to be rather convenient for the leading infrared logarithm approximation; see the seminal paper by Woodard~\cite{2005NuPhS.148..108W} and et seq.~\cite{2005NuPhB.724..295T,2015PhRvD..91j3537O,2018PhRvD..97l3531K}. Through our proposed trick to "hang up" the mass, one can calculate the correlation functions of a free massive scalar field and proceed with quantum corrections relying only on the known infrared part of the correlation function of a free massless one. Remarkably, our free massive field's two-point correlation function~\eqref{correlator_massive_free} coincides with that of the Ornstein{--}Uhlenbeck mean-reverting stochastic process~\cite{Uhlenbeck:1930zz,1989fpem.book.....R,1994hsmp.book.....G}. By its virtue, that correlation function over time tends to the equilibrium state, which depends only on the difference of times and turns out to be de Sitter-invariant. That attractor feature was also established from other considerations~\cite{1994PhRvD..50.6357S,2000PhRvD..62l4019A} and later rediscovered in~\cite{2019arXiv191100022G}.

Using the Yang{--}Feldman-type equation, we compute the two-point correlation function up to $\lambda^3$ order and the four-point correlation function up to $\lambda^2$ order. Our outcomes for the two-point correlation function are in agreement in the late-time limit with those obtained in~\cite{2022EPJC...82..345K} via the Schwinger{--}Keldysh (or ''in-in'') diagrammatic technique~\cite{Schwinger:1960qe,Keldysh:1964ud}. Our full expressions differ from~\cite{2022EPJC...82..345K} due to the initial setting, namely, the choice of de Sitter-invariant Bunch{--}Davies vacuum for a massive scalar field made in~\cite{2022EPJC...82..345K}. We also compare with expressions coming from the Starobinsky{--}Yokoyama stochastic approach~\cite{Starobinsky:1986fx,1994PhRvD..50.6357S} and the Hartree{--}Fock (Gaussian) approximation.

Starobinsky's stochastic approach was the first non-perturbative development for the vacuum expectation value of the coarse-grained, long-wavelength scalar field fluctuations in de Sitter space. Such an approach matches the long-wavelength part of the quantum field to the classical stochastic field with a probability distribution function that satisfies the Fokker{--}Planck equation. The astonishing feature of the Starobinsky{--}Yokoyama stochastic approach~\cite{1994PhRvD..50.6357S} is that resummed non-perturbative quantities are free of secular growth, which appears for the two-point correlation function of a massless scalar field already at the tree level~\cite{Vilenkin:1982wt,Linde:1982uu}. 
The quantum field perturbative series's resummation methods for the computation of the correlation functions were stimulated by the success of the stochastic approach. Some kind of quantum field theory's renormalization group technique~\cite{1980vtkp.book.....B} can be employed in quite different areas of physics and mathematics~\cite{Priroda,2001PhR...352..219S}. In particular, it was applied in~\cite{Chen:1995ena} in order to treat secular growth terms arising at the iterative solution of some complicated differential equations. Contrary, we have only some pieces of that perturbative information and do not have a general equation form. In the spirit of~\cite{Chen:1995ena}, a renormalization group-inspired autonomous equation for treating perturbative series for a massless scalar field was constructed in~\cite{2020PhRvD.102f5010K}. Following this idea, we have constructed an autonomous equation for the two-point correlation function of a massive scalar field using our calculated perturbative series. Integrating its approximate version, one can obtain a non-analytic expression with respect to the self-interaction coupling constant $\lambda$ that reproduces the correct perturbative series up to the two-loop level.

This paper is organized as follows: section~\ref{Y-F_chapter} presents step by step the Yang{--}Feldman-type equation, from which all expressions for calculation are derived. We compute the two-point correlation function up to $\lambda^3$ order and the four-point correlation function up to~$\lambda^2$ order in the long-wavelength approximation. Section~\ref{Correspondence_chapter} establishes the correspondence between the Schwinger{--}Keldysh diagrammatic technique and our outcomes at the late-time limit. Further, we compare our results with the Starobinsky{--}Yokoyama stochastic approach and the Hartree{--}Fock approximation in section~\ref{Comparison_chapter}. In section~\ref{Autonomous_chapter}, we obtain an autonomous equation for the two-point correlation function and its approximate solution that are non-analytic with respect to the self-interaction coupling constant~$\lambda$. We conclude the paper with a recap of the main outcomes and possible outlooks for our elaboration in section~\ref{Conclusion}. The details can be found in the corresponding appendices~\ref{AddMass}{--}\ref{Starob_Hartee_details}.

\section{Yang{--}Feldman-type equation for a massive scalar field in de Sitter space and two-point and four-point correlation functions}\label{Y-F_chapter}
In this section, we present reasoning that leads to an iterative series for the Yang{--}Feldman-type equation for a massive scalar field, through which further calculations are performed.

The Yang{--}Feldman formalism~\cite{Yang:1950vi} recursively defines the interacting field as a formal power series in the coupling constant through the free field. In this approach, a solution to the Klein{--}Gordon equation for a massless scalar field 
is placed~by
\begin{equation}\label{Y-F}
\phi\left(t, \x \right) = \phi_0\left(t, \x \right) - \int d^4 x^{\prime} \sqrt{-g(x^{\prime})}\, G_{\text{R}} \left(t, \x \, ; t^{\prime}, \x^{\,\prime}\right) {V^\prime}_\phi \bigl( \phi\left(t^\prime, \x^{\,\prime} \right) \bigr), 
\end{equation} where $\phi_0\left(t, \x \right)$ is a solution of the homogeneous equation, $\Box \phi_0\left(t, \x \right) = 0$, and the retarded Green's function 
is any solution to
\begin{equation}\label{Retarded_Green_eq}
\Box \, G_{\text{R}}\left(t, \x \, ; t^{\prime}, \x^{\,\prime} \right) = 
\frac{\, \delta(t-t^\prime) \, \delta(\x-\x^{\,\prime})}{\sqrt{-g(x^{\prime})}} \end{equation} with retarded boundary conditions, $G_{\text{R}}\left(t, \x \, ; t^{\prime}, \x^{\,\prime}\right) =0$ for $t \leq t^{\prime}$. 
One can express the solution to equation~\eqref{Retarded_Green_eq} as
\begin{equation}\label{Retarded_Green_sol}
G_{\text{R}}\left(t, \x \, ; t^{\prime}, \x^{\,\prime}\right) = i \Theta(t-t^\prime) \, \langle \bigl[\phi_0\left(t, \x \right), \phi_0 \left(t^\prime, \x^{\,\prime} \right) \bigr]  \rangle.
\end{equation}
The representation of the scalar field in the commutator above upon canonically normalized creation and annihilation operators is the following:
\begin{equation}
\phi_0\left(t, \x \right) = \int \frac{d^3 \kk}{\,\, (2 \pi)^{3/2}} \Bigl(u_{k} (t) \,  \e^{i\kk \x} \, \hat{a}_{\kk} \, + \, u_{k}^* (t) \, \e^{- i\kk \x} \, \hat{a}^{\dagger}_{\kk} \, \Bigr),
\end{equation}
here the modes $u_k(t)$ in de Sitter spacetime~\eqref{S0}, 
are solutions of the linear equation 
\begin{equation}\label{linear_modes}
\ddot{u}_k + 3H\dot{u}_k  + k^2 \e^{-2Ht} u_k =0 \end{equation}
and must be normalized through the Wronskian
\begin{equation}\label{Wronskian}
W[u_k(t), u^*_k(t)] = \dot{u}_k u_k^* - u_k \dot{u}_k^* = - i \e^{-3Ht}
\end{equation} as a consequence of the canonical commutation relations. Straightforwardly, the expression for retarded Green's function~\eqref{Retarded_Green_sol} has the form
\begin{equation}\label{Retarded_Green_modes}
G_{\text{R}}\left(t, \x \, ; t^{\prime}, \x^{\,\prime}\right) = i \Theta(t-t^\prime)  \int \frac{d^3 \kk}{\,\,\, (2 \pi)^{3}} \, \e^{i\kk \, (\x-\x^{\,\prime})} \, \Bigl(u_{k} (t)  u_{k}^* (t^\prime)  - u_{k}^* (t) \, u_{k} (t^\prime) \Bigr) .    
\end{equation}

We are interested in the contribution of the very soft, long-wavelength (l-w) modes whose wave~numbers are small, $k\leq H\e^{Ht}$. Thus, in equation~\eqref{linear_modes}, one can neglect the last term $\sim k^2$, leading to the general solution 
\begin{equation}\label{modes_k0}
u^{\text{l-w}}_k(t)= c_1 + c_2 \, \e^{-3Ht}. \end{equation}
By employing the Wronskian~\eqref{Wronskian} 
and expressions~\eqref{Retarded_Green_modes} and \eqref{modes_k0}, we obtain
\begin{equation}\label{Green_retarded_lw}
G^{\, \text{l-w}}_{\text{R}} \left(t, \x \, ; t^{\prime}, \x^{\,\prime}\right) = \frac{\Theta(t-t^\prime)}{3H} \Bigl( \e^{-3Ht^\prime} - \e^{-3Ht}\Bigr) \, \delta(\x-\x^{\,\prime}).
\end{equation} 
Substituting the obtained expression into~\eqref{Y-F}, one can notice the multiplication by the measure. The first exponent in brackets in~\eqref{Green_retarded_lw} always prevails, forasmuch as the second is ''switched on'' only at the upper limit of integration and has been irrelevant (with an eye to escape from an additional logarithm) for the very soft, long-wavelength mode approximation. Therefore, at the leading logarithm approximation, equation~\eqref{Y-F} takes a simple form
\begin{equation}\label{Y-F_k0}
\phi\left(t, \x \right) = \phi_0\left(t, \x \right) -\frac{1}{3H} \int\limits_{0}^{t} dt^\prime \, 
{V^\prime}_\phi \bigl( \phi\left(t^\prime, \x \right) \bigr).
\end{equation} 
Note that we did not use any particular choice of the vacuum.\footnote{Howbeit, one can arrive at the form of the retarded Green’s function~\eqref{Green_retarded_lw} owing to the explicit form for the basis functions of the chosen vacuum in the Fock space, the so-called Bunch{--}Davies vacuum~\cite{Bunch:1978yq}; see the seminal work~\cite{2005NuPhS.148..108W}. To do this, use the explicit form for Bunch{--}Davies basis functions and expand them up to the third order in $k$, taking into account that we are interested only in the contribution of the long-wavelength modes $k\leq H\e^{Ht}$.}  One can now use equation~\eqref{Y-F_k0} iteratively to calculate the correlation functions of the massless field $\phi\left(t, \x \right)$ relying only on the known two-point correlation function of the free massless field $\phi_0\left(t, \x \right)$.

\vspace{0.25cm}

Using equation~\eqref{Y-F_k0}, one can define the free massive field $ \tilde{\phi}\left(t, \x \right)$ via the free massless one as  
\begin{equation}\label{FreeMassiveField}
\boxed{\,\, \tilde{\phi}\left(t, \x \right)  = \phi_0 \left(t, \x \right) - \frac{m^2}{3H} \, \e^{-\tfrac{m^2t}{3H}} \int\limits_{0}^{t} dt^\prime \, \e^{\tfrac{m^2t^\prime}{3H}} \, \phi_0 \left(t^\prime, \x \right) ; \,\, }\end{equation} see appendix~\ref{AddMass}, namely~\eqref{phi_tilde_sol}. This introduced relation, in addition to the massive Yang{--}Feldman-type equation, is very important for us. We employ it to calculate the two-point correlation function of the free massive scalar field, relying only on the known infrared part of the free massless one. 

Furthermore, one can find the corresponding analog to equation~\eqref{Y-F_k0}:
\begin{equation}\label{Y-F-massive}
\boxed{\,\, \phi\left(t, \x \right) = \tilde{\phi}\left(t, \x \right) - \frac{\lambda}{3H} \, \e^{-\tfrac{m^2t}{3H}} \int\limits_{0}^{t} dt^\prime \, \e^{\tfrac{m^2t^\prime}{3H}} \, {\phi}^{\,3}\left(t^\prime, \x \right); \,\,\, }
\end{equation} see~\eqref{equivalent_form} and the nearby reasoning. Let us notice that equations~\eqref{Y-F_k0} and \eqref{Y-F-massive} have the same structure with two differences: the free massless field $\phi_0\left(t, \x \right)$ is substituted by the free massive one $\tilde{\phi}\left(t, \x \right)$, and the retarded Green’s function for the massive field acquires an additional exponential factor. 
The iterated Yang{--}Feldman-type equation~\eqref{Y-F-massive} can be written out up to the few first terms of this expansion as follows: 
\begin{align}\label{Y-F-massive_iterated}
\phi\left(t, \x \right) = \tilde{\phi}\left(t, \x \right) & - \frac{\lambda}{3H} \e^{-\tfrac{m^2t}{3H}} \int\limits_{0}^{t} d t^{\prime} \, \e^{\tfrac{m^2t^\prime}{3H}} \tilde{\phi}^{3} \left(t^{\prime}, \x \right) \\ \nonumber 
& + \frac{\lambda^2}{3H^2} \e^{-\tfrac{m^2t}{3H}} \int\limits_{0}^{t} d t^{\prime} \, \tilde{\phi}^{2} \left(t^{\prime}, \x \right) \int\limits_{0}^{t^\prime} d t^{\prime\prime} \e^{\tfrac{m^2t^{\prime\prime}}{3H}} \tilde{\phi}^{3} \left(t^{\prime\prime}, \x \right) + ...  \end{align}
It allows us to calculate correlation functions of the massive scalar field $\phi\left(t, \x \right)$ in terms of the known  correlation function of the free massless one $\phi_0\left(t, \x \right)$. 

\vspace{0.35cm}
\subsection{Two-point function}
Let us start from the two-point correlation function for the free massive field $\tilde{\phi}\left(t, \x\right)$, where the spatial arguments of the two spacetime points coincide, but the time coordinates are different
\begin{equation}\label{correlator_massive_free} \bigl\langle \tilde{\phi} \left(t_1, \x \right) \tilde{\phi} \left(t_2, \x \right) \bigr\rangle \equiv \, \boxed{\, \left. \bigl\langle \tilde{\phi}\left(t_1\right) \tilde{\phi}\left(t_2 \right) \bigr\rangle
= \frac{3H^4}{8\pi^2m^2} \left( \e^{-\tfrac{m^2}{3H}\left|t_1-t_2\right|} - \e^{-\tfrac{m^2}{3H}\left(t_1+t_2\right)}\right). \right. } \end{equation} 
Hereafter, we omit the argument $\x$ since it is the same in all contributions and we have used the known result for the long-wavelength infrared part of the free massless field~\cite{2015PhRvD..91j3537O}:
\begin{equation}\label{correlator_massless_free} \bigl\langle \phi_0\left(t_1, \x \right) \phi_0 \left(t_2, \x \right) \bigr\rangle \equiv \bigl\langle \phi_0 \left(t_1 \right) \phi_0 \left(t_2 \right) \bigr\rangle  = \dfrac{H^3}{4\pi^2} \cdot \begin{cases} \, t_2 \, ,   &  t_2 \leq t_1 \, ; \\  \, t_1 \, ,  & t_2 \geq t_1 \, ; \end{cases} \end{equation} 
see details in appendix~\ref{Details}, namely~\eqref{correlator_massive_free_details_1}{--}\eqref{correlator_massive_free_details_3}. In the equal-time case, this simplifies to the well-known result~\cite{Vilenkin:1982wt,Linde:1982uu}: $ \bigl\langle \phi^2_0\left(t \right) \bigr\rangle = \dfrac{H^3t}{4\pi^2} $. Henceforth,~\eqref{correlator_massive_free} is our basic input for the computation of perturbative series using the iterated Yang{--}Feldman-type equation~\eqref{Y-F-massive_iterated}. Before proceeding with the results, let us make a comment. 

Our obtained correlation function~\eqref{correlator_massive_free} exactly coincides with that of the Ornstein{--}Uhlenbeck stochastic process~\cite{Uhlenbeck:1930zz}, which is the unique Gaussian and Markov process and has a stationary~state~\cite{1989fpem.book.....R,1994hsmp.book.....G}. 
The notable feature is its tendency to drift towards the average with the mean-reversion rate~$m^2/3H$. In our elaboration, correlation functions have a smooth massless limit, which coincides with the expressions obtained for a massless scalar field in the literature. Such a construction can be considered a theory of the massive scalar field with a vacuum "inherited" from the massless one. 
One drops de Sitter invariance in order to get a smooth transition between the massive and massless fields. Even so, by virtue of the correlation function~\eqref{correlator_massive_free}, over time this correlation function
tends to the equilibrium state, which depends only on the difference of times and turns out to be de Sitter-invariant. 

Building on~\eqref{correlator_massive_free}, one can organize the loop series using the Yang{--}Feldman-type equation~\eqref{Y-F-massive} or iterated~\eqref{Y-F-massive_iterated}~as
\begin{equation}\label{loop_series_Two_Point} \begin{gathered}
\bigl\langle \phi \left(t_1\right) \phi \left(t_2\right) \bigr\rangle = \bigl\langle \tilde{\phi} \left(t_1 \right) \tilde{\phi} \left(t_2 \right) \bigr\rangle  + \bigl\langle \phi \left(t_1 \right) \phi \left(t_2 \right) \bigr\rangle_{1-\text{loop}} + \bigl\langle \phi \left(t_1 \right) \phi \left(t_2 \right) \bigr\rangle_{2-\text{loop}} \\ 
\qquad \qquad \qquad \quad \,\,\,\, + \, \bigl\langle \phi \left(t_1 \right) \phi \left(t_2 \right) \bigr\rangle_{3-\text{loop}} + O\left(\lambda^4\right) \end{gathered} \end{equation} 
with the following one-loop contribution
\begin{align}\label{OneLoop_gen}
\bigl \langle &\phi \left(t_1\right) \phi \left(t_2 \right) \bigr\rangle_{1-\text{loop}} = -\frac{27 \lambda H^8}{128 \, \pi^4 m^6}  \Biggl(2\, \e^{-\tfrac{m^2}{3H} \left|t_1-t_2\right|} + \frac{4m^2}{3H} \biggl(\left|t_1-t_2\right| -\left(t_1+t_2 \, \right) \biggr)\, \e^{-\tfrac{m^2}{3H} \left(t_1+t_2 \right)} \nonumber \\ & \qquad \qquad \quad - \e^{\, \tfrac{m^2}{3H}\bigl(\left|\,t_1-t_2\right| - 2 \left(t_1+t_2\right)\bigr)}  + \frac{2m^2}{3H}\left|t_1-t_2 \right| \left( \e^{-\tfrac{m^2}{3H} \left|t_1-t_2\right|}  - \e^{-\tfrac{m^2}{3H} \left(t_1+t_2\right)} \right) \\ \nonumber & \qquad \qquad \quad - \e^{-\tfrac{m^2}{3H}\bigl(\left|t_1-t_2 \right| + 2 \left(t_1+t_2 \right)\bigr)} + \e^{-\tfrac{m^2}{3H}\bigl(2\left|t_1-t_2 \right|+ \left(t_1 + t_2 \right) \bigr)} - \e^{-\tfrac{m^2}{3H}\left(t_1+t_2 \right)}\Biggr), \end{align} 
the two-loop contribution 
\begin{align}\label{TwoLoop_gen}
\bigl \langle &\phi \left(t_1\right) \phi \left(t_2 \right) \bigr\rangle_{2-\text{loop}} = \frac{81 \lambda^2 H^{12}}{2048 \, \pi^6 m^{10}} \Biggl(\biggl(30 + \frac{12m^2}{H} \left|t_1-t_2\right| + \frac{2m^4}{3H^2} \left|t_1-t_2\right|^2 \biggr) \e^{-\tfrac{m^2}{3H}\left|t_1-t_2\right|}  \nonumber \\
& 
+ 2 \e^{-\tfrac{m^2}{H}\left|t_1-t_2\right|} - 5 \,  \e^{-\tfrac{m^2}{H}\left(t_1+t_2\right)} + \biggl( 36 + \frac{2m^2}{H}\Bigl( 9 \left|t_1-t_2\right| - 14 \left(t_1+t_2\right)\Bigr) \\ \nonumber 
& 
- \frac{2m^4}{3H^2} \Bigl(\left|t_1-t_2\right| - 2 \left(t_1+t_2\right)\Bigr)^2 \biggr) \e^{-\tfrac{m^2}{3H}\left(t_1+t_2 \right)} + \frac{15}{2}\e^{-\tfrac{m^2}{3H}\bigl(3\left|t_1-t_2\right| + 2 \left(t_1+t_2 \right)\bigr)} \\ \nonumber 
& 
+ \biggl(48 + \frac{2m^2}{H}  \Bigl( 7\left|t_1 -t_2 \right| + 2 \left(t_1+t_2\right)\Bigr) \biggr) \, \e^{-\tfrac{m^2}{3H} \bigl(2\left|t_1-t_2\right| +\left(t_1+t_2\right)\bigr)} - \frac{15}{2} \e^{-\tfrac{m^2}{3H}\bigl(2\left|t_1-t_2\right| + 3 \left(t_1+t_2\right)\bigr)} \\ \nonumber 
& 
- \biggl(45 + \frac{2m^2}{H}\Bigl(\left|t_1-t_2\right| +8 \left(t_1+t_2\right)\Bigr) \biggr) \e^{-\tfrac{m^2}{3H}\bigl(\left|t_1 -t_2 \right| + 2 \left(t_1 +t_2 \right)\bigr)} - \frac{15}{2} \e^{\, \tfrac{m^2}{3H}\bigl(2\left|t_1-t_2\right| - 3 \left(t_1+t_2\right)\bigr)} \\ \nonumber 
& 
- \biggl( \frac{117}{2} - \frac{2m^2}{H}\Bigl(7\left|t_1-t_2\right| - 8 \left(t_1+t_2\right) \Bigr) \biggr) \e^{\,\, \tfrac{m^2}{3H}\bigl(\left|t_1-t_2\right| - 2 \left(t_1+t_2\right)\bigr)}\Biggr), 
\end{align} and the three-loop contribution
\begin{align} \label{ThreeLoop_gen}
& \bigl \langle  \phi \left(t_1\right) \phi \left(t_2 \right) \bigr\rangle_{3-\text{loop}} = - \frac{729 \lambda^3 H^{16}}{4096 \, \pi^8 m^{14}} \Biggl(\biggl(26 + \frac{85m^2}{6H} \left|t_1-t_2\right| + \frac{5m^4}{6H^2} \left|t_1-t_2\right|^2  \\ \nonumber 
& 
+ \frac{m^6}{54H^3} \left|t_1-t_2\right|^3 \biggr) \e^{-\tfrac{m^2}{3H}\left|t_1-t_2\right|} + \biggl(7 + \frac{3m^2}{2H} \left|t_1-t_2 \right|\biggr) \e^{-\tfrac{m^2}{H}\left|t_1-t_2 \right|}\\ \nonumber 
& 
+ \frac{5}{4} \e^{-\tfrac{m^2}{3H}\bigl(4 \left|t_1 - t_2 \right| + \left(t_1 + t_2 \right)\bigr)} + \biggl( \frac{223}{2} + \frac{11m^2}{H} \Bigl(4 \left|t_1-t_2\right| + \left(t_1+t_2 \right)\Bigr)  \\ \nonumber 
& 
+ \frac{m^4}{12H^2} \Bigl(7 \left|t_1-t_2\right| + 2\left(t_1+t_2 \right)\Bigr)^2 \biggr) \e^{-\tfrac{m^2}{3H}\bigl(2\left|t_1-t_2\right| + \left(t_1+t_2\right)\bigr)} + \biggl( \frac{2135}{12} + \frac{m^2}{3H} \Bigl(96 \left|t_1-t_2\right|  \nonumber \\  
\nonumber 
& 
- 119 \left(t_1+t_2 \right)\Bigr) - \frac{m^4}{12H^2} \Bigl(17 \left|t_1-t_2\right| -26 \left(t_1+t_2\right)\Bigr)\Bigl(\left|t_1-t_2\right| -2 \left(t_1+t_2\right)\Bigr)  \\ 
\nonumber 
& 
+  \frac{m^6}{54H^3} \Bigl( \left|t_1-t_2\right| -2 \left(t_1 +t_2 \right)\Bigr)^3 \biggr) \e^{-\tfrac{m^2}{3H}\left(t_1+t_2 \right)} + \biggl( \frac{345}{8} + \frac{5m^2}{8H} \Bigl(17 \left|t_1-t_2\right| + 8 \left(t_1+t_2 \right)\Bigr) \biggr) \times \\ \nonumber 
& \quad 
\times \e^{-\tfrac{m^2}{3H}\bigl(3 \left|t_1-t_2\right| + 2\left(t_1+t_2\right)\bigr)} - \biggl(\frac{339}{4} + \frac{m^2}{4H} \Bigl(3 \left|t_1-t_2\right| + 208 \left(t_1+t_2\right)\Bigr)  \\ \nonumber 
& 
+ \frac{m^4}{12H^2}\Bigl( \left|t_1-t_2\right| + 8\left(t_1+t_2\right)\Bigr)^2 \biggr) \, \e^{-\tfrac{m^2}{3H}\bigl( \left|t_1-t_2 \right| + 2 \left(t_1+t_2\right)\bigr)} - \biggl( \frac{1203}{8} - \frac{m^2}{8H} \Bigl(441 \left|t_1-t_2\right|  \\ \nonumber 
& 
- 488 \left(t_1+t_2\right)\Bigr) + \frac{m^4}{12H^2} \Bigl(7 \left|t_1-t_2\right| - 8 \left(t_1+t_2\right)\Bigr)^2\biggr)  \e^{\, \tfrac{m^2}{3H}\bigl( \left|t_1 -t_2\right| -2 \left(t_1+t_2\right)\bigr)} \\ \nonumber 
& 
- \biggl(\frac{505}{16} - \frac{15m^2}{4H} \Bigl(\left|t_1-t_2\right| -2 \left(t_1+t_2\right)\Bigr) \biggr) \e^{-\tfrac{m^2}{H}\left(t_1+t_2\right)} + \frac{175}{48} \e^{-\tfrac{m^2}{3H}\bigl(4 \left|t_1-t_2\right| + 3\left(t_1+t_2\right)\bigr)}  \\ \nonumber 
& 
- \biggl(\frac{675}{16} + \frac{5m^2}{8H} \Bigl(7 \left|t_1 -t_2 \right| + 18 \left(t_1+t_2\right)\Bigr) \biggr) \e^{-\tfrac{m^2}{3H}\bigl(2 \left|t_1-t_2\right| + 3 \left(t_1 + t_2 \right)\bigr)} - \frac{35}{16} \, \e^{\, \tfrac{m^2}{3H}\bigl(\left|t_1-t_2 \right| -4 \left(t_1 +t_2 \right)\bigr)} \\ \nonumber 
& 
- \biggl(\frac{2395}{48} - \frac{5m^2}{8H} \Bigl(17 \left|t_1 -t_2 \right| -18  \left(t_1 +t_2 \right)\Bigr)\biggr) \e^{\, \tfrac{m^2}{3H}\bigl(2 \left|t_1 -t_2 \right| - 3 \left(t_1 +t_2 \right)\bigr)} \\ \nonumber 
&  
- \, \frac{35}{16} \, \e^{-\tfrac{m^2}{3H}\bigl(\left|t_1 -t_2 \right| + 4 \left(t_1 + t_2 \right)\bigr)} - \frac{175}{48} \e^{-\tfrac{m^2}{3H}\bigl(3 \left|t_1-t_2\right| + 4 \left(t_1 + t_2 \right)\bigr)} - \frac{175}{48} \e^{\, \tfrac{m^2}{3H} \bigl(3 \left|t_1-t_2\right| - 4  \left(t_1+ t_2 \right)\bigr)} \Biggr). \end{align}
At equal times, 
i.e., $t_1=t_2\equiv t$, expressions~\eqref{OneLoop_gen}{--}\eqref{ThreeLoop_gen} for series~\eqref{loop_series_Two_Point} become
\begin{align}\label{TwoPoint_equal_times_full_series}
&\bigl\langle \phi^2 \left(t \right) \bigr\rangle =  \frac{3H^4}{8\pi^2m^2} \left( 1 - \e^{-\tfrac{2m^2t}{3H}}\right)  -\frac{27 \lambda H^8}{64\pi^4 m^6} \biggl(1 -\frac{4m^2t}{3H} \e^{-\tfrac{2m^2t}{3H}} - \e^{-\tfrac{4m^2t}{3H}} \biggr) \\ \nonumber 
& \quad + \frac{81 \lambda^2 H^{12}}{64 \, \pi^6 m^{10}} \Biggl(1 + \biggl(\frac{21}{8} - \frac{3m^2t}{2H} - \frac{m^4t^2}{3H^2}\biggr)\e^{-\tfrac{2m^2t}{3H}} - \biggl(3 + \frac{2m^2t}{H} \biggr)\e^{-\tfrac{4m^2t}{3H}} - \frac{5}{8} \e^{-\tfrac{2m^2t}{H}} \Biggr) \\ \nonumber
& \quad - \frac{2187 \lambda^3 H^{16}}{4096 \, \pi^8 m^{14}} \Biggl( 11 + \biggl(\frac{872}{9} - \frac{172m^2t}{9H} - \frac{16m^4t^2}{3H^2}-\frac{32m^6t^3}{81H^3}\biggr)\e^{-\tfrac{2m^2t}{3H}} \\ \nonumber 
&  \quad - \biggl(64 + \frac{72m^2t}{H} + \frac{128m^4t^2}{9H^2}  \biggr)\e^{-\tfrac{4m^2t}{3H}} - \biggl(40 + \frac{20m^2t}{H} \biggr)\e^{-\tfrac{2m^2t}{H}}  - \frac{35}{9}\e^{-\tfrac{8m^2t}{3H}} \Biggr) + O(\lambda^4),
\end{align}
and at the late-time limit this series is 
\begin{equation}\label{loop_series_t_inf}
\bigl\langle \phi^2 \left(t \right) \bigr\rangle \xrightarrow[]{\, t \rightarrow \infty \,} \frac{3H^4}{8\pi^2m^2} - \frac{27 \lambda H^8}{64 \, \pi^4 m^6} + \frac{81 \lambda^2 H^{12}}{64 \, \pi^6 m^{10}} - \frac{24057 \lambda^3 H^{16}}{4096 \, \pi^8 m^{14}}+ O\left(\lambda^4\right).
\end{equation}
At the non-coinciding times and late-time limit, we have with the use of the obtained results~\eqref{OneLoop_gen}{--}\eqref{ThreeLoop_gen} for series~\eqref{loop_series_Two_Point} the following
\begin{align} \label{loop_series_late_time} 
& \bigl\langle \phi \left(t_1\right) \phi \left(t_2\right) \bigr\rangle \xrightarrow[\,\,\, \text{times}\,\,\,]{\text{late}}  
\frac{3H^4}{8\pi^2m^2} \Biggl(\e^{-\tfrac{m^2}{3H}\left|t_1-t_2\right|} - \frac{9\lambda H^4}{8\pi^2m^4} \biggl(1+\frac{m^2}{3H}\left|t_1 -t_2 \right| \biggr) \e^{-\tfrac{m^2}{3H}\left|t_1-t_2\right|} \\ \nonumber  
& \qquad \qquad \quad + \frac{27\lambda^2 H^8}{128\, \pi^4 m^8} \biggl(15+\frac{6m^2}{H}\left|t_1 -t_2 \right| + \frac{m^4}{3H^2}\left|t_1 - t_2 \right|^2 + \e^{-\tfrac{2m^2}{3H}\left|t_1 - t_2 \right|} \biggr)\e^{-\tfrac{m^2}{3H}\left|t_1-t_2\right|} \\ \nonumber  
& \qquad \qquad \quad  - \frac{243 \lambda^3 H^{12}}{512 \, \pi^6 m^{12}} \biggl(26+\frac{85m^2}{6H}\left|t_1-t_2\right| + \frac{5m^4}{6H^2}\left|t_1-t_2\right|^2 +  \frac{m^6}{54H^3}\left|t_1 - t_2 \right|^3 \\ \nonumber 
& \qquad \qquad \qquad \qquad \qquad \qquad + \Bigl(7 +  \frac{3m^2}{2H}\left|t_1 -t_2 \right| \Bigr) \e^{-\tfrac{2m^2}{3H}\left|t_1 -t_2 \right|} \biggr) \e^{-\tfrac{m^2}{3H}\left|t_1-t_2\right|} \Biggr)  + O\left(\lambda^4\right); \end{align}
while the massless limit of series \eqref{loop_series_Two_Point} is 
\begin{align}\label{loop_series_massless}
&\bigl \langle \phi \left(t_1 \right) \phi \left(t_2 \right) \bigr\rangle \Bigr|_{t_2 \, \leq \, t_1} \xrightarrow[]{\,\,m \rightarrow 0 \,\,} \, \frac{H^3 t_2}{4\pi^2}  - \frac{\lambda H^5}{96 \,\pi^4}\Bigl( 3\, t_1^2 \, t_2 + t_2^3 \Bigr) + \frac{\lambda^2 H^7}{1536\, \pi^6}\left( 11 \, t_1^4 \, t_2  + 2 \, t_1^2 \, t_2^3  + \frac{31}{5} \, t_2^5 \right) \nonumber \\
& \qquad -\frac{\lambda^3 H^9}{184320 \, \pi^8} \left( 471 \, t_1^6 \, t_2 + 55 \, t_1^4 \, t_2^3 + 93 \, t_1^2 \, t_2^5 + 160 \, t_1 \, t_2^6 + \frac{1331}{7} \, t_2^7 \right) + O \left(\lambda^4\right), 
\end{align} or, once again, at the coinciding times
\begin{equation}\label{loop_series_massless_coincide}
\bigl\langle \phi^2 \left(t \right) \bigr\rangle \xrightarrow[]{m \rightarrow 0} \frac{H^3t}{4\pi^2} - \frac{\lambda H^5 t^3}{24 \, \pi^4} + \frac{\lambda^2 H^7 t^5}{80 \, \pi^6} - \frac{53 \, \lambda^3 H^9 t^7}{10080 \, \pi^8} + O\left(\lambda^4\right).
\end{equation}

Here we have only shown the final and complete results of computations with a few valuable limits. All details can be found in appendix~\ref{Details}. The comparison of these series with other approaches will be done in the next two sections~\ref{Correspondence_chapter}~and~\ref{Comparison_chapter}.


\subsection{Four-point function}
In this subsection, we present the same results for the four-point correlation function up to $\lambda^2$ order. The details are also presented in appendix~\ref{Details}. Suppose that the time ordering is $t_1 \geq t_2 \geq t_3 \geq t_4$. Then, one has the series for the four-point correlation function
\begin{align}\label{loop_series_FourPoint} \ \bigl\langle \phi \left(t_1\right) \phi \left(t_2\right)  \phi \left(t_3\right) \phi \left(t_4\right) \bigr\rangle & =  \bigl\langle \phi \left(t_1\right) \phi \left(t_2\right)  \phi \left(t_3\right) \phi \left(t_4\right) \bigr\rangle_{0}  +  \bigl\langle \phi \left(t_1\right) \phi \left(t_2\right)  \phi \left(t_3\right) \phi \left(t_4\right) \bigr\rangle_\lambda
\nonumber \\
& \qquad + \bigl\langle \phi \left(t_1\right) \phi \left(t_2\right)  \phi \left(t_3\right) \phi \left(t_4\right) \bigr\rangle_{\lambda^2} 
+ O\left(\lambda^3\right) \end{align} 
with the following contributions: 
\begin{align}\label{four_point_tree}
& \bigl\langle \phi \left(t_1\right) \phi \left(t_2\right)  \phi \left(t_3\right) \phi \left(t_4\right) \bigr\rangle_{0} = 
\frac{9H^8}{64 \, \pi^4m^4} \, \e^{-\tfrac{m^2}{3H}\left(t_1+t_2+t_3+t_4\right)} \biggl( \, \e^{\tfrac{2m^2}{3H}\left(t_2+t_4\right)} \\ \nonumber 
& \qquad \qquad \qquad \qquad \qquad \qquad + 2 \, \e^{\tfrac{2m^2}{3H}\left(t_3+t_4\right)} - \e^{\tfrac{2m^2t_2}{3H}}- 2 \, \e^{\tfrac{2m^2t_3}{3H}} - 3 \, \e^{\tfrac{2m^2t_4}{3H}} + 3 \biggr) \xrightarrow[]{\,\,\, m \rightarrow 0 \,\,\,}  \\  \label{4point_massles_tree}
& \qquad \qquad \qquad \qquad \qquad \xrightarrow[]{\,\,\, m \rightarrow 0 \,\,\,} \, \frac{H^6}{16 \, \pi^4} \Bigl( t_2 t_4 + 2\, t_3 t_4 \Bigr); 
\end{align} while at the linear $\lambda$ order it reads
\begin{align}
\label{four_point_OneLoop}
& \bigl\langle \phi \left(t_1\right) \phi \left(t_2\right)  \phi \left(t_3\right) \phi \left(t_4\right) \bigr\rangle_\lambda 
= - \frac{81 \lambda H^{12}}{1024 \, \pi^6m^8} \Biggl( \Bigl(4 + \frac{2m^2}{3H} \bigl(t_1-t_2+t_3-t_4 \bigr)\Bigr)  \e^{\tfrac{2m^2}{3H}\left(t_2+t_4\right)} \nonumber \\ \nonumber
& \quad 
+ \Bigl(16 + \frac{4m^2}{3H} \bigl(t_1+3t_2-3t_3-t_4 \bigr)\Bigr) \, \e^{\tfrac{2m^2}{3H}\left(t_3+t_4\right)} - 2 \, \e^{\tfrac{4m^2 t_4}{3H}} - 2 \, \e^{-\tfrac{2m^2}{3H}\left(t_1 - t_2 - t_3 - t_4\right)} \\ \nonumber
& \quad 
- \Bigl( 3 + \frac{2m^2}{3H} \bigl(t_1-t_2+t_3+3t_4 \bigr)\Bigr) \, \e^{\tfrac{2m^2 t_2}{3H}} - \Bigl(14 + \frac{4m^2}{3H} \bigl(t_1+3t_2-3t_3+3t_4 \bigr)\Bigr) \, \e^{\tfrac{2m^2 t_3}{3H}} \\ \nonumber 
& \quad 
+ \e^{\tfrac{2m^2}{3H}\left(t_2-t_3+t_4\right)}  - \Bigl( 33 + \frac{2m^2}{H} \bigl(t_1+3t_2+5t_3-5t_4 \bigr)\Bigr) \, \e^{\tfrac{2m^2 t_4}{3H}} + 2 \, \e^{-\tfrac{2m^2}{3H}\left(t_1-t_2-t_3\right)} \\  
& \quad 
+ 3 \, \e^{-\tfrac{2m^2}{3H}\left(t_1-t_2-t_4\right)} + 4 \, \e^{-\tfrac{2m^2}{3H}\left(t_1-t_3-t_4\right)} + 4 \, \e^{-\tfrac{2m^2}{3H}\left(t_2-t_3-t_4\right)} - \e^{\tfrac{2m^2}{3H}\left(t_2-t_3\right)}  \\ \nonumber 
& \quad 
- \e^{\tfrac{2m^2}{3H}\left(t_2-t_4\right)} - 2 \, \e^{\tfrac{2m^2}{3H}\left(t_3 -t_4 \right)} + 30 + \frac{2m^2}{H}\bigl(t_1+3t_2+5t_3+7t_4 \bigr) -  3 \, \e^{-\tfrac{2m^2}{3H}\left(t_1-t_2\right)} \\ 
& \quad  
- 4 \, \e^{-\tfrac{2m^2}{3H}\left(t_1-t_3\right)} - 5 \, \e^{-\tfrac{2m^2}{3H}\left(t_1-t_4\right)} - 4 \, \e^{-\tfrac{2m^2}{3H}\left(t_2-t_3\right)} - 5 \, \e^{-\tfrac{2m^2}{3H}\left(t_2-t_4\right)} - 5 \, \e^{-\tfrac{2m^2}{3H}\left(t_3-t_4\right)} \nonumber \\
& \quad 
+ 5 \, \e^{-\tfrac{2m^2 t_1}{3H}} + 5 \, \e^{-\tfrac{2m^2 t_2}{3H}} + 5 \, \e^{-\tfrac{2m^2 t_3}{3H}} + 5 \, \e^{-\tfrac{2m^2 t_4}{3H}} \Biggr) \e^{-\tfrac{m^2}{3H}\left(t_1+t_2+t_3+t_4\right)} \,\, \xrightarrow[]{\,\,\,\, m \rightarrow 0 \,\,\,\,}  
\nonumber \\ 
\label{OneLoop_FourPoint_massless}
&  \xrightarrow[]{ \,\,\,\,  m \rightarrow 0 \,\,\,\, } \,\,  - \, \frac{\lambda H^8}{384 \, \pi^6} \, \Bigl(3 \, t_1^2 \, t_2 t_4 +6 \, t_1^2 \, t_3 t_4 + 12 \, t_1 t_2 t_3 t_4 + t_2^3 \, t_4 + 6 \, t_2^2\,  t_3 t_4 + 3 \, t_2t_3^2 \,t_4 \\ \nonumber 
& \qquad \qquad \qquad \qquad \qquad + t_2t_4^3 + 2 \, t_3^3 \,t_4 + 2t_3t_4^3 \Bigr); \end{align}
and at the $\lambda^2$ order the contribution is \begin{align} 
\label{four_point_TwoLoop_full}
& \bigl\langle \phi \left(t_1\right) \phi \left(t_2\right)  \phi \left(t_3\right) \phi \left(t_4\right) \bigr\rangle_{\lambda^2} = \frac{729 \, \lambda^2 H^{16}}{4096 \, \pi^8m^{12}} \, \e^{-\tfrac{m^2}{3H}\left(t_1+t_2+t_3+t_4\right)} \Biggl( \, \frac{1}{6} \, \e^{\tfrac{2m^2}{3H}\left(t_2-t_3+2t_4\right)} \\ \nonumber  
& \quad 
+ \e^{ \, \tfrac{m^2}{3H}\left(t_2+t_3+2t_4\right)} + \biggl(6 + \frac{m^2}{3H} \bigl(t_1-t_2+t_3-t_4 \bigr)\biggr)\biggl(1 + \frac{m^2}{6H} \bigl(t_1-t_2+t_3-t_4 \bigr)\biggr) \e^{\tfrac{2m^2}{3H}\left(t_2+t_4\right)} \\ \nonumber 
& \quad
- \biggl(1 + \frac{2m^2}{3H} \bigl(t_3-t_4 \bigr)\biggr) \, \e^{\, \tfrac{m^2}{3H}\left(t_2-t_3+ 4\, t_4\right)} + \biggl(\frac{91}{2} + \frac{4m^2}{3H} \bigl(3t_1+11t_2-11t_3-3t_4 \bigr) 
\end{align} 
\begin{align} \nonumber 
& \quad 
+ \frac{m^4}{9H^2} \bigl(t_1+3t_2-3t_3-t_4 \bigr)^2\biggr) \, \e^{\tfrac{2m^2}{3H}\left(t_3+t_4\right)} - \biggl(\frac{21}{2} + \frac{m^2}{3H} \bigl(t_1+3t_2+3t_3-7t_4 \bigr)\biggr) \, \e^{\tfrac{4m^2 t_4}{3H}} \nonumber \\ \nonumber 
& \quad 
+ \frac{1}{6} \, \e^{-\tfrac{2m^2}{3H}\left(t_1-2\, t_2-t_4\right)} - \biggl(\frac{23}{2} + \frac{m^2}{3H} \bigl(9t_1-5t_2-3t_3-t_4 \bigr)\biggr) \, \e^{-\tfrac{2m^2}{3H}\left(t_1-t_2-t_3-t_4\right)} \nonumber \\ \nonumber 
& \quad 
+ \frac{1}{2} \, \e^{-\tfrac{2m^2}{3H}\left(t_1-t_2- 2\,t_4\right)} + \frac{2}{3} \, \e^{-\tfrac{2m^2}{3H}\left(t_1-t_3- 2\,t_4\right)} + \frac{2}{3} \, \e^{-\tfrac{2m^2}{3H}\left(t_1- 2\, t_3- t_4\right)} + \frac{2}{3} \, \e^{-\tfrac{2m^2}{3H}\left(t_2-t_3- 2\,t_4\right)} \\ \nonumber 
& \quad 
+ \frac{2}{3} \, \e^{-\tfrac{2m^2}{3H}\left(t_2- 2\, t_3- t_4\right)} - \biggl( \frac{m^2}{6H}\bigl(7t_1-7t_2+7t_3+29t_4 \bigr) +  \frac{m^4}{18H^2}\bigl(t_1-t_2+t_3+3t_4 \bigr)^2 \biggr) \, \e^{\tfrac{2m^2 t_2}{3H}} \\ \nonumber 
& \quad 
- \, \e^{ \,\tfrac{m^2}{3H}\left(t_2+t_3\right)} + \biggl(\frac{9}{2} + \frac{m^2}{6H} \bigl(t_1-t_2+9t_3-5t_4 \bigr)\biggr) \e^{\tfrac{2m^2}{3H}\left(t_2-t_3+t_4\right)} - \,  \e^{\, \tfrac{m^2}{3H}\left(t_2- 3 \,t_3+ 4\,t_4\right)} \\ \nonumber 
& \quad 
+ \biggl(2 - \frac{4m^4}{9H^2} \bigl(t_3-t_4\bigr)^2 \biggr) \e^{\, \tfrac{m^2}{3H}\left(t_2-t_3+2\,t_4\right)}  - \biggl( \frac{63}{2} + \frac{m^2}{3H}\bigl(11t_1 + 41t_2 -41t_3+41t_4 \bigr)\\ \nonumber 
& \quad  +  \frac{m^4}{9H^2}\bigl(t_1 + 3t_2 -3t_3+3t_4 \bigr)^2 \biggr) \, \e^{\tfrac{2m^2 t_3}{3H}}  - \biggl( 146 + \frac{m^2}{2H}\bigl(15t_1+53t_2 + 107t_3 - 107t_4 \bigr) \\ \nonumber 
& \quad 
+ \frac{m^4}{18H^2} \Bigl(3\bigl(t_1 + 3t_2\bigr)^2 + \bigl( 30t_1 + 90 t_2 + 67 t_3-134t_4 \bigr) \, t_3 - \bigl(30t_1 + 90t_2-67t_4 \bigr) \, t_4 \Bigr) \biggr) \, \e^{\tfrac{2m^2 t_4}{3H}} \\ \nonumber 
& \quad 
- \frac{5}{2} \, \e^{-\tfrac{2m^2}{3H}\left(2\, t_1 - t_2- t_3 - t_4\right)} + \biggl(11 + \frac{m^2}{3H} \bigl(9t_1-5t_2-3t_3+3t_4 \bigr)\biggr) \, \e^{-\tfrac{2m^2}{3H}\left(t_1-t_2-t_3\right)}  \\ \nonumber 
& \quad 
- \frac{1}{6} \, \e^{-\tfrac{2m^2}{3H}\left(t_1-2\, t_2\right)} + \biggl( \frac{39}{2} + \frac{m^2}{2H} \bigl(9t_1-5t_2+5t_3-5t_4 \bigr)\biggr) \, \e^{-\tfrac{2m^2}{3H}\left(t_1-t_2-t_4\right)} \\ \nonumber 
& \quad 
+ \biggl( 30 + \frac{2m^2}{3H} \bigl(9t_1+7t_2-7t_3-5t_4 \bigr)\biggr) \, \e^{-\tfrac{2m^2}{3H}\left(t_1-t_3-t_4\right)} - \frac{2}{3} \, \e^{-\tfrac{2m^2}{3H}\left(t_1 - 2\, t_3 \right)}  - \frac{5}{3} \, \e^{-\tfrac{2m^2}{3H}\left(t_1 - 2\, t_4\right)} \\ \nonumber 
& \quad 
+ \biggl( \frac{69}{2} + \frac{2m^2}{3H} \bigl(t_1+15t_2-7t_3-5t_4 \bigr)\biggr) \, \e^{-\tfrac{2m^2}{3H}\left(t_2-t_3-t_4\right)} - \frac{2}{3} \, \e^{-\tfrac{2m^2}{3H}\left(t_2 - 2\, t_3 \right)}  - \frac{5}{3} \, \e^{-\tfrac{2m^2}{3H}\left(t_2 - 2\, t_4 \right)} \\ \nonumber 
& \quad - \frac{2}{3} \, \e^{-\tfrac{2m^2}{3H}\left(t_3 - 2\, t_4 \right)} - \biggl( \frac{17}{4} + \frac{m^2}{6H} \bigl(t_1-t_2+9t_3+7t_4 \bigr)\biggr) \, \e^{\tfrac{2m^2}{3H}\left(t_2-t_3\right)} + \frac{5}{8} \, \e^{\tfrac{2m^2}{3H}\left(t_2 - 2\, t_3 + t_4 \right)} \\ \nonumber 
& \quad 
- \frac{4m^2}{3H} \bigl(t_3-t_4 \bigr) \, \e^{\, \tfrac{m^2}{3H}\left(t_2- 3\, t_3 + 2\, t_4\right)} - \biggl( \frac{43}{8} + \frac{m^2}{6H} \bigl(t_1-t_2+t_3+15t_4 \bigr)\biggr) \, \e^{\tfrac{2m^2}{3H}\left(t_2-t_4\right)} \\ \nonumber 
& \quad 
- \biggl( \frac{51}{4} + \frac{m^2}{3H} \bigl(t_1+3t_2-3t_3+15t_4 \bigr)\biggr) \, \e^{\tfrac{2m^2}{3H}\left(t_3-t_4\right)} + \frac{4m^2}{3H} \bigl(t_3-t_4 \bigr)\biggl(1+ \frac{m^2}{3H} \bigl(t_3-t_4 \bigr)\biggr) \, \e^{\, \tfrac{m^2}{3H}\left(t_2-t_3\right)} \\ \nonumber 
& \quad 
+ \frac{377}{4} + \frac{m^2}{3H} \bigl(21t_1+75t_2+149t_3+259t_4 \bigr) + \frac{m^4}{18H^2} \biggl( \bigl(3t_1+18t_2 + 30 t_3 + 42 t_4\bigr) t_1 \\ \nonumber 
& \quad 
+ \bigl(27t_2+90t_3+ 126t_4 \bigr) t_2 + \bigl(67t_3+226 t_4 \bigr) t_3 +139 t_4^2 \biggr) + \frac{5}{2} \, \e^{-\tfrac{2m^2}{3H}\left(2\, t_1 - t_2 - t_3 \right)} \\ \nonumber 
& \quad 
+ \frac{25}{8} \, \e^{-\tfrac{2m^2}{3H}\left(2\, t_1 - t_2 - t_4 \right)} + \frac{15}{4} \, \e^{-\tfrac{2m^2}{3H}\left(2\, t_1 - t_3 - t_4 \right)} + \frac{5}{2} \, \e^{-\tfrac{2m^2}{3H}\left(t_1 + t_2 - t_3 - t_4\right)} + \frac{5}{4} \, \e^{-\tfrac{2m^2}{3H}\left(t_1 - t_2 + t_3 - t_4 \right)} \\ \nonumber 
& \quad 
- \biggl( \frac{75}{4} + \frac{m^2}{2H} \bigl(9t_1-5t_2+5t_3+7t_4 \bigr)\biggr) \, \e^{-\tfrac{2m^2}{3H}\left(t_1-t_2\right)} + \frac{1}{2} \, \e^{-\tfrac{2m^2}{3H}\left(t_1 - t_2-t_3 + t_4\right)} \\ \nonumber 
& \quad 
- \biggl(29 + \frac{2m^2}{3H} \bigl(9t_1+7t_2-7t_3+7t_4 \bigr)\biggr) \, \e^{-\tfrac{2m^2}{3H}\left(t_1-t_3\right)} + \frac{15}{4} \, \e^{-\tfrac{2m^2}{3H}\left(2\,t_2-t_3-t_4\right)} \\ \nonumber 
& \quad 
- \biggl(\frac{165}{4} + \frac{5m^2}{6H} \bigl(9t_1+7t_2+9t_3-9t_4 \bigr)\biggr) \, \e^{-\tfrac{2m^2}{3H}\left(t_1-t_4\right)} - \biggl(\frac{67}{2} + \frac{2m^2}{3H} \bigl(t_1+15t_2-7t_3+7t_4 \bigr)\biggr) \times  \\ \nonumber 
& \qquad \quad 
\times \e^{-\tfrac{2m^2}{3H}\left(t_2-t_3\right)} - \biggl(\frac{375}{8} + \frac{5m^2}{6H} \bigl(t_1+15t_2+9t_3-9t_4 \bigr)\biggr) \, \e^{-\tfrac{2m^2}{3H}\left(t_2-t_4\right)} 
\end{align} 
\begin{align} \nonumber 
& \quad 
- \biggl(\frac{105}{2} + \frac{m^2}{6H} \bigl(5t_1+15t_2+97t_3-37t_4 \bigr)\biggr) \, \e^{-\tfrac{2m^2}{3H}\left(t_3-t_4\right)} - \frac{5}{8} \, \e^{\tfrac{2m^2}{3H}\left(t_2-2\, t_3\right)} - \frac{1}{2} \, \e^{\, \tfrac{m^2}{3H}\left(t_2- 5\, t_3 + 2\, t_4\right)}  \\ \nonumber 
& \quad 
+  \biggl(2 + \frac{4m^2}{3H} \bigl(t_3-t_4 \bigr)\biggr) \, \e^{\, \tfrac{m^2}{3H}\left(t_2-3\, t_3\right)} - \frac{5}{12} \, \e^{\tfrac{2m^2}{3H}\left(t_2- t_3 - t_4\right)} - \biggl(\frac{3}{2} + \frac{2m^2}{3H} \bigl(t_3-t_4 \bigr)\biggr) \, \e^{\, \tfrac{m^2}{3H}\left(t_2-t_3 - 2\, t_4\right)} \\ \nonumber 
& \quad 
- \frac{5}{8} \, \e^{\tfrac{2m^2}{3H}\left(t_2- 2\, t_4\right)} - \frac{5}{4} \, \e^{\tfrac{2m^2}{3H}\left(t_3 - 2\, t_4\right)} + \biggl(40 + \frac{5m^2}{6H} \bigl(9 t_1+7t_2 + 9t_3 + 11t_4 \bigr)\biggr) \, \e^{-\tfrac{2m^2t_1}{3H}}\\ \nonumber 
& \quad 
- \frac{25}{8} \, \e^{-\tfrac{2m^2}{3H}\left(2 t_1 - t_2\right)} - \frac{15}{4} \, \e^{-\tfrac{2m^2}{3H}\left( 2t_1 - t_3\right)} - \frac{35}{8} \, \e^{-\tfrac{2m^2}{3H}\left(2t_1 - t_4\right)} - \frac{5}{2} \, \e^{-\tfrac{2m^2}{3H}\left(t_1 + t_2 - t_3\right)} \\ \nonumber 
& \quad 
- \frac{35}{12} \, \e^{-\tfrac{2m^2}{3H}\left(t_1 + t_2 - t_4\right)} - \frac{35}{12} \, \e^{-\tfrac{2m^2}{3H}\left(t_1 + t_3 - t_4\right)} - \frac{5}{4} \, \e^{-\tfrac{2m^2}{3H}\left(t_1 - t_2 + t_3\right)} - \frac{5}{4} \, \e^{-\tfrac{2m^2}{3H}\left(t_1 - t_2 + t_4\right)} \\ \nonumber 
& \quad 
- \frac{5}{3} \, \e^{-\tfrac{2m^2}{3H}\left(t_1 - t_3 + t_4\right)} - \frac{15}{4} \, \e^{-\tfrac{2m^2}{3H}\left(2t_2 - t_3\right)} + \biggl(\frac{365}{8} + \frac{5m^2}{6H} \bigl(t_1+15t_2+9t_3+11t_4 \bigr)\biggr) \, \e^{-\tfrac{2m^2t_2}{3H}} \\ \nonumber 
& \quad 
- \frac{35}{8} \, \e^{-\tfrac{2m^2}{3H}\left(2t_2 - t_4\right)} + \biggl(\frac{197}{4} + \frac{m^2}{6H} \bigl(5t_1+15t_2+97t_3+63t_4 \bigr)\biggr) \, \e^{-\tfrac{2m^2t_3}{3H}} \\ \nonumber 
& \quad 
- \frac{35}{12} \, \e^{-\tfrac{2m^2}{3H}\left(t_2 + t_3 - t_4\right)} - \frac{5}{3} \, \e^{-\tfrac{2m^2}{3H}\left(t_2 - t_3 + t_4\right)} + \biggl(\frac{467}{8} + \frac{m^2}{6H} \bigl(5t_1+15t_2+29t_3+131t_4 \bigr)\biggr) \times \\ \nonumber
& \qquad \quad 
\times \e^{-\tfrac{2m^2t_4}{3H}} - \frac{31}{8} \, \e^{-\tfrac{2m^2}{3H}\left(2t_3 - t_4\right)} + \frac{1}{2} \, \e^{\, \tfrac{m^2}{3H}\left(t_2 - 5\,t_3 \right)} - \e^{\, \tfrac{m^2}{3H}\left(t_2 - 3\,t_3  - 2\, t_4\right)} + \frac{1}{2} \, \e^{\, \tfrac{m^2}{3H}\left(t_2 - t_3 - 4\, t_4 \right)} \qquad \qquad \qquad \qquad \nonumber \\ \nonumber 
& \quad  + \frac{35}{12} \, \e^{- \tfrac{2m^2}{3H}\left(t_1 +t_2 \right)}  + \frac{35}{12} \, \e^{- \tfrac{2m^2}{3H}\left(t_1 +t_3 \right)} +  \frac{35}{12} \, \e^{- \tfrac{2m^2}{3H}\left(t_1 +t_4 \right)}  +  \frac{35}{12} \, \e^{- \tfrac{2m^2}{3H}\left(t_2 +t_3 \right)} +  \frac{35}{12} \, \e^{- \tfrac{2m^2}{3H}\left(t_2 +t_4 \right)}  \\ \nonumber 
& \quad + \frac{47}{12} \, \e^{- \tfrac{2m^2}{3H}\left(t_3 +t_4 \right)} + \frac{35}{8} \, \e^{- \tfrac{4m^2 t_1}{3H}} + \frac{35}{8} \, \e^{- \tfrac{4m^2 t_2}{3H}} + \frac{31}{8} \, \e^{- \tfrac{4m^2 t_3}{3H}} +  \frac{31}{8} \, \e^{- \tfrac{4m^2 t_4}{3H}} \Biggr) \xrightarrow[]{\,\,\,\,m \rightarrow 0\,\,\,\,} \\
\label{TwoLoop_FourPoint_massless}
&  \xrightarrow[]{ \,\,\,\,  m \rightarrow 0 \,\,\,\, } \,\, \frac{\lambda^2 H^{10}}{6144 \, \pi^8} \, \biggl( 11 \, t_1^4 t_2 t_4 + 22 \, t_1^4 t_3t_4 + 88 \, t_1^3 t_2 t_3 t_4 + 2\, t_1^2 t_2^3 t_4 + 12 \, t_1^2 t_2^2 t_3 t_4  \\ \nonumber 
& \qquad \qquad  \qquad \qquad \,\, + 6 \, t_1^2 t_2 t_3^2 t_4 + 2 \, t_1^2 t_2 t_4^3 + 4 \, t_1^2 t_3^3 t_4 + 4 \, t_1^2 t_3 t_4^3 + 8 \, t_1 t_2^3 t_3 t_4 + 8 \, t_1 t_2 t_3^3 t_4  \\ \nonumber 
& \qquad \qquad \qquad \qquad \,\, + 8 \, t_1 t_2 t_3 t_4^3 + \frac{31}{5} \, t_2^5 t_4 + 62 \, t_2^4 t_3 t_4 + 2 \, t_2^3 t_3^2 t_4 + \frac{2}{3} \, t_2^3 t_4^3 + 4 \, t_2^2 t_3^3 t_4 + 4 \, t_2^2 t_3 t_4^3  \\ \nonumber 
& \qquad \qquad \qquad \qquad \,\, + 11 \, t_2 t_3^4 t_4 + 2\, t_2 t_3^2 t_4^3 + \frac{31}{5} \, t_2 t_4^5 + \frac{182}{5} \, t_3^5 t_4 + \frac{4}{3} \, t_3^3 t_4^3 + \frac{62}{5} \, t_3 t_4^5 + 16 \, t_4^6\biggr)\, ; 
\end{align} and at equal times, this series appears to be
\begin{align} \label{four_point_tree_equal_times}
& \bigl\langle \phi^4 \left(t\right) \bigr\rangle
= \frac{27H^8}{64 \, \pi^4m^4} \, \biggl( 1 - \e^{-\tfrac{2m^2 t}{3H} } \biggr)^2 
\, - \, \frac{81 \lambda H^{12}}{64 \, \pi^6 m^8} \Biggl(1- \left(\frac{9}{4} + \frac{m^2 t}{H} \right) \e^{-\tfrac{2m^2 t}{3H}} 
\\ \nonumber 
& \,\, + \frac{2m^2t}{H} \e^{-\tfrac{4m^2 t}{3H}} + \frac{5}{4} \e^{-\tfrac{2m^2 t}{H}}\Biggr) + \frac{729 \lambda^2 H^{16}}{4096 \, \pi^8 m^{12}} \Biggl( \, 33 \, - \biggl( 86 + \frac{48m^2 t}{H} + \frac{16 m^4 t^2}{3H^2}\biggr) \e^{-\tfrac{2m^2t}{3H}} \\ \nonumber 
& \,\, - \biggl( 132 - \frac{88m^2 t}{H}  - \frac{128m^4 t^2}{3H^2}\biggr) \e^{-\tfrac{4m^2t}{3H}} + \biggl( 150 + \frac{120m^2 t}{H} \biggr) \e^{-\tfrac{2m^2t}{H}} + 35 \, \e^{-\tfrac{8m^2t}{3H}}\Biggr)  + O\left(\lambda^3\right); \\ \label{massle_four_point_full}
& \quad \xrightarrow[]{\,\,\, m \rightarrow 0 \,\,\,} \,\,\,\, \frac{3 H^6 t^2}{16 \, \pi^4} - \,\, \frac{3 \lambda H^8 t^4}{32 \, \pi^6} + \, \frac{53 \, \lambda^2 H^{10} \, t^6}{960 \, \pi^8} + O\left(\lambda^3\right);
\\ \label{stochastic_limit_Four_point_function}
& \quad \xrightarrow[]{\,\,\, t \rightarrow \infty \,\,\,} \,\, \frac{27H^8}{64 \, \pi^4m^4} - \, \frac{81 \lambda H^{12}}{64 \, \pi^6 m^8} + \, \frac{24057 \lambda^2 H^{16}}{4096 \, \pi^8 m^{12}} + O\left(\lambda^3\right). \end{align}

All the obtained perturbative massive series are finite and have a smooth massless limit. We computed the two-point correlation function of the free massive scalar field, relying solely on the known two-point correlation function of the free massless one. Further, we proceeded to calculate the four-point function and loop quantum corrections to the two-point and four-point correlation functions through the Yang{--}Feldman-type equation.

\section{Correspondence between the Yang{--}Feldman-type equation and the \\ Schwinger{--}Keldysh diagrammatic technique}\label{Correspondence_chapter}
In view of the previous section, the calculation of correlation functions in the Yang{--}Feldman technique is reduced to an integration of some products of the two-point correlation function of the free field with appropriate combinatorial coefficients. 

In this section, we intend to trace out the correspondence between integral structures used in the study of the Yang{--}Feldman-type equation and the Feynman/Schwinger{--}Keldysh diagrammatic techniques. These two types of diagrams have the same topological structure, but in the Schwinger{--}Keldysh case, one encounters two types of vertices and four types of propagators. To spot what kind of correspondence may exist, we shall rely on the outputs for a massive scalar field’s two-point correlation function calculated at the two-loop level via the Schwinger{--}Keldysh diagrammatic technique in~\cite{2022EPJC...82..345K} and for a massless one at equal times taken from~\cite{2020PhRvD.102f5010K}.

All integral structures for the two-point correlation function at the two-loop level in the Yang{--}Feldman-type equation are the following:
\begin{align}\label{1_twoLoop_structure}
\mathcal{I}_{1, \text{a}} &= \frac{2\lambda^2}{H^2} \e^{-\tfrac{m^2t_1}{3H}} \int\limits_{0}^{t_1} dt^{\prime} \int\limits_{0}^{t^{\prime}} dt^{\prime\prime} \e^{\tfrac{m^2 t^{\prime\prime}}{3H}} \bigl\langle \tilde{\phi} \left(t^{\prime} \right) \tilde{\phi} \left(t_2\right) \bigr\rangle \bigl\langle \tilde{\phi} \left(t^{\prime}\right) \tilde{\phi} \left(t^{\prime\prime} \right) \bigr\rangle \bigl\langle \tilde{\phi}^2 \left(t^{\prime\prime} \right) \bigr\rangle \, ; \\ \nonumber 
& \quad \mathcal{I}_{1, \text{b}} = \mathcal{I}_{1, \text{a}} \,\,\, \text{with} \,\,\, \bigl( t_1 \leftrightarrow t_2 \bigr); 
\\
\label{2_twoLoop_structure}
\mathcal{I}_{2, \text{a}} &= \frac{2\lambda^2}{H^2} \e^{-\tfrac{m^2t_1}{3H}} \int\limits_{0}^{t_1} dt^{\prime} \int\limits_{0}^{t^{\prime}} dt^{\prime\prime} \e^{\tfrac{m^2 t^{\prime\prime}}{3H}}  \bigl\langle \tilde{\phi} \left(t^{\prime\prime} \right)\tilde{\phi} \left(t_2\right) \bigr\rangle \Bigl(\bigl\langle \tilde{\phi} \left(t^{\prime}\right) \tilde{\phi} \left(t^{\prime\prime} \right) \bigr\rangle\Bigr)^2 ; \\ \nonumber 
& \quad 
\mathcal{I}_{2, \text{b}} = \mathcal{I}_{2, \text{a}} \,\,\, \text{with} \,\,\, \bigl( t_1 \leftrightarrow t_2 \bigr); \\
\label{3_twoLoop_structure}
\mathcal{I}_{3, \text{a}}  &= \frac{\lambda^2}{H^2} \e^{-\tfrac{m^2t_1}{3H}} \int\limits_{0}^{t_1} dt^{\prime} \int\limits_{0}^{t^{\prime}} dt^{\prime\prime} \e^{\tfrac{m^2 t^{\prime\prime}}{3H}} \bigl\langle \tilde{\phi} \left(t^{\prime\prime} \right) \tilde{\phi} \left(t_2\right) \bigr\rangle \bigl\langle \tilde{\phi}^2 \left(t^{\prime}\right) \bigr\rangle \bigl\langle \tilde{\phi}^2 \left(t^{\prime\prime} \right) \bigr\rangle \, ; \\ \nonumber 
& \quad \mathcal{I}_{3, \text{b}} = \mathcal{I}_{3, \text{a}} \,\,\, \text{with} \,\,\, \bigl( t_1 \leftrightarrow t_2 \bigl);  \\ 
\label{4_twoLoop_structure}
\mathcal{I}_4 &= \frac{2\lambda^2}{3H^2} \e^{-\tfrac{m^2}{3H} \left(t_1+t_2 \right)} \int\limits_{0}^{t_1} dt^{\prime} \e^{\tfrac{m^2 t^{\prime}}{3H}} \int\limits_{0}^{t_2} dt^{\prime\prime} \e^{\tfrac{m^2 t^{\prime\prime}}{3H}}  \Bigl(\bigl\langle \tilde{\phi} \left(t^{\prime}\right) \tilde{\phi} \left(t^{\prime\prime} \right) \bigr\rangle\Bigr)^3 ; \\
\label{5_twoLoop_structure}
\mathcal{I}_5 &=  \frac{\lambda^2}{H^2} \e^{-\tfrac{m^2}{3H} \left(t_1+t_2 \right)} \int\limits_{0}^{t_1} dt^{\prime} \e^{\tfrac{m^2 t^{\prime}}{3H}} \int\limits_{0}^{t_2} dt^{\prime\prime} \e^{\tfrac{m^2 t^{\prime\prime}}{3H}} \bigl\langle \tilde{\phi}^2\left(t^{\prime}\right) \bigr\rangle \bigl\langle \tilde{\phi}\left(t^{\prime}\right) \tilde{\phi} \left(t^{\prime\prime}\right) \bigr\rangle \bigl\langle \tilde{\phi}^2 \left(t^{\prime\prime}\right) \bigr\rangle \, ;
\end{align}
see~\eqref{TwoLoop_TwoPoint_FULLexpression} in appendix~\ref{Details}. Alongside, all the possible two-loop level diagram structures are presented in figure~\ref{picture_Two_Loop} below. Our correspondence hypothesis in this section is based on the following assignment assumption: In order to obtain one of the diagram topologies, the points might be connected either by an explicit correlation function in structures~\eqref{1_twoLoop_structure}{--}\eqref{5_twoLoop_structure} or by limits of integration variables. 

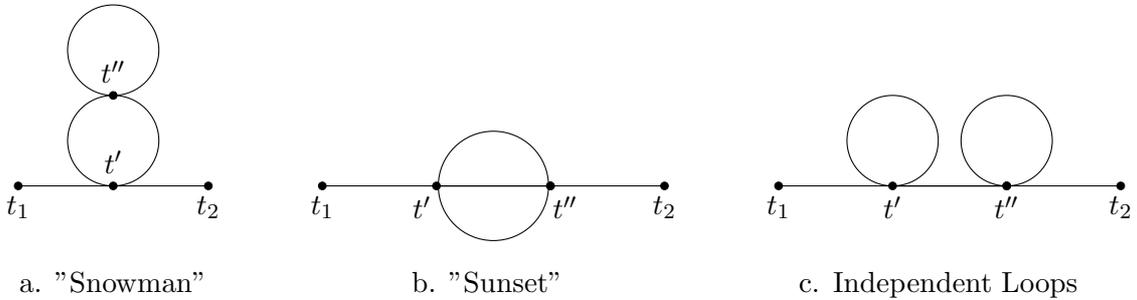
\begin{figure}[th!]
\centering
\begin{tikzpicture}[baseline=(current bounding box.north)] \begin{feynman}
\vertex(in1); \vertex[right=1.25cm of in1] (v1); \vertex[above=1.2cm of v1] (u1); \vertex[right=1.25cm of v1] (out1); \vertex[above=1.2cm of v1] (w1); \diagram{(in1) -- (v1) -- (out1)};
\draw[fill=black] (v1) circle(0.5mm); \draw[fill=black] (u1) circle(0.5mm);
\draw[fill=black] (in1) circle(0.5mm) node[anchor=north]{$t_1$}; \draw[fill=black] (out1) circle(0.5mm) node[anchor=north]{$t_2$};
\draw (1.25,1.8) circle (0.6cm) node[anchor=north]{$t^{\prime\prime}$}; 
\draw (1.25,-1.0) node[anchor=north]{a. "Snowman"};
\draw (1.25,0.6) circle (0.6cm) node[anchor=north]{$t^\prime$};
\vertex[right=1.5cm of out1] (in2); \vertex[right=1.5cm of in2] (v2); \vertex[right=1.5cm of v2] (u2); \vertex[right=1.5cm of v2] (out2); \vertex[right=1.5cm of u2] (w2); \diagram{(in2) -- (u2) -- (v2) -- (out2) -- (w2)};
\draw[fill=black] (v2) circle(0.5mm) node[anchor=north]{$t^\prime \quad$}; \draw[fill=black] (u2) circle(0.5mm) node[anchor=north]{$\quad t^{\prime\prime}$};
\draw[fill=black] (in2) circle(0.5mm) node[anchor=north]{$t_1$}; \draw[fill=black] (out2) circle(0.5mm); \draw[fill=black] (w2) circle(0.5mm) node[anchor=north]{$t_2$}; \draw (6.17,-1.0) node[anchor=north]{b. "Sunset"}; \draw (6.25,0) circle (0.725cm);
\vertex[right=1.5cm of w2] (in3); \vertex[right=1.5cm of in3] (v3); \vertex[right=1.5cm of v3] (u3);
\vertex[right=1.5cm of v3] (out3); \vertex[right=1.5cm of u3] (w3);
\diagram{(in3) -- (u3) -- (v3) -- (out3) -- (w3)};
\draw[fill=black] (v3) circle(0.5mm) node[anchor=north]{$t^\prime$}; \draw[fill=black] (u3) circle(0.5mm) node[anchor=north]{$t^{\prime\prime}$};
\draw[fill=black] (in3) circle(0.5mm) node[anchor=north]{$t_1$}; \draw[fill=black] (out3) circle(0.5mm); \draw[fill=black] (w3) circle(0.5mm) node[anchor=north]{$t_2$}; 
\draw (12.1,-1.0) node[anchor=north]{c. Independent Loops}; 
\draw (11.5,0.6) circle (0.6cm); \draw (13.0,0.6) circle (0.6cm); 
\end{feynman} \end{tikzpicture}  \caption{All the possible two-loop-level diagram structures.} \label{picture_Two_Loop}
\end{figure}

Let us look at the first structure~\eqref{1_twoLoop_structure}. One can notice $t^{\prime\prime}$ is integrated from $0$ to $t^{\prime}$, while $t^{\prime}$ is integrated from $0$ to $t_1$ (or $t_2$). To reflect the fact that $t^{\prime\prime}$ is integrated until~$t^{\prime}$, we shall connect the points $t^{\prime}$ and $t^{\prime\prime}$ and add the line connecting the point $t^{\prime}$ with the external points $t_1$ (or $t_2$), corresponding to the absent free massive field $\tilde{\phi}(t_1)$ (or $\tilde{\phi}(t_2)$). One has another external point $t_2$ (or $t_1$), where the field $\tilde{\phi}(t_2)$ (or $\tilde{\phi}(t_1)$) is present and connected with the field at the point $t^\prime$ through the correlation function under the integral. There is also the correlation function $\bigl\langle \tilde{\phi} \left(t^{\prime}\right) \tilde{\phi} \left(t^{\prime\prime} \right) \bigr\rangle$ that connects $t^{\prime}$ and $t^{\prime\prime}$ {---} one more line. Finally, we have a closed loop attached to the point $t^{\prime\prime}$. As a result, we obtain~a topology of the so-called "Snowman" diagram; see Figure~\ref{picture_Two_Loop}a. Computing the contribution of these structures, one has
\begin{align}\label{snowman}
& \bigl\langle \phi \left(t_1\right) \phi \left(t_2\right)  \bigr\rangle^\text{Snowman}_{2-\text{loop}} = \, \mathcal{I}_{1, \text{a}} + \, \mathcal{I}_{1, \text{b}} = \frac{243 \, \lambda^2 H^{12}}{1024 \, \pi^6 m^{10}} \Biggl(\biggl(2 + \frac{2m^2}{3H}|t_1 - t_2| \biggr) \, \e^{-\tfrac{m^2}{3H}\left|t_1-t_2 \right|} \\ \nonumber 
& \qquad - \biggl(1 - \frac{2m^2}{3H} \Bigl(|t_1 - t_2| - 2 (t_1 + t_2) \Bigr) + \frac{2m^4}{9H^2} \Bigl(|t_1 - t_2| - t_1 - t_2) \Bigr)^2 \biggr) \, \e^{-\tfrac{m^2}{3H}\left(t_1+t_2 \right)} \\ \nonumber 
&  \qquad + \biggl(2 + \frac{2m^2}{3H} \Bigl(|t_1 - t_2| +t_1 + t_2 \Bigr) \biggr)\, \e^{-\tfrac{m^2}{3H}\bigl( 2\left|t_1-t_2 \right| + \left(t_1 +t_2 \right)\bigr)} + \frac{1}{2} \, \e^{-\tfrac{m^2}{3H}\bigl(3\left|t_1-t_2 \right| + 2\left(t_1 +t_2 \right)\bigr)} \\ \nonumber 
&  \qquad - \biggl(2 + \frac{2m^2}{3H} \Bigl(|t_1 - t_2| +t_1 + t_2 \Bigr) \biggr)\, \e^{-\tfrac{m^2}{3H}\bigl( \left|t_1-t_2 \right| + 2 \left(t_1 +t_2 \right)\bigr)} - \frac{1}{2} \, \e^{-\tfrac{m^2}{3H}\bigl(2\left|t_1-t_2 \right| + 3\left(t_1 +t_2 \right)\bigr)} \\ \nonumber 
&  \qquad - \biggl(\frac{1}{2} - \frac{2m^2}{3H} \Bigl(|t_1 - t_2| - t_1 - t_2 \Bigr) \biggr)\, \e^{\, \tfrac{m^2}{3H}\bigl( \left|t_1-t_2 \right| -2 \left(t_1 +t_2 \right)\bigr)} - \frac{1}{2} \, \e^{ \, \tfrac{m^2}{3H}\bigl(2\left|t_1-t_2 \right| - 3\left(t_1 +t_2 \right)\bigr)}  \Biggr) \\ \label{snowman_latetimes}
&  \xrightarrow[\,\,\,\,\, \text{times}\,\,\,\,\,]{\text{late}} \,\,\, \frac{243 \lambda^2 H^{12}}{512 \, \pi^6 m^{10}} \biggl(1 + \frac{m^2}{3H}|t_1 - t_2| \biggr) \, \e^{-\tfrac{m^2}{3H}\left|t_1-t_2 \right|} ;
\\ \label{snowman_massless}
&  \xrightarrow[t_2 \leq t_1]{\,\,\,\,\, m \rightarrow 0 \,\,\,\,\,} \,\,\, \frac{\lambda^2 H^{7}}{1920 \, \pi^6 } \Bigl( 5\, t_1^4 \, t_2 + 3\, t_2^5\Bigr).
\end{align}

In the case of the second structure~\eqref{2_twoLoop_structure}, we have two correlation functions connecting the points $t^\prime$ and $t^{\prime\prime}$ and also the third line, which connects these two points due to the fact that $t^{\prime\prime}$ is integrated until $t^\prime$. An integration limit of $t^\prime$ is till $t_1$ (or $t_2$), and the correlation function inside also connects $t^{\prime\prime}$ with external points $t_2$ (or $t_1$). As a result, one has the so-called "Sunset" diagram; see figure~\ref{picture_Two_Loop}b. Moreover, structure~\eqref{4_twoLoop_structure} has the same topology: the integration of variables $t^\prime$ and $t^{\prime\prime}$ until the times $t_1$ and $t_2$, so an integration substitutes the “external” fields $\tilde{\phi}(t_1)$ or $\tilde{\phi}(t_2)$, which are absent in this structure. One can draw from these points $t^\prime$ and $t^{\prime\prime}$ two lines connecting these points with two external points $t_1$ and $t_2$ corresponding to the absent fields $\tilde{\phi}(t_1)$ and $\tilde{\phi}(t_2)$, and these two internal points, $t^\prime$ and $t^{\prime\prime}$, are connected by three identical correlation functions. Therefore, their sum leads to
\begin{align}\label{sunset}
& \bigl\langle \phi \left(t_1\right) \phi \left(t_2\right) \bigr\rangle^\text{Sunset}_{2-\text{loop}} = \, \mathcal{I}_{2, \text{a}} + \, \mathcal{I}_{2, \text{b}} + \, \mathcal{I}_{4} = \frac{243 \, \lambda^2 H^{12}}{1024 \, \pi^6 m^{10}} \Biggl(\biggl(1 + \frac{2m^2}{3H}|t_1 - t_2| \biggr) \, \e^{-\tfrac{m^2}{3H}\left|t_1-t_2 \right|} \nonumber \\
& \quad 
+ \frac{1}{3} \, \e^{-\tfrac{m^2}{H}\left|t_1-t_2 \right|} + \biggl(7 + \frac{2m^2}{3H} \Bigl( 3|t_1 - t_2| - 4 (t_1 + t_2) \Bigr) \biggr) \, \e^{-\tfrac{m^2}{3H}\left(t_1+t_2 \right)}  \\ \nonumber 
&  \quad 
+ \biggl(5 + \frac{4m^2}{3H} |t_1 - t_2| \biggr) \, \e^{-\tfrac{m^2}{3H}\bigl(2\left|t_1-t_2 \right| + \left(t_1 +t_2 \right)\bigr)} - \biggl(5 + \frac{4m^2}{3H} (t_1 + t_2) \biggr) \, \e^{-\tfrac{m^2}{3H}\bigl(\left|t_1-t_2 \right| + 2\left(t_1 +t_2 \right)\bigr)} \\ \nonumber 
&  \quad 
- \frac{1}{3} \, \e^{-\tfrac{m^2}{H}\left(t_1 +t_2 \right)} + \frac{1}{2} \, \e^{-\tfrac{m^2}{3H}\bigl(3\left|t_1-t_2 \right| + 2\left(t_1 +t_2 \right)\bigr)} - \frac{1}{2} \, \e^{-\tfrac{m^2}{3H}\bigl(2\left|t_1-t_2 \right| + 3\left(t_1 +t_2 \right)\bigr)} \\ \nonumber 
& \quad 
-  \biggl(\frac{15}{2} - \frac{4m^2}{3H} \Bigl( |t_1- t_2| - (t_1 + t_2) \Bigr) \biggr) \, \e^{\tfrac{m^2}{3H}\bigl(\left|t_1-t_2 \right| - 2\left(t_1 +t_2 \right)\bigr)} - \frac{1}{2} \, \e^{\tfrac{m^2}{3H}\bigl(2\left|t_1-t_2 \right| -3\left(t_1 +t_2 \right)\bigr)} \Biggr) \\ 
\label{sunset_latetimes}
& 
\xrightarrow[\,\,\, \text{times}\,\,\,]{\text{late}} \, \frac{243 \lambda^2 H^{12}}{1024 \, \pi^6 m^{10}} \Biggl(\biggl(1 + \frac{2m^2}{3H}|t_1 - t_2| \biggr) \, \e^{-\tfrac{m^2}{3H}\left|t_1-t_2 \right|} + \frac{1}{3} \, \e^{-\tfrac{m^2}{H}\left|t_1-t_2 \right|} \Biggr);
\\ \label{sunset_massless}
& 
\xrightarrow[t_2 \leq t_1]{\,\,\,\,\, m \rightarrow 0 \,\,\,\,\,} \, \frac{\lambda^2 H^{7}}{1920 \, \pi^6 } \Bigl( 5\, t_1^4 \, t_2 + 3\, t_2^5\Bigr).
\end{align}

In the last case of the third structure~\eqref{3_twoLoop_structure}, we have the closed loops attached to both the internal points. There is also the correlation function connecting these two points, and there are, as in the case of the first structure, two lines connecting the moments $t^\prime$ and $t^{\prime\prime}$ with the external fields $\tilde{\phi}(t_1)$ and $\tilde{\phi}(t_2)$. In the fifth structure~\eqref{5_twoLoop_structure}, we also have two closed loops attached to both the internal points $t^\prime$ and $t^{\prime\prime}$, and, hence, we obtain again two independent loops, or the so-called "Double Seagull" diagram; see figure~\ref{picture_Two_Loop}c. After computation, we get for this contribution
\begin{align}\label{seagull}
& \bigl\langle \phi \left(t_1\right) \phi \left(t_2\right) \bigr\rangle^\text{Ind. Loops}_{2-\text{loop}}  = \, \mathcal{I}_{3, \text{a}} + \, \mathcal{I}_{3, \text{b}} + \, \mathcal{I}_{5} = \\ \nonumber 
& 
= \frac{243 \, \lambda^2 H^{12}}{1024 \, \pi^6 m^{10}} \Biggl(\biggl(2 + \frac{2m^2}{3H}|t_1 - t_2| + \frac{m^4}{9H^2}|t_1 - t_2|^2 \biggr) \, \e^{-\tfrac{m^2}{3H}\left|t_1-t_2 \right|}  \\ \nonumber 
& \quad 
+ \biggl(\frac{m^2}{3H} \Bigl( |t_1 - t_2| - 2 (t_1 + t_2) \Bigr) + \frac{m^4}{9H^2} \Bigl( |t_1 - t_2|^2 - 2 (t_1 + t_2)^2 \Bigr) \biggr) \, \e^{-\tfrac{m^2}{3H}\left(t_1+t_2 \right)} \\ \nonumber 
&  \quad 
+ \, \frac{1}{4} \, \e^{-\tfrac{m^2}{3H}\bigl(3\left|t_1-t_2 \right| + 2\left(t_1 +t_2 \right)\bigr)} + \biggl(1 + \frac{m^2}{3H} |t_1 - t_2| \biggr) \, \e^{-\tfrac{m^2}{3H}\bigl(2\left|t_1-t_2 \right| + \left(t_1 +t_2 \right)\bigr)}  - \frac{1}{2} \, \e^{-\tfrac{m^2}{H}\left(t_1+t_2 \right)} \\ \nonumber 
&  \quad  
- \biggl(\frac{1}{2} - \frac{m^2}{3H} \Bigl( |t_1 - t_2| - 2 \, (t_1 + t_2) \Bigr) \biggr) \, \e^{-\tfrac{m^2}{3H}\bigl(\left|t_1-t_2 \right| + 2\left(t_1 +t_2 \right)\bigr)} - \frac{1}{4} \, \e^{-\tfrac{m^2}{3H}\bigl(2\left|t_1-t_2 \right| + 3\left(t_1 +t_2 \right)\bigr)} \\ \nonumber 
& \quad 
- \biggl(\frac{7}{4} - \frac{m^2}{3H} \Bigl( |t_1- t_2| - 2\, (t_1 + t_2) \Bigr) \biggr) \, \e^{\tfrac{m^2}{3H}\bigl(\left|t_1-t_2 \right| - 2\left(t_1 +t_2 \right)\bigr)} - \frac{1}{4} \, \e^{\tfrac{m^2}{3H}\bigl(2\left|t_1-t_2 \right| -3\left(t_1 +t_2 \right)\bigr)} \Biggr) 
\end{align} 
\begin{align} \label{seagull_latetimes}
& 
\xrightarrow[\,\,\,\,\, \text{times}\,\,\,\,\,]{\text{late}} \,\,\, \frac{243 \lambda^2 H^{12}}{1024 \, \pi^6 m^{10}} \biggl(2 + \frac{2m^2}{3H}|t_1 - t_2| + \frac{m^4}{9H^2} |t_1 - t_2|^2 \biggr) \, \e^{-\tfrac{m^2}{3H}\left|t_1-t_2 \right|}; \\ \label{seagull_massless}
& 
\xrightarrow[t_2 \leq t_1]{\,\,\,\,\, m \rightarrow 0 \,\,\,\,\,} \,\,\, \frac{\lambda^2 H^{7}}{64 \, \pi^6 } \biggl(\, \frac{1}{8} \, t_1^4 \, t_2 + \frac{1}{12}\, t_1^2 \, t_2^3  + \frac{7}{120} \, t_2^5 \, \biggr).
\end{align} 

Our outcomes at the late-time limit~\eqref{snowman_latetimes}, \eqref{sunset_latetimes}, and \eqref{seagull_latetimes} precisely coincide with those obtained using the Schwinger{--}Keldysh diagrammatic technique for a massive scalar field in~\cite{2022EPJC...82..345K}, while massless limits~\eqref{snowman_massless}, \eqref{sunset_massless}, and \eqref{seagull_massless} evaluated at equal times coincide with~\cite{2020PhRvD.102f5010K}. For sure, the results at the one-loop level also match at late~times. Our full expressions~\eqref{snowman},~\eqref{sunset}, and~\eqref{seagull} for each diagram contribution differ from~\cite{2022EPJC...82..345K}. That is due to the initial choice of the de Sitter-invariant Bunch{--}Davies vacuum state for a massive scalar field made in~\cite{2022EPJC...82..345K}. Nonetheless, by virtue of our "building block"~\eqref{correlator_massive_free}, the obtained results tend to the equilibrium, de Sitter-invariant state at the late-time limit and turn out to be in agreement with~\cite{2022EPJC...82..345K}.

Hence, instead of developing the diagrammatic technique, one can divine the topology from integral structures arising in the Yang{--}Feldman-type iterated equation. It is possible to proceed with this reasoning up to higher-order corrections. 
In the case of the four-point correlation function, there is another assignment between the notion of the loops and $\lambda$ orders: the connected contribution from the linear~$\lambda$ order in~\eqref{four_point_OneLoop_perm} belongs to the single vertex diagram, while the connected contribution from~\eqref{four_point_TwoLoop} refers to the one-loop contribution.

\section{Comparison with the Starobinsky{--}Yokoyama stochastic approach and the Hartree{--}Fock approximation}\label{Comparison_chapter}

Since the obtained massive correlation function~\eqref{correlator_massive_free} already coincides with Ornstein{--}Uhlenbeck mean-reverting stochastic process, our results must be in agreement at the late-time limit with the results of Starobinsky{--}Yokoyama stochastic approach~\cite{Starobinsky:1986fx,1994PhRvD..50.6357S}, which operates with a near equilibrium state. Nevertheless, in this section, we present the basic expressions of this approach for the sake of completeness, deduce the recurrent expression for any n-point correlation function, and also compare our results with the Hartree-Fock approximation.

\subsection{Starobinsky{--}Yokoyama stochastic approach and its recurrent formula}
Starobinsky's stochastic approach~\cite{Starobinsky:1986fx,1994PhRvD..50.6357S} matches the long-wavelength part of the quantum field $\phi(t,\x)$ to the classical stochastic field~$\varphi(t,\x)$ with a probability distribution function $\rho\bigl[\varphi(t,\x)\bigr]$ that satisfies the Fokker{--}Planck equation: 
\begin{equation}\label{Fokker-Planck}
\partial_t \, \rho\bigl[\varphi(t,\x)\bigr] = \frac{1}{3H} \, \partial_\varphi \Bigl(\rho\bigl[\varphi(t,\x)\bigr] V^\prime_{\varphi}\bigl(\varphi(t,\x)\bigr)\Bigr) + \frac{H^3}{8\pi^2} \, \partial^2_\varphi \Bigl( \rho\bigl[\varphi(t,\x)\bigr] \Bigr), 
\end{equation} where $V\bigl(\varphi(t,\x)\bigr)$ is the same potential that was considered in the quantum field theoretical approach. 
Any solution to the Fokker{--}Planck equation tends to the static one at late times
\begin{equation}\label{rho_lambda_small}
\rho\bigl[\varphi(t,\x)\bigr] \quad \xrightarrow[\,\,\, \text{times}\,\,\,]{\text{late}} \quad \rho_{\text{st}} \left[\varphi\right] = \frac{1}{\mathcal{N}} \,\, \e^{-\tfrac{8\pi^2}{3H^4}V\left(\varphi\right)}. 
\end{equation}
Therefore, the expectation values of this stochastic variable $\bigl\langle \varphi^{2} \bigr\rangle$ and $\bigl\langle \varphi^{4} \bigr\rangle$ calculated at the small self-interaction coupling constant $\lambda$ expansion with~\eqref{rho_lambda_small} are
\begin{equation} \bigl\langle \varphi^{\, 2\n} \bigr\rangle = 
\,\,\,  \begin{cases} \,\,\, \dfrac{3H^4}{8\pi^2m^2}-\dfrac{27\lambda H^8}{64 \, \pi^4 m^6}+\dfrac{81\lambda^2 H^{12}}{64\, \pi^6 m^{10}}-\dfrac{24057\lambda^3 H^{16}}{4096\, \pi^8 m^{14}} + O\left( \lambda^4 \right), & \n=1; \\ & \\ \,\,\, \dfrac{27H^8}{64 \, \pi^4 m^4}-\dfrac{81\lambda H^{12}}{64 \, \pi^6 m^8}+\dfrac{24057\lambda^2 H^{16}}{4096\, \pi^8 m^{12}} + O\left( \lambda^3 \right) \, , & \n=2 \, ; \end{cases} \end{equation}
which exactly coincide with~\eqref{loop_series_t_inf} and~\eqref{stochastic_limit_Four_point_function} correspondingly; see~\eqref{rho_approx}{--}\eqref{phi4_approx}. 

Another way to avoid relying on straightforward integration is to reduce the expressions to the modified Bessel functions of the second kind $\mathcal{K}_\nu (z)$. As it was shown in~\cite{2012JCAP...10..052E,2022EPJC...82..345K}, 
\begin{equation}\label{N_def}
\mathcal{N}  = \int\limits_{-\infty}^{+\infty} d \varphi \, \e^{-\tfrac{8\pi^2}{3H^4}\left(\tfrac{m^2 \varphi^2}{2} + \tfrac{\lambda \varphi^4}{2} \right)}  = \frac{m}{\sqrt{2\lambda}} \, \text{exp}\left( \frac{\pi^2 m^4}{3 \lambda H^4} \right) \mathcal{K}_{1/4} \left( \frac{\pi^2 m^4}{3 \lambda H^4} \right).
\end{equation} One can notice that\footnote{Let us also point out that this expression resembles the exact zero-modes’ propagator in Euclidean de Sitter space~\cite{2013PhRvD..87f4018B}.} 
\begin{equation}\label{phi2_z}
\bigl\langle \varphi^2  \bigr\rangle =  - \frac{3H^4}{4 \pi^2} \frac{1}{\mathcal{N}} \frac{d \mathcal{N}}{d m^2} \quad \Rightarrow \quad \bigl\langle \varphi^2  \bigr\rangle = \frac{m^2}{2\lambda} \Biggl( \frac{\mathcal{K}_{3/4}(z)}{\mathcal{K}_{1/4}(z)} \, - 1 \Biggr),  \quad \text{where} \quad z \equiv \frac{\pi^2m^4}{3\lambda H^4},
\end{equation}
and we have used the recurrence relation~\cite{1994tisp.book.....G}:
\begin{equation}\label{recurrence_Bessel}
\frac{d}{d z} \mathcal{K}_\nu (z) = - \frac{\nu}{z}
\, \mathcal{K}_\nu (z) - \mathcal{K}_{\nu-1} (z) \qquad \text{and} \qquad \mathcal{K}_{-\nu} (z) = \mathcal{K}_\nu (z). \end{equation} 
One can proceed further and get the expression anew with the help of~\eqref{recurrence_Bessel} for $\bigl\langle \varphi^4  \bigr\rangle$ as
\begin{equation}\label{phi4}
\bigl\langle \varphi^4  \bigr\rangle = \left( - \frac{3H^4}{4 \pi^2} \right)^2 \frac{1}{\mathcal{N}} \, \frac{d^{\,2} \mathcal{N}}{d \left(m^2\right)^2} \qquad \Rightarrow \qquad  \bigl\langle \varphi^4  \bigr\rangle = \frac{3H^4}{8\pi^2\lambda} + \frac{m^4}{2\lambda^2} \Biggl( 1- \frac{\mathcal{K}_{3/4}(z)}{\mathcal{K}_{1/4}(z)}\Biggr). 
\end{equation}

Let us also point out the general structure of any 2n'th expectation value coming from this reasoning. Thanks to~\eqref{recurrence_Bessel}, one will always have 
\begin{equation}
\frac{d^{\, \n} \mathcal{N} \,\,\,}{d \left(m^2\right)^\n} = \alpha_\n 
\, \e^z \, \mathcal{K}_{1/4}(z) + \beta_\n 
\, \e^z \, \mathcal{K}_{3/4}(z),
\end{equation} where from definition~\eqref{N_def} $\alpha_0 = m/\sqrt{2\lambda} \, $ and $\beta_0 =0$, while
\begin{equation}
\alpha_1 = -\beta_1 = \frac{\sqrt{2} \, \pi^2 m^3}{3\, \lambda^{3/2} H^4} \qquad  \text{and} \qquad 
\alpha_2 = \frac{\sqrt{2}\, \pi^2 m}{3 \lambda^{3/2} H^4} + \frac{4\sqrt{2}\, \pi^4 m^5}{9 \lambda^{5/2} H^8}; \quad \beta_2 = - \frac{4\sqrt{2}\, \pi^4 m^5}{9 \lambda^{5/2} H^8}.
\end{equation} Consequently, the defined coefficients satisfy the recurrence relations
\begin{equation}\begin{gathered}\label{recurrence_coeff}
\alpha_{\n+1} = \frac{d \alpha_\n \,\,}{dm^2} - \frac{\alpha_\n }{2m^2} + \frac{2\pi^2m^2 }{3\lambda H^4} \bigl( \alpha_\n - \beta_\n \bigr), \\ 
\beta_{\n+1} = \frac{d \beta_\n \,\,\,}{dm^2} - \frac{3 \beta_\n }{2m^2} - \frac{2\pi^2m^2 }{3\lambda H^4} \bigl( \alpha_\n - \beta_\n \bigr)
\end{gathered} \end{equation}
and result in 
\begin{equation}\label{n-point}
\bigl\langle \varphi^{2\n}  \bigr\rangle = \left( - \frac{3H^4}{4 \pi^2} \right)^\n \frac{1}{\mathcal{N}} \frac{d^{\, \n} \mathcal{N}}{d \left(m^2\right)^\n} \,\,\Rightarrow \,\,  \bigl\langle \varphi^{2\n}  \bigr\rangle = \frac{\sqrt{2\lambda}}{\,\,m} \left( - \frac{3H^4}{4 \pi^2} \right)^\n  \Biggl( \alpha_\n 
+ \beta_\n 
\, \frac{\mathcal{K}_{3/4}(z)}{\mathcal{K}_{1/4}(z)}\Biggr).
\end{equation}
Naturally, at large $z\gg1$, an expansion of the obtained expressions~\eqref{phi2_z} and~\eqref{phi4} becomes
\begin{align}
\bigl\langle \varphi^2 \bigr\rangle &= \,\,\, \frac{m^2}{2\lambda} \Biggl( \frac{K_{3/4}(z)}{K_{1/4}(z)} - 1 \Biggr) \,\,\, \rightarrow  \,\,\, \frac{m^2}{8\lambda z} -  \frac{3m^2}{64\lambda z^2} +  \frac{3m^2}{64\lambda z^3} - \frac{297m^2}{4096\lambda z^4} + \, ...  \,\, ; \\ 
\bigl\langle \varphi^4 \bigr\rangle &= \, \frac{3H^4}{8\pi^2\lambda} + \frac{m^4}{2\lambda^2} \Biggl( 1- \frac{K_{3/4}(z)}{K_{1/4}(z)}\Biggr) \,\, \rightarrow \,\, \frac{3H^4}{8\pi^2\lambda} -  \frac{m^4}{8\lambda^2 z} +  \frac{3m^4}{64\lambda^2 z^2} \\ \nonumber 
& \qquad \qquad \qquad \qquad \qquad \qquad \qquad \qquad \quad \,\,\,\, - \frac{3m^4}{64\lambda^2 z^3} \, + \frac{297m^4}{4096\lambda^2 z^4} + \, ... \,\, ; 
\end{align} and it reproduces with assigned $z$ through~\eqref{phi2_z} our series~\eqref{loop_series_t_inf} and~\eqref{stochastic_limit_Four_point_function} obtained above.

One more comparison of our perturbative results can be done with the stochastic expectation value of
\begin{align}
\label{massles_stochastic_Woodard}
\bigl\langle \varphi^{2\n} \left(t\right) \bigr\rangle
= \bigl( 2\n-1 & \bigr) !! \left(\frac{H^3 t}{4\pi^2}\right)^\n 
\Biggl( 1 - \frac{\lambda \, \n }{12 \pi^2} \, \bigl(\n+1\bigr) \, H^2 t^2 
\nonumber \\ & 
+ \frac{\lambda^2 \, \n}{10080 \, \pi^4} \, \Bigl(35 \, \n^3 + 170 \, \n^2 + 225 \, \n + 74 \Bigr) \, H^4 t^4 \\ \nonumber 
& 
- \frac{\, \lambda^3 \, \n \, \bigl(\n+1\bigr) \bigl(\n+2\bigr)}{362880 \, \pi^6} \, \Bigl(35 \, \n^3 + 300 \, \n^2 + 685 \, \n + 252 \Bigr) \, H^6 t^6\Biggr) +  O \bigl(\lambda^4\bigr),
\end{align} 
extracted in~\cite{2005NuPhB.724..295T} using the Fokker{--}Planck equation. Here, we broadened this expression up to the~next $\lambda^3$ order through the same reasoning as in~\cite{2005NuPhB.724..295T}; see details in appendix~\ref{Woodard}. Our massless limits of two- and four-point correlation functions at equal times,~\eqref{loop_series_massless_coincide} and~\eqref{massle_four_point_full}, are identical to~\eqref{massles_stochastic_Woodard}. Following this approach, one can also trace out the perturbative series on $m^2$; 
see~\eqref{Woodard_pert_series}{--}~\eqref{massive_pert_series_Woodard_4pt}. Through that, it is clear our $ \bigl\langle \phi^2 (t) \bigr\rangle_0$ and~$ \bigl\langle \phi^4(t) \bigr\rangle_0$ in~\eqref{TwoPoint_equal_times_full_series} and~\eqref{four_point_tree_equal_times} are precisely the resummed expressions of these series.

\subsection{Hartree-Fock approximation}
To compare our outcomes for $\bigl\langle \phi^2 \left(t\right) \bigr\rangle$ to those obtained in the Hartree-Fock (Gaussian) approximation, 
one can consider the Klein-Gordon equation 
\begin{equation}
\Box \, \phi(t) = -m^2 \phi(t) - \lambda \phi^3(t).
\end{equation} Multiplying both sides of the equation above by $\phi(t)$, integrating the left-hand  side by parts, taking expectation values of the field operators, and using the Hartree-Fock (Gaussian) approximation, namely $\bigl\langle \phi^4 \left(t\right) \bigr\rangle= 3\bigl\langle \phi^2 \left(t\right) \bigr\rangle^2$, one arrives at
\begin{equation}\label{Hartree-Fock_1}
\frac{1}{2} \, \Box  \,  \bigl\langle \phi^2 \bigr\rangle - \bigl\langle \phi^{, \mu} \phi_{, \mu} \bigr\rangle = -m^2 \bigl\langle \phi^2 \bigr\rangle - 3 \lambda \bigl\langle \phi^2  \bigr\rangle^2.
\end{equation} 
In the case of a massless and non-interacting scalar field, r.h.s.~\eqref{Hartree-Fock_1} $=0$, the contribution to the long-wave part of l.h.s.~\eqref{Hartree-Fock_1}, which is $3H \partial_t$, prevails, and $\Box \, \bigl\langle \phi^2_0 \left(t\right) \bigr\rangle = 3H^4/4 \pi^2$. We dare to assume that in the case of the small mass and self-interaction coupling constant, this approximation is still correct
\begin{equation}\label{HFA_equation}
\frac{d \bigl\langle \phi^2(t) \bigr\rangle}{dt} = \frac{H^3}{4\pi^2} - \frac{2m^2}{3H} \bigl\langle \phi^2 (t)\bigr\rangle - \frac{2\lambda}{H} \bigl\langle \phi^2 (t) \bigr\rangle^2.
\end{equation}
The solution to the equation above is 
\begin{equation}\label{Hartree_Fock_solution}
\bigl\langle \phi^2 \left(t\right) \bigr\rangle
= \frac{\dfrac{3H^4}{4 \pi^2 m^2} \left(1-\text{exp}\left(-\dfrac{2m^2 t}{3H}\sqrt{1+\dfrac{9\lambda H^4}{2\pi^2 m^4}}\, \right)\right)}{1+\sqrt{1+\dfrac{9\lambda H^4}{2\pi^2 m^4}}- \left(1-\sqrt{1+\dfrac{9\lambda H^4}{2\pi^2 m^4}} \right) \text{exp}\left( -\dfrac{2m^2t}{3H}\sqrt{1+\cfrac{9\lambda H^4}{2\pi^2 m^4}} \right)}; 
\end{equation} see details in appendix~\ref{HF_appendix}. 

Expanding this solution in the series along a small coupling constant $\lambda$, we get
\begin{align}
\bigl\langle \phi^2 \left(t\right) \bigr\rangle \equiv \bigl\langle \phi^2 \left(t\right) \bigr\rangle_{\text{HF}} \xrightarrow[\quad ]{} \, & \, \frac{3H^4}{8\pi^2 m^2} \left( 1- \e^{-\tfrac{2m^2t}{3H}}\right) - \frac{27 \lambda H^8}{64 \, \pi^4 m^6} \left( 1- \frac{4m^2 t}{3H} \, \e^{-\tfrac{2m^2t}{3H}} - \e^{-\tfrac{4m^2t}{3H}}\right) \,\,\, \nonumber \\
& \,\, + \frac{243 \lambda^2 H^{12}}{512 \, \pi^6 m^{10}} \Biggl( 2 + \left(1 - \frac{4m^2 t}{3H} \, - \frac{8m^4 t^2}{9H^2}\right) \, \e^{-\tfrac{2m^2t}{3H}} \\ \nonumber 
& \qquad \qquad \qquad \quad \,\, - \left(2 + \frac{8m^2 t}{3H} \right) \, \e^{-\tfrac{4m^2t}{3H}} - \,  \e^{-\tfrac{2m^2t}{H}}\Biggr) + O \left( \lambda^3\right).
\end{align} 
At the tree and at the one-loop levels, this approximation gives correct results~\eqref{TwoPoint_equal_times_full_series}, while already at the two-loop level it does not. Through our outcomes for each of the two-loop diagrams in the previous section~\ref{Correspondence_chapter}, one can notice that the Hartree-Fock (Gaussian) approximation only resums the "Snowman" diagram~\eqref{snowman} and "Double Seagull"~\eqref{seagull}, leaving aside the "Sunset"~\eqref{sunset} one. The same statement on the ''Cactus''-type diagrams holds for the three-loop result.

\section{An autonomous equation for the two-point correlation function}\label{Autonomous_chapter} 
As we faced an absentee "Sunset" diagram contribution, in this section we go beyond the Hartree{--}Fock (or Gaussian) approximation. Here, we construct through the known perturbative series an autonomous first-order differential equation. The solution of such an equation is the non-analytic in the self-interaction coupling constant $\lambda$ while providing the correct perturbative series up to $O(\lambda^3)$, including the "Sunset" diagram. Previously, an autonomous equation for a massless scalar field in de Sitter space was elaborated in~\cite{2020PhRvD.102f5010K}, and we follow that~spirit.

The idea is to get an equation such that its expanded solution reproduces the correct perturbative series up to the two-loop level. We will rely on our outcomes from section~\ref{Y-F_chapter}, i.e., series~\eqref{TwoPoint_equal_times_full_series}. At the tree level from~\eqref{TwoPoint_equal_times_full_series} one has
\begin{equation}\label{aut_zero_sol}
\bigl\langle \phi^2 \left(t\right) \bigr\rangle \equiv f(t) 
= \frac{3H^4}{8\pi^2 m^2} \left( 1- \e^{-\tfrac{2m^2t}{3H}}\right).
\end{equation}
This tree-level expression above is a solution to the following autonomous equation:
\begin{equation}\label{aut_zero_eq}
\frac{d}{dt} \bigl(f (t)
\bigr)= \frac{H^3}{4\pi^2} - \frac{2m^2}{3H}f \, (t) 
\end{equation} One takes the time derivatives from both l.h.s.~and~r.h.s. of~\eqref{aut_zero_sol} and extract the exponent through $f(t)
$ to establish that. 

Hereafter, we repeat this prescription in order to find an autonomous equation at the one-loop level. We replace the function in~\eqref{aut_zero_eq} with $f(t)
=\bigl\langle \phi^2 \left(t\right) \bigr\rangle$, taken up to 
linear order with respect to the self-interaction coupling constant $\lambda$. In that case, comparing again l.h.s. and~r.h.s. of~\eqref{aut_zero_sol}, the correcting term $\Delta f_1 (t)
$ needs to be added
\begin{equation} \begin{gathered}
\frac{d}{dt} \bigl(f(t)
\bigr)= \frac{H^3}{4\pi^2} - \frac{2m^2}{3H}\, f(t)
+ \Delta f_1 (t)
, \quad \text{where} \qquad \qquad \\
\Delta f_1 (t)
= - \frac{ 9 \lambda H^7}{32 \, \pi^4 m^4} \biggl(1 - \e^{-\tfrac{2m^2 t}{3H}}\biggr)^2 = - \frac{2\lambda}{H} \bigl(f(t)
\bigr)^2 . 
\end{gathered} \end{equation}
As expected, we already know what this equation is: that is~\eqref{HFA_equation} in the Hartree{--}Fock approximation above, and the solution to such an autonomous equation is~\eqref{Hartree_Fock_solution}. 

Proceeding with that routine, one is also able to find the correction term for the two-loop level, namely,
\begin{align}\label{aut_second_eq}
\frac{d}{dt} \bigl(f(t)
\bigr) = \frac{H^3}{4\pi^2} & - \frac{2m^2}{3H}  \, f(t)
- \frac{2\lambda}{H} \, \bigl(f(t)
\bigr)^2  -\frac{27\lambda^2 H^{11}}{16\, \pi^6 m^8}\Biggl(\frac{2 \pi^2m^2}{H^4} \, f(t)
- \frac{8 \pi^4m^4}{H^8} \bigl(f(t)
\bigr)^2  \\ \nonumber
&  + \, \frac{128 \, \pi^6m^6}{27 H^{12}} \, \bigl(f(t)
\bigr)^3 + \frac{3}{4} \biggl(1 - \, \frac{8\pi^2 m^2}{3H^4}\, f(t)
\biggr)^2 \text{ln}\biggl(1 - \, \frac{8\pi^2 m^2}{3H^4}\, f(t)
\biggr) \Biggr).
\end{align}
This autonomous equation above in the massless limit reduces to its analog from the preceding paper~\cite{2020PhRvD.102f5010K}:
\begin{equation}
\frac{d}{dt} \bigl(f(t)
\bigr) =  \frac{H^3}{4\pi^2} - \frac{2\lambda}{H} \, \bigl(f(t)
\bigr)^2 + \frac{16 \, \pi^2 \lambda^2}{3H^5} \, \bigl(f(t)
\bigr)^4.
\end{equation}

For our purposes, to catch the contribution of the "Sunset" diagram, it is sufficient to represent a solution to the obtained equation~\eqref{aut_second_eq} in the following form:
\begin{equation}
f(t)
= \bigl\langle \phi^2 \left(t\right) \bigr\rangle_{\text{HF}} + \delta f (t)
\equiv \frac{3H^4}{4\pi^2m^2} \, f_{\mathcal{Z}}(t)
+ \delta f (t), 
\end{equation} where we have introduced \begin{equation}\label{fZ_Z_z}
f_{\mathcal{Z}} (t)
= \cfrac{1-\e^{-\tfrac{2m^2}{3H}\mathcal{Z}t}}{1+\mathcal{Z}- \left(1-\mathcal{Z} \right) \e^{-\tfrac{2m^2}{3H}\mathcal{Z}t}} \quad \,\, \text{with} \quad \,\, \mathcal{Z} \equiv \sqrt{\, 1+\frac{3}{2z} \,\,} \, ; \quad 
z \stackrel{ \,\eqref{phi2_z} \,}{=}\frac{\pi^2 m^4}{3\lambda H^4}.
\end{equation} Suppose that the correcting 
term in our autonomous equation~\eqref{aut_second_eq} is to be small 
then its linearized version is
\begin{align}
\label{linear_correction2}
\frac{d}{dt} \Bigl(\delta f (t)
\Bigr) & = - \biggl( \frac{2m^2}{3H}  +\frac{3\lambda H^3}{\pi^2 m^2} f_{\mathcal{Z}}(t)
\biggr) \,  \delta f (t) 
\\ \nonumber 
& \qquad \,\,\, -  \frac{ 81\lambda^2 H^{11}}{64\, \pi^6 \, m^8}  
\Biggl(  2\, f_{\mathcal{Z}}  (t)
- 6\,  f^{\,2}_{\mathcal{Z}}  (t)
+ \frac{8}{3} \, f^{\,3}_{\mathcal{Z}} (t)
+ \Bigl(1-2f_{\mathcal{Z}} (t)
\Bigr)^2 \text{ln}\Bigl(1-2f_{\mathcal{Z}} (t)
\Bigr) \Biggr).
\end{align}
Further, the solution to this equation above, with assigned $\mathcal{Z}_{\pm} \equiv \mathcal{Z} \pm 1$ to present the expression in a more compact form, is
\begin{align}\label{Full_aut}
\delta f (t) = & \, \frac{243 \, \lambda^{2} H^{12}}{32 \, \pi^{6} m^{10}} \, \biggl(\mathcal{Z}_{+}+\mathcal{Z}_{-}\, \e^{-\tfrac{2 m^{2} }{3 H} \mathcal{Z} t}\biggr)^{-2} \Biggl(-\frac{3 \mathcal{Z}^{2}-3 \mathcal{Z}-2}{6 \mathcal{Z} \mathcal{Z}_{+}} \\ \nonumber 
& -\frac{\mathcal{Z}_{-}^{2}}{4 \mathcal{Z}} \, \ln\left(\frac{\mathcal{Z}_{-}+ \mathcal{Z}_{+} \, \e^{-\tfrac{2 m^{2}}{3 H}\mathcal{Z} t}}{\mathcal{Z}_{+}+\mathcal{Z}_{-}\,\e^{-\tfrac{2 m^{2}}{3 H}\mathcal{Z} t}}\right) +
\left(\frac{\mathcal{Z}_{+}\mathcal{Z}_{-}}{2 \mathcal{Z}} \Biggl(\mathrm{Li}_2\left(1+\frac{\mathcal{Z}_{-}}{\mathcal{Z}_{+}}\e^{-\tfrac{2 m^{2}}{3 H}\mathcal{Z} t}\right) \right. \\ \nonumber 
& -\mathrm{Li}_2\left(1+\frac{\mathcal{Z}_{+}}{\mathcal{Z}_{-}}\e^{-\tfrac{2 m^{2}}{3 H}\mathcal{Z} t}\right) - \mathrm{Li}_2\left(\frac{2 \mathcal{Z}}{\mathcal{Z}_{+}}\right) + \mathrm{Li}_2\left(\frac{2 \mathcal{Z}}{\mathcal{Z}_{-}}\right)\Biggr) \\ \nonumber 
& -\frac{2 \left(\mathcal{Z}^{2}-3\right)\left(3 \mathcal{Z}^{2}+1\right)}{3 \left(\mathcal{Z}_{+}\mathcal{Z}_{-}\right)^{2}} \, \ln\left(\frac{\mathcal{Z}_{+}+\mathcal{Z}_{-}\,\e^{-\tfrac{2 m^{2}}{3 H}\mathcal{Z} t}}{2\mathcal{Z}}\right) -\frac{m^{2} t}{3 H} \mathcal{Z}_{+}\mathcal{Z}_{-}\, \ln\left(\frac{\mathcal{Z}_{-}}{\mathcal{Z}_{+}}\right) \\ 
\nonumber 
& -\frac{2 m^{2} t \left(3 \mathcal{Z}^{2}+1\right)}{9 H \left(\mathcal{Z}_{+}\mathcal{Z}_{-}\right)^{2}} \left(\mathcal{Z}^{3}-3 \mathcal{Z}+2\right) \, + \, \frac{3 \mathcal{Z}^{2}+1}{3 \mathcal{Z}_{+}\mathcal{Z}_{-}} \, \Biggr)\e^{-\tfrac{2 m^{2}}{3 H}\mathcal{Z} t} \\ \nonumber &   +\Biggl(-\frac{3 \mathcal{Z}^{2}+3 \mathcal{Z}-2}{6 \mathcal{Z} \mathcal{Z}_{-}} \left.+\frac{\mathcal{Z}_{+}^{2}}{4 \mathcal{Z}} \, \ln\Biggl(\frac{\mathcal{Z}_{-}+\mathcal{Z}_{+}\,\e^{-\tfrac{2 m^{2}}{3 H}\mathcal{Z} t}}{\mathcal{Z}_{+}+\mathcal{Z}_{-}\,\e^{-\tfrac{2 m^{2}}{3 H}\mathcal{Z} t}}\Biggr)\Biggr)\e^{-\tfrac{4 m^{2}}{3 H}\mathcal{Z} t}\right).
\end{align} 
see for details~\eqref{linear_W}{--}\eqref{first_mult}, and $\text{Li}_2$ is the dilogarithm function.
Immediately, in the small~$\lambda$ regime, we have precisely the contribution of the 'Sunset' diagram~\eqref{sunset} at equal times, while the limiting value at $t \rightarrow \infty$ from the first line of the obtained expression~\eqref{Full_aut} reads 
\begin{align}
\delta f (t)
& \xrightarrow[]{\,\,\, t\rightarrow \infty \,\,\,}  \,\, \frac{243 \, \lambda^2 H^{12}}{32 \,\pi^6 m^{10}} \left( -\frac{3\mathcal{Z}^2 - 3\mathcal{Z} - 2}{6\mathcal{Z} \bigl(\mathcal{Z}+1 \bigr)^3} - \frac{\bigl(\mathcal{Z}-1\bigr)^2}{4\mathcal{Z}\bigl(\mathcal{Z}+1 \bigr)^2} \,\, \text{ln}\left(\frac{\mathcal{Z}-1}{\mathcal{Z}+1}\right) \right) \\ & \qquad \,\, \approx \,\,\, \frac{81 \lambda^2 H^{12}}{256 \, \pi^6 m^{10}} + O \left( \lambda^3 \right). \qquad \qquad \qquad \qquad \qquad \qquad 
\end{align}
That matches~\eqref{sunset_latetimes}, which was absent in the Hartree{--}Fock (Gaussian) approximation. 

The full 
result for the two-point correlation function 
at the late-time limit is
\begin{align}\label{Aut_late}
& \bigl\langle \phi^2 \left(t \right) \bigr\rangle^{\text{aut}}  = \bigl\langle \phi^2 \left(t\right) \bigr\rangle_{\text{HF}} + \delta f (t)
\,\, \xrightarrow[]{ \, t \, \rightarrow \, \infty \,} \frac{H^2}{\pi \sqrt{\lambda}} \left( \frac{\sqrt{12z+18} - \sqrt{12z}}{12} \right. \\ \nonumber 
& 
+ \frac{\bigl(3\sqrt{4z^2+6z} - 2z - 9 \bigr)\bigl(\sqrt{12z+18} - \sqrt{12z} \bigr)^3}{1728z \sqrt{4z^2 + 6z}} 
+ \left.\frac{3\bigl(\sqrt{2z+3} - \sqrt{2z}\bigr)^4}{64z^2 \sqrt{12z + 18}} \text{ln}\left(\frac{\sqrt{2z+3} + \sqrt{2z}}{\sqrt{2z+3} - \sqrt{2z}}\right)\right) 
\\
& \qquad \qquad \,\, \xrightarrow[]{ \,\,\, m \, \rightarrow \, 0 \,} \,\, \frac{7 \sqrt{2} \, H^2}{ 24 \, \pi \sqrt{\lambda \,}\,} \, .  
\end{align} Expanding over the small self-interaction coupling constant $\lambda$, this expression naturally leads to~\eqref{loop_series_t_inf} up to~$O(\lambda^3)$. 

Besides, it almost coincides with the Starobinsky{--}Yokoyama stochastic approach result. Through expressions~\eqref{phi2_z} 
and~\eqref{Aut_late}, one can establish that
\begin{equation}\label{compare_aut_star}
\frac{\bigl\langle \phi^2 \bigr\rangle^{\text{aut}}_{ t \rightarrow \, \infty}}{\bigl\langle \varphi^2  \bigr\rangle_{\text{Stoch}}}
\,\,\, \xrightarrow[]{ \, z \, \rightarrow \, 0 \, } \,\,\, \frac{7\pi}{6\sqrt{6\,} \, \Gamma^2\left(\frac{3}{4} \right)} \approx 0.9964 \,  \qquad \text{and} \qquad \frac{\bigl\langle \phi^2 \bigr\rangle^{\text{aut}}_{ t \rightarrow \, \infty}}{\bigl\langle \varphi^2  \bigr\rangle_{\text{Stoch}}} \,\,\, \xrightarrow[]{ \, z \, \rightarrow \, \infty \, } \,\,\, 1. 
\end{equation} From some numerical estimations, this relation grows up to the maximal value, that is $\approx 1.0055$ at $z\approx 0.145$, then decreases to its asymptotic value. Therefore, the deviation amplitude $\bigl\langle \phi^2 \bigr\rangle^{\text{aut}}_{ t \rightarrow \, \infty}$ from the Starobinsky{--}Yokoyama stochastic approach does not exceed more than $0.6 \, \%$ in the whole interval of a new dimensionless parameter $0 \leq  \tfrac{\pi^2 m^4}{3\lambda H^4} < \infty$.

To conclude this section, let us note the autonomous equation technique was applied in \cite{Bhattacharya:2022aqi,Bhattacharya:2022wjl},
while a slightly different but close in the spirit resummation technique 
was implemented in~\cite{Bhattacharya:2023yhx,Bhattacharya:2023xvd,
Bhattacharya:2023jxo}.

\section{Conclusion and Outlook}\label{Conclusion}
Our present paper regards a rather particular theory~\eqref{S0} of a massive scalar field living on de Sitter background. In contrast to the standard theory of a massive scalar field based on the de Sitter-invariant vacuum, we develop the vacuum-independent reasoning that may not possess de Sitter invariance but results in a smooth massless limit of the correlation function's infrared part. We started from the known Yang{--}Feldman equation~\cite{2005NuPhS.148..108W,2005NuPhB.724..295T} for a massless scalar field~\eqref{Y-F_k0} and proposed a trick to "hang up" the mass there; see~\eqref{FreeMassiveField}{--}\eqref{Y-F-massive_iterated} and~\eqref{Y-F-type_general}{--}\eqref{trick_end}. Our elaboration allows us to calculate correlation functions of the free massive scalar field and to proceed with quantum corrections, relying only on the known two-point correlation function's infrared part of the free massless one. Such a development can be considered a theory of a massive scalar field with the vacuum "inherited" from a massless one. By employing Yang{--}Feldman-type equation~\eqref{Y-F-massive_iterated}, we computed the two-point correlation function up to $\lambda^3$ order and the four-point correlation function up to $\lambda^2$ order in the long-wavelength approximation; see section~\ref{Y-F_chapter} and appendix~\ref{Details}. 
Thereto, from our introduced relation~\eqref{FreeMassiveField}, the resulting two-point correlation function~\eqref{correlator_massive_free} for the free massive scalar field coincides with that of the Ornstein{--}Uhlenbeck stochastic process: the unique Gaussian and Markov process, which has a stationary state~\cite{Uhlenbeck:1930zz,1989fpem.book.....R,1994hsmp.book.....G}. The notable feature is its drift towards the average with the mean-reversion rate~$m^2/3H$. In our approach, correlation functions have massless limits, and initially one does not have de Sitter invariance in order to get that smooth transition between the massive and massless fields. Even so, by virtue of our two-point correlation function~\eqref{correlator_massive_free}, over time it tends to the equilibrium state, which depends only on the difference of times and turns out to be de Sitter-invariant. That attractor property is in agreement with findings from~\cite{1994PhRvD..50.6357S,2000PhRvD..62l4019A} and a more recent one~\cite{2019arXiv191100022G}.

Further, we established the correspondence between an iterated solution of the Yang{--}Feldman-type equation and the Schwinger{--}Keldysh diagrammatic technique in section~\ref{Correspondence_chapter}. Instead of computing diagrams, one can try to guess the corresponding topology from the integral structures arising in the Yang{--}Feldman formalism. Our calculation results are consistent at the late-time limit with the diagrammatic computations~\cite{2022EPJC...82..345K}.

Since the obtained massive correlation function~\eqref{correlator_massive_free} already coincides with the Ornstein{--}Uhlenbeck mean-reverting stochastic process, our results must be in agreement at the late-time limit with the Starobinsky{--}Yokoyama stochastic approach~\cite{Starobinsky:1986fx,1994PhRvD..50.6357S}, which operates with a near equilibrium state. 
Nevertheless, we compared our outcomes to those for the completeness of the paper. In addition, we deduced the recurrent expression for any n-point correlation function in the framework of the stochastic approach; see~\eqref{recurrence_coeff}~and~\eqref{n-point}. We have also shown that the Hartree-Fock (Gaussian) approximation gives the correct perturbative series at the tree and one-loop levels, while at the two-loop level it resums only the so-called "Snowman"~and "Double Seagull" diagrams, leaving aside the "Sunset" one. The same statement on the ''Cactus''-type diagrams resummation by the Hartree{--}Fock (or Gaussian) approximation holds for our three-loop results. 

Finally, we have constructed an autonomous equation for the two-point correlation function in section~\ref{Autonomous_chapter}, relying only on the obtained perturbative series~\eqref{TwoPoint_equal_times_full_series}. Integrating its approximate version, one can get the non-analytic expression with respect to the self-interaction coupling constant $\lambda$ that reproduces the correct perturbative series up to the two-loop level. At the late-time limit, our expression~\eqref{Aut_late} almost coincides with the Starobinsky{--}Yokoyama stochastic approach in the whole interval of a new dimensionless parameter~$0 \leq \tfrac{\pi^2 m^4}{3\lambda H^4} < \infty$; see~\eqref{compare_aut_star}. 

Let us end up with some prospects. Our trick for "hanging up" the mass could also be employed to the two-point correlation function of the free massless scalar field where the spatial arguments of the spacetime points do not coincide~\cite{2015PhRvD..91j3537O} for $t_1 \, \geq \, t_2$:
\begin{equation}\label{free_massless_noncoincide}
\bigl\langle \phi_0 \left(t_1, \x_1 \right) \phi_0 \left(t_2, \x_2 \right) \bigr\rangle 
= \frac{H^2}{4\pi^2} \left(H t_2 + \sum\limits_{\n=1}^\infty \frac{(-1)^\n H^{2\n} | \x_1 - \x_2|^{2\n}}{2 \n (2\n+1)!} \Bigl(\e^{2\n Ht_2} - 1 \Bigr)\right).
\end{equation}
Therefore, for the free massive field $\tilde{\phi}\left(t, \x\right)$ through our~\eqref{FreeMassiveField} and~\eqref{correlator_massive_free_details_1} one can get
\begin{align}\label{free_massive_noncoincide_2}
& \bigl\langle \tilde{\phi} \left(t_1 , \x_1 \right) \tilde{\phi} \left(t_2, \x_2 \right) \bigr\rangle  = \frac{3H^4}{8\pi^2m^2} \left( \e^{-\tfrac{m^2}{3H}\left|t_1-t_2\right|} - \e^{-\tfrac{m^2}{3H}\left(t_1+t_2\right)}\right) \\ \nonumber
& \quad \qquad + \frac{H^2}{4\pi^2} \sum\limits_{\n=1}^\infty \frac{(-1)^\n H^{2\n} | \x_1 - \x_2 |^{2\n}}{2 \n \, (2\n+1)! \left(1 + \tfrac{m^2}{3\n H^2} \right)} \left(\e^{- \n H \bigl(\, \left|t_1-t_2 \right| - \left(t_1+t_2 \right)\bigr) -\tfrac{m^2}{3H}\left|t_1-t_2\right|} - \e^{-\tfrac{m^2}{3H}\left(t_1 + t_2 \right)} \right)  
\end{align} and to continue with perturbative corrections. 

In principle, one can proceed with the computation 
through our elaboration in the flat Friedmann power law, $a(t) \propto t^s$, background~\cite{1985PhRvD..32.1316L}. With the use of analogous expressions to~\eqref{Wronskian}{--}\eqref{modes_k0}, we arrive at the retarded Green's function~\eqref{Green_retarded_lw} as
\begin{equation} G^{\, \text{l-w}}_{\text{R}} \left(t, \x \, ; t^{\prime}, \x^{\,\prime}\right) = \frac{\Theta(t-t^\prime)}{3s-1} \Bigl( (t^\prime \,)^{\, -3s+1} - t^{\, -3s+1}\Bigr) \, \delta(\x-\x^{\,\prime}). \end{equation} 
The Yang{--}Feldman equation in that setting is
\begin{equation}
\phi\left(t, \x \right) = \phi_0\left(t, \x \right)- \frac{\lambda}{3s-1} \, 
\int\limits_{0}^{t} d t^{\prime} \, \, t^\prime \, \phi^{3} \left(t^{\prime}, \x \right),  
\end{equation} and the expression of the free massive scalar field through the free massless one is 
\begin{equation}
\tilde{\phi}\left(t, \x \right) = \phi_0\left(t, \x \right)- \frac{m^2}{3s-1} \, \e^{-\tfrac{m^2t^2}{6s-2}} \int\limits_{0}^{t} d t^{\prime} \, \, \e^{\, \tfrac{\, \, m^2{t^\prime}^{2}}{6s-2}} \, t^\prime \, \phi_{0} \left(t^{\prime}, \x \right).  
\end{equation}
Thus, the analogue to our Yang{--}Feldman-type equation is the following:
\begin{equation}
\phi\left(t, \x \right) = \tilde{\phi}\left(t, \x \right) - \frac{\lambda}{3s-1} \, \e^{-\tfrac{m^2t^2}{6s-2}} \int\limits_{0}^{t} d t^{\prime} \, \, \, \e^{\, \tfrac{\, \, m^2{t^\prime}^{2}}{6s-2}} \, t^\prime \, \phi^{3} \left(t^{\prime}, \x \right). \end{equation} 
What is needed is the correlation function's infrared part of the massless scalar field, and everything is almost prepared for calculations in appendix~\ref{Details}.

Besides, it would also be interesting to go beyond the leading-logarithm approximation 
and furthermore to non-perturbative techniques in this direction. 
Through the modified form of the Fokker{--}Planck equation, some steps towards it were carried out in~\cite{2019arXiv191100022G,2024JHEP...04..004C}. We~plan to tackle that in the near future from other perspectives.



\appendix
\renewcommand{\theequation}{\thesection.\arabic{equation}}
\section{A trick to "hang up" the mass to the Yang{--}Feldman equation} 
\label{AddMass}
In this appendix, we will show how to get the massive Yang{--}Feldman-type equation instead of the massless one. 

Let us consider a slightly more general problem, starting with the following equation
\begin{equation}\label{Y-F-type_general}
\phi\left(t, \x \right) = \phi_0\left(t, \x \right) + \int\limits_{0}^{t} dt^\prime \, F \bigl( \phi\left(t^\prime, \x \right) \bigr),
\end{equation} 
here $F\bigl( \phi\left(t^\prime, \x \right) \bigr)$ is the general function of a scalar field 
$F\bigl( \phi \left(t^\prime, \x \right) \bigr) = \alpha \phi\left(t^\prime, \x \right) + W \bigl(\phi\left(t^\prime, \x \right)\bigr)$ and $\phi_0\left(t^\prime, \x \right)$ is a given function. Introducing the new scalar field $\tilde{\phi}\left(t, \x \right)$, which satisfies the equation 
\begin{equation}\label{phi_tilde_def}
\tilde{\phi}\left(t, \x \right)  = \phi_0 \left(t, \x \right) + \alpha \int\limits_{0}^{t} dt^\prime \, \tilde{\phi} \left(t^\prime, \x \right),
\end{equation} 
and taking the time derivative, one can find the exact solution to an ordinary inhomogeneous first-order linear differential equation:
\begin{equation}\label{phi_tilde_sol}
\tilde{\phi} \left(t, \x \right)  = \phi_0 \left(t, \x \right) + \alpha \e^{\alpha t} \int\limits_{0}^{t} dt^\prime \, \e^{-\alpha t^\prime} \phi_0 \left(t^\prime, \x \right). \end{equation} 
One can also express the known function $\phi_0 \left(t, \x \right)$ in terms of the new free massive one $\tilde{\phi}\left(t, \x \right)$ from \eqref{phi_tilde_def} and 
substitute this expression into~\eqref{Y-F-type_general}, resulting in
\begin{equation}\label{phi_tilde}
\phi\left(t, \x \right) = \tilde{\phi}\left(t, \x \right) + \alpha \int\limits_{0}^{t} dt^\prime \, \Bigl( \phi\left(t^\prime, \x \right) - \tilde{\phi}\left(t^\prime, \x \right) \Bigr) + \int\limits_{0}^{t} dt^\prime \, W \bigl(\phi\left(t^\prime, \x \right)\bigr).
\end{equation}
Hereafter, we introduce a new equation
\begin{equation}\label{equivalent_form}
\phi\left(t, \x \right) = \tilde{\phi}\left(t, \x \right) + \e^{\alpha t} \int\limits_{0}^{t} dt^\prime \, \e^{-\alpha t^\prime} W \bigl(\phi\left(t^\prime, \x \right)\bigr),
\end{equation} which is equivalent to~\eqref{phi_tilde}. To establish this, we get $\phi\left(t, \x \right) - \tilde{\phi}\left(t, \x \right)$ from~\eqref{equivalent_form}, substitute into \eqref{phi_tilde}, and after changing the order of integration as
\begin{equation}\label{trick_end}
\alpha \int\limits_{0}^{t} dt^\prime \,   \e^{\alpha t^\prime}  \int\limits_{0}^{t^\prime} dt^{\prime\prime} \, \e^{-\alpha t^{\prime\prime}} W \bigl(\phi\left(t^{\prime\prime}, \x \right)\bigr) = \alpha \int\limits_{0}^{t} dt^{\prime\prime} \, \e^{-\alpha t^{\prime\prime}} W \bigl(\phi\left(t^{\prime\prime}, \x \right)\bigr) \int\limits_{t^{\prime\prime}}^{t} dt^\prime \,   \e^{\alpha t^\prime}, 
\end{equation} 
one can see that~\eqref{equivalent_form} reduces to~\eqref{phi_tilde}. 

\section{All details for two-point and four-point correlation functions}\label{Details}
In this part of the appendix, we shall present all the expressions that have been used in order to calculate perturbative series~\eqref{loop_series_Two_Point} and \eqref{loop_series_FourPoint} in section~\ref{Y-F_chapter}. 

Through the use of our relation~\eqref{FreeMassiveField}, one can write down the two-point correlation function for the free massive field $\tilde{\phi} \, (t, \x)$, where the spatial arguments of the spacetime points coincide while the time coordinates are different as follows:
\begin{align}\label{correlator_massive_free_details_1}
\bigl\langle \tilde{\phi}
\left(t_1\right) \tilde{\phi}\left(t_2 \right) \bigr\rangle = \bigl\langle \phi_0\left(t_1\right) \phi_0 \left(t_2\right) \bigr\rangle & - \frac{m^2}{3H} \, \e^{-\tfrac{m^2t_1}{3H}} \int\limits_{0}^{t_1} dt^{\prime} \e^{\tfrac{m^2 t^{\prime}}{3H}}\bigl\langle \phi_0 \left(t^{\prime}\right) \phi_0\left(t_2\right) \bigr\rangle \nonumber \\ & - \frac{m^2}{3H} \, \e^{-\tfrac{m^2t_2}{3H}} \int\limits_{0}^{t_2} dt^{\prime} \e^{\tfrac{m^2 t^{\prime}}{3H}} \bigl\langle \phi_0 \left(t_1\right) \phi_0 \left(t^{\prime} \right) \bigr\rangle \\ \nonumber 
& + \frac{m^4}{9H^2} \,\e^{- \tfrac{m^2}{3H} \, (t_1+t_2)} \int\limits_{0}^{t_1} dt^{\prime} \e^{\tfrac{m^2 t^{\prime}}{3H}}\int\limits_{0}^{t_2} dt^{\prime\prime} \e^{\tfrac{m^2 t^{\prime\prime}}{3H}} \bigl\langle \phi_0 \left(t^{\prime}\right) \phi_0 \left(t^{\prime\prime} \right) \bigr\rangle. \end{align} 
Therefore, the usage in this expression of the known result for the free massless field's long-wavelength part, presented above in~\eqref{correlator_massless_free}, leads to
\begin{align}\label{correlator_massive_free_details_2}
& \left. \bigl \langle \tilde{\phi} \left(t_1\right) \tilde{\phi} \left(t_2 \right) \bigr\rangle \right|_{t_2 \leq t_1} = \frac{H^3}{4\pi^2} \Biggl(t_2 - \frac{m^2}{3H} \,  \e^{-\tfrac{m^2t_1}{3H}} \biggl( t_2 \int\limits_{t_2}^{t_1} dt^{\prime} \e^{\tfrac{m^2 t^{\prime}}{3H}} + \int\limits_{0}^{t_2} dt^{\prime} \e^{\tfrac{m^2 t^{\prime}}{3H}} t^{\prime} \biggr) \\ \nonumber & \qquad \qquad  - \frac{m^2}{3H} \,  \e^{-\tfrac{m^2t_2}{3H}} \int\limits_{0}^{t_2} dt^{\prime} \e^{\tfrac{m^2 t^{\prime}}{3H}} t^{\prime}  + \frac{m^4}{9H^2} \,\e^{- \tfrac{m^2}{3H} \, (t_1+t_2)} \biggl( \,\,  \int\limits_{0}^{t_2} dt^{\prime} \e^{\tfrac{m^2 t^{\prime}}{3H}} t^{\prime}\int\limits_{t^{\prime}}^{t_2} dt^{\prime\prime} \e^{\tfrac{m^2 t^{\prime\prime}}{3H}} \\ \nonumber & \qquad \qquad 
+ \int\limits_{t_2}^{t_1} dt^{\prime} \e^{\tfrac{m^2 t^{\prime}}{3H}} \int\limits_{0}^{t_2} dt^{\prime\prime} \e^{\tfrac{m^2 t^{\prime\prime}}{3H}} t^{\prime\prime}+ \int\limits_{0}^{t_2} dt^{\prime} \e^{\tfrac{m^2 t^{\prime}}{3H}} \int\limits_{0}^{t^{\prime}} dt^{\prime\prime} \e^{\tfrac{m^2 t^{\prime\prime}}{3H}} t^{\prime\prime} \biggr) \Biggr) \nonumber \\  \nonumber & \qquad \qquad \qquad \qquad \,\,\, 
= \frac{3H^4}{8\pi^2m^2} \left( \e^{-\tfrac{m^2}{3H}(t_1-t_2)} - \e^{-\tfrac{m^2}{3H}(t_1+t_2)}\right); \end{align} 
and, obviously, for another time ordering:
\begin{align} \label{correlator_massive_free_details_3}
& \left. \bigl \langle \tilde{\phi} \left(t_1\right) \tilde{\phi} \left(t_2 \right) \bigr\rangle \right|_{t_2 \geq t_1} = \frac{H^3}{4\pi^2} \Biggl(t_1  - \frac{m^2}{3H} \, \e^{-\tfrac{m^2t_1}{3H}} \int\limits_{0}^{t_1} dt^{\prime} \e^{\tfrac{m^2 t^{\prime}}{3H}} t^{\prime} \\ \nonumber  & \quad - \frac{m^2}{3H} \, \e^{-\tfrac{m^2t_2}{3H}} \biggl( \int\limits_{0}^{t_1} dt^{\prime} \e^{\tfrac{m^2 t^{\prime}}{3H}} t^{\prime} + t_1 \int\limits_{t_1}^{t_2} dt^{\prime} \e^{\tfrac{m^2 t^{\prime}}{3H}} \biggr) \\ \nonumber  
& \quad  
+ \frac{m^4}{9H^2} \,\e^{- \tfrac{m^2}{3H} \, (t_1+t_2)} \biggl( \int\limits_{0}^{t_1} dt^{\prime} \e^{\tfrac{m^2 t^{\prime}}{3H}} t^{\prime}\int\limits_{t^{\prime}}^{t_1} dt^{\prime\prime} \e^{\tfrac{m^2 t^{\prime\prime}}{3H}} 
+ \int\limits_{0}^{t_1} dt^{\prime} \e^{\tfrac{m^2 t^{\prime}}{3H}} t^{\prime}\int\limits_{t_1}^{t_2} dt^{\prime\prime} \e^{\tfrac{m^2 t^{\prime\prime}}{3H}} \\ \nonumber  
& \qquad \qquad \qquad \qquad \quad 
+ \int\limits_{0}^{t_1} dt^{\prime} \e^{\tfrac{m^2 t^{\prime}}{3H}} \int\limits_{0}^{t^{\prime}} dt^{\prime\prime} \e^{\tfrac{m^2 t^{\prime\prime}}{3H}} t^{\prime\prime} \biggr) \Biggr)
= \frac{3H^4}{8\pi^2m^2} \left( \e^{\tfrac{m^2}{3H}(t_1 - t_2)} - \e^{-\tfrac{m^2}{3H}(t_1+t_2)}\right).  
\end{align} As one can see, in the general case, the only need is to add the modulus to the first term, resulting in our building block~\eqref{correlator_massive_free}. Hereafter, we apply the same appropriate ordering "tricks" to calculate higher corrections to~correlation~functions.
Once again, the basic idea of the Yang{--}Feldman formalism used in this paper is to recursively define the interacting field as a formal power series in the coupling constant $\lambda$. Whence, at the linear order in $\lambda$, through the Yang{--}Feldman-type equation~\eqref{Y-F-massive_iterated}, we get
\begin{align}\label{correlator_massive_OneLoop_details_1}
& \bigl\langle \phi \left(t_1 \right) \phi\left(t_2 \right) \bigr\rangle_{1-\text{loop}} =  - \frac{\lambda}{3H} \, \e^{-\tfrac{m^2t_1}{3H}} \int\limits_{0}^{t_1} dt^{\prime} \e^{\tfrac{m^2 t^{\prime}}{3H}}\bigl\langle \tilde{\phi}\left(t_2\right) \tilde{\phi}^3 \left(t^{\prime}\right) \bigr\rangle  \\ \nonumber  
& \qquad \qquad \qquad \qquad \quad \,\,
- \frac{\lambda}{3H} \,  
\e^{-\tfrac{m^2t_2}{3H}} \int\limits_{0}^{t_2} dt^{\prime} \e^{\tfrac{m^2 t^{\prime}}{3H}} \bigl\langle \tilde{\phi} \left(t_1\right) \tilde{\phi}^3 \left(t^{\prime} \right) \bigr\rangle \nonumber  \\
& \qquad \qquad \qquad \qquad \,  = - \frac{\lambda}{H} 
\,\e^{-\tfrac{m^2t_1}{3H}} \int\limits_{0}^{t_1} dt^{\prime} \e^{\tfrac{m^2 t^{\prime}}{3H}}\bigl\langle \tilde{\phi} \left(t^{\prime}\right) \tilde{\phi}\left(t_2\right) \bigr\rangle \bigl\langle  \tilde{\phi}^2 \left(t^{\prime}\right) \bigr\rangle \\ \nonumber  & \qquad \qquad \qquad \qquad \quad \,\, 
- \frac{\lambda}{H} \,
\e^{-\tfrac{m^2t_2}{3H}} \int\limits_{0}^{t_2} dt^{\prime} \e^{\tfrac{m^2 t^{\prime}}{3H}} \bigl\langle \tilde{\phi} \left(t_1\right) \tilde{\phi} \left(t^{\prime} \right) \bigr\rangle \bigl\langle \tilde{\phi}^2 \left(t^{\prime} \right) \bigr\rangle; 
\end{align}
correspondingly, at the two-loop level or $\lambda^2$, we have \begin{align}\label{TwoLoop_TwoPoint_FULLexpression_without_comb}
& \bigl\langle \phi \left(t_1 \right)  \phi \left(t_2 \right) \bigr\rangle_{2-\text{loop}} = \frac{\lambda^2}{3H^2} \e^{-\tfrac{m^2t_1}{3H}} \int\limits_{0}^{t_1} dt^{\prime} \int\limits_{0}^{t^{\prime}} dt^{\prime\prime} \e^{\tfrac{m^2 t^{\prime\prime}}{3H}} \, \bigl\langle \tilde{\phi} \left(t_2\right) \tilde{\phi}^2 \left(t^{\prime} \right) \tilde{\phi}^3 \left(t^{\prime\prime} \right) \bigr\rangle \nonumber \\
& \qquad \qquad \qquad \qquad + \frac{\lambda^2}{3H^2} \e^{-\tfrac{m^2t_2}{3H}} \int\limits_{0}^{t_2} dt^{\prime} \int\limits_{0}^{t^{\prime}} dt^{\prime\prime} \e^{\tfrac{m^2 t^{\prime\prime}}{3H}} \, \bigl\langle \tilde{\phi} \left(t_1\right) \tilde{\phi}^2 \left(t^{\prime} \right) \tilde{\phi}^3 \left(t^{\prime\prime} \right)  \bigr\rangle \\ \nonumber
& \qquad \qquad \qquad \qquad+  \frac{\lambda^2}{9H^2} \e^{-\tfrac{m^2}{3H} \left(t_1+t_2\right)} \int\limits_{0}^{t_1} dt^{\prime} \e^{\tfrac{m^2 t^{\prime}}{3H}} \int\limits_{0}^{t_2} dt^{\prime\prime} \e^{\tfrac{m^2 t^{\prime\prime}}{3H}} \bigl\langle \tilde{\phi}^3 \left(t^{\prime}\right) \tilde{\phi}^3 \left(t^{\prime\prime} \right) \bigr\rangle 
\end{align} \begin{align} 
\nonumber 
\label{TwoLoop_TwoPoint_FULLexpression}
&  = \frac{\lambda^2}{H^2} \e^{-\tfrac{m^2t_1}{3H}} \int\limits_{0}^{t_1} dt^{\prime} \int\limits_{0}^{t^{\prime}} dt^{\prime\prime} \e^{\tfrac{m^2 t^{\prime\prime}}{3H}} \biggl(2 \bigl\langle \tilde{\phi} \left(t^{\prime} \right) \tilde{\phi} \left(t_2\right) \bigr\rangle \bigl\langle \tilde{\phi} \left(t^{\prime}\right) \tilde{\phi} \left(t^{\prime\prime} \right) \bigr\rangle \bigl\langle \tilde{\phi}^2 \left(t^{\prime\prime} \right) \bigr\rangle \nonumber \\
& \qquad \qquad \quad \qquad \,\, + 2 \bigl\langle  \tilde{\phi} \left(t^{\prime\prime} \right) \tilde{\phi} \left(t_2\right) \bigr\rangle \Bigl(\bigl\langle \tilde{\phi} \left(t^{\prime}\right) \tilde{\phi} \left(t^{\prime\prime} \right) \bigr\rangle\Bigr)^2 + \bigl\langle \tilde{\phi} \left(t^{\prime\prime} \right) \tilde{\phi} \left(t_2\right) \bigr\rangle \bigl\langle \tilde{\phi}^2 \left(t^{\prime}\right) \bigr\rangle \bigl\langle \tilde{\phi}^2 \left(t^{\prime\prime} \right) \bigr\rangle  \biggr) \nonumber \\
& + \frac{\lambda^2}{H^2} \e^{-\tfrac{m^2t_2}{3H}} \int\limits_{0}^{t_2} dt^{\prime} \int\limits_{0}^{t^{\prime}} dt^{\prime\prime} \e^{\tfrac{m^2 t^{\prime\prime}}{3H}} \biggl(2 \bigl\langle \tilde{\phi} \left(t_1\right) \tilde{\phi} \left(t^{\prime} \right) \bigr\rangle \bigl\langle \tilde{\phi} \left(t^{\prime}\right) \tilde{\phi} \left(t^{\prime\prime} \right) \bigr\rangle \bigl\langle \tilde{\phi}^2 \left(t^{\prime\prime} \right) \bigr\rangle \\ \nonumber 
& \qquad \qquad \qquad \qquad + 2 \bigl\langle \tilde{\phi} \left(t_1\right) \tilde{\phi} \left(t^{\prime\prime} \right) \bigr\rangle \Bigl(\bigl\langle \tilde{\phi} \left(t^{\prime}\right) \tilde{\phi} \left(t^{\prime\prime} \right) \bigr\rangle\Bigr)^2 + \bigl\langle \tilde{\phi} \left(t_1\right) \tilde{\phi} \left(t^{\prime\prime} \right) \bigr\rangle \bigl\langle \tilde{\phi}^2 \left(t^{\prime}\right) \bigr\rangle \bigl\langle \tilde{\phi}^2 \left(t^{\prime\prime} \right) \bigr\rangle \biggr) \\ \nonumber 
& +  \frac{\lambda^2}{H^2} \e^{-\tfrac{m^2}{3H} \left(t_1+t_2\right)} \int\limits_{0}^{t_1} dt^{\prime} \e^{\tfrac{m^2 t^{\prime}}{3H}} \int\limits_{0}^{t_2} dt^{\prime\prime} \e^{\tfrac{m^2 t^{\prime\prime}}{3H}} \biggl( \frac{2}{3} \Bigl(\bigl\langle \tilde{\phi} \left(t^{\prime}\right) \tilde{\phi} \left(t^{\prime\prime} \right) \bigr\rangle\Bigr)^3 + \bigl\langle \tilde{\phi}^2 \left(t^{\prime}\right) \bigr\rangle \bigl\langle \tilde{\phi}\left(t^{\prime}\right) \tilde{\phi} \left(t^{\prime\prime}\right) \bigr\rangle \bigl\langle \tilde{\phi}^2\left(t^{\prime\prime}\right) \bigr\rangle \biggr); \end{align}
and, finally, at the three-loop level, the expression is the following:
\begin{align}
& \bigl\langle \phi \left(t_1 \right) \phi \left(t_2 \right) \bigr\rangle_{3-\text{loop}} = -\frac{\lambda^3}{H^3} \e^{-\tfrac{m^2t_2}{3H}} \int\limits_{0}^{t_2} dt^{\prime} \int\limits_{0}^{t^{\prime}} dt^{\prime\prime} \int\limits_{0}^{t^{\prime\prime}} dt^{\prime\prime\prime} \e^{\tfrac{m^2 t^{\prime\prime\prime}}{3H}} \biggl( \bigl\langle \tilde{\phi} \left(t_1\right) \tilde{\phi} \left(t^{\prime\prime\prime} \right) \bigr\rangle \times \\ \nonumber 
& \qquad 
\quad 
\times \bigl\langle \tilde{\phi}^2 \left(t^{\prime} \right) \bigr\rangle \bigl\langle \tilde{\phi}^2 \left(t^{\prime\prime} \right) \bigr\rangle \bigl\langle \tilde{\phi}^2 \left(t^{\prime\prime\prime} \right) \bigr\rangle + 4\,  \bigl\langle \tilde{\phi} \left(t_1\right) \tilde{\phi} \left(t^{\prime} \right) \bigr\rangle \bigl\langle \tilde{\phi} \left(t^{\prime}\right) \tilde{\phi} \left(t^{\prime\prime} \right) \bigr\rangle \bigl\langle \tilde{\phi} \left(t^{\prime\prime}\right) \tilde{\phi} \left(t^{\prime\prime\prime} \right) \bigr\rangle \bigl\langle  \tilde{\phi}^2\left(t^{\prime\prime\prime} \right) \bigr\rangle \\ \nonumber
& \qquad 
\quad + 2\, \bigl\langle \tilde{\phi} \left(t_1\right) \tilde{\phi} \left(t^{\prime} \right) \bigr\rangle \bigl\langle \tilde{\phi} \left(t^{\prime}\right) \tilde{\phi} \left(t^{\prime\prime\prime} \right) \bigr\rangle \bigl\langle \tilde{\phi}^2 \left(t^{\prime\prime}\right) \bigr\rangle \bigl\langle \tilde{\phi}^2 \left(t^{\prime\prime\prime} \right) \bigr\rangle \\ \nonumber
& \qquad 
\quad + 4 \, \bigl\langle \tilde{\phi} \left(t_1\right) \tilde{\phi} \left(t^{\prime} \right) \bigr\rangle \bigl\langle \tilde{\phi} \left(t^{\prime}\right) \tilde{\phi} \left(t^{\prime\prime\prime} \right) \bigr\rangle \Bigl( \bigl\langle \tilde{\phi} \left(t^{\prime\prime}\right) \tilde{\phi} \left(t^{\prime\prime\prime} \right) \bigr\rangle\Bigr)^2 \\ \nonumber
& \qquad 
\quad + 2\, \bigl\langle \tilde{\phi} \left(t_1\right) \tilde{\phi} \left(t^{\prime\prime} \right) \bigr\rangle \bigl\langle \tilde{\phi}^2 \left(t^{\prime}\right) \bigr\rangle \bigl\langle \tilde{\phi} \left(t^{\prime\prime}\right) \tilde{\phi} \left(t^{\prime\prime\prime} \right) \bigr\rangle \bigl\langle  \tilde{\phi}^2\left(t^{\prime\prime\prime} \right) \bigr\rangle \\ \nonumber
& \qquad 
\quad + 4\, \bigl\langle \tilde{\phi} \left(t_1\right) \tilde{\phi} \left(t^{\prime\prime} \right) \bigr\rangle \bigl\langle \tilde{\phi} \left(t^{\prime}\right) \tilde{\phi} \left(t^{\prime\prime}\right) \bigr\rangle \bigl\langle \tilde{\phi} \left(t^{\prime}\right) \tilde{\phi} \left(t^{\prime\prime\prime} \right) \bigr\rangle \bigl\langle  \tilde{\phi}^2\left(t^{\prime\prime\prime} \right) \bigr\rangle \\ \nonumber
& \qquad 
\quad + 4 \,\bigl\langle \tilde{\phi} \left(t_1\right) \tilde{\phi} \left(t^{\prime\prime} \right) \bigr\rangle \Bigl(\bigl\langle \tilde{\phi} \left(t^{\prime}\right) \tilde{\phi} \left(t^{\prime\prime\prime}\right) \bigr\rangle \Bigr)^2 \bigl\langle \tilde{\phi} \left(t^{\prime\prime}\right) \tilde{\phi} \left(t^{\prime\prime\prime} \right) \bigr\rangle \\ \nonumber
& \qquad 
\quad + 2\, \bigl\langle \tilde{\phi} \left(t_1\right) \tilde{\phi} \left(t^{\prime\prime\prime} \right) \bigr\rangle \Bigl(\bigl\langle \tilde{\phi} \left(t^{\prime\prime}\right) \tilde{\phi} \left(t^{\prime\prime\prime}\right) \bigr\rangle \Bigr)^2 \bigl\langle \tilde{\phi} ^2\left(t^{\prime}\right) \bigr\rangle \\ \nonumber
& \qquad 
\quad + 2\, \bigl\langle \tilde{\phi} \left(t_1\right) \tilde{\phi} \left(t^{\prime\prime\prime} \right) \bigr\rangle \Bigl(\bigl\langle \tilde{\phi} \left(t^{\prime}\right) \tilde{\phi} \left(t^{\prime\prime}\right) \bigr\rangle \Bigr)^2 \bigl\langle \tilde{\phi} ^2\left(t^{\prime\prime\prime}\right) \bigr\rangle \\ \nonumber
& \qquad 
\quad + 8\, \bigl\langle \tilde{\phi} \left(t_1\right) \tilde{\phi} \left(t^{\prime\prime\prime} \right) \bigr\rangle \bigl\langle \tilde{\phi} \left(t^{\prime}\right) \tilde{\phi} \left(t^{\prime\prime} \right) \bigr\rangle \bigl\langle \tilde{\phi} \left(t^{\prime}\right) \tilde{\phi} \left(t^{\prime\prime\prime} \right) \bigr\rangle \bigl\langle \tilde{\phi} \left(t^{\prime\prime}\right) \tilde{\phi} \left(t^{\prime\prime\prime} \right) \bigr\rangle \\ \nonumber
& \qquad 
\quad + 2 \, \bigl\langle \tilde{\phi} \left(t_1\right) \tilde{\phi} \left(t^{\prime\prime\prime} \right) \bigr\rangle \Bigl(\bigl\langle \tilde{\phi} \left(t^{\prime}\right) \tilde{\phi} \left(t^{\prime\prime\prime} \right) \bigr\rangle \Bigr)^2 \bigl\langle \tilde{\phi}^2 \left(t^{\prime\prime}\right) \bigr\rangle \biggr) 
\\ \nonumber 
& - \frac{\lambda^3}{H^3} \e^{-\tfrac{m^2t_2}{3H}} \int\limits_{0}^{t_2} dt^{\prime} \e^{\tfrac{-m^2 t^{\prime}}{3H}} \int\limits_{0}^{t^{\prime}} dt^{\prime\prime}  \e^{\tfrac{m^2 t^{\prime\prime}}{3H}} \int\limits_{0}^{t^{\prime}} dt^{\prime\prime\prime}  \e^{\tfrac{m^2 t^{\prime\prime\prime}}{3H}} \biggl( \bigl\langle \tilde{\phi} \left(t_1\right) \tilde{\phi} \left(t^{\prime} \right) \bigr\rangle \bigl\langle \tilde{\phi}^2 \left(t^{\prime\prime} \right) \bigr\rangle \times \\ \nonumber
& \qquad \qquad \qquad \qquad \quad \times \bigl\langle \tilde{\phi} \left(t^{\prime\prime}\right) \tilde{\phi} \left(t^{\prime\prime\prime} \right) \bigr\rangle \bigl\langle \tilde{\phi}^2 \left(t^{\prime\prime\prime} \right) \bigr\rangle + \frac{2}{3} \bigl\langle \tilde{\phi} \left(t_1\right) \tilde{\phi} \left(t^{\prime} \right) \bigr\rangle \Bigl(\bigl\langle \tilde{\phi} \left(t^{\prime\prime}\right) \tilde{\phi} \left(t^{\prime\prime\prime} \right) \bigr\rangle \Bigr)^3  \\ \nonumber 
& \qquad \qquad \qquad \qquad \quad 
+ 2 \bigl\langle \tilde{\phi} \left(t_1\right) \tilde{\phi} \left(t^{\prime\prime} \right) \bigr\rangle \bigl\langle \tilde{\phi} \left(t^{\prime}\right) \tilde{\phi} \left(t^{\prime\prime} \right) \bigr\rangle \bigl\langle \tilde{\phi} \left(t^{\prime\prime}\right) \tilde{\phi} \left(t^{\prime\prime\prime} \right) \bigr\rangle \bigl\langle \tilde{\phi}^2 \left(t^{\prime\prime\prime} \right) \bigr\rangle \\ \nonumber
& \qquad \qquad \qquad \qquad \quad 
+ \bigl\langle \tilde{\phi} \left(t_1\right) \tilde{\phi} \left(t^{\prime\prime} \right) \bigr\rangle \bigl\langle \tilde{\phi} \left(t^{\prime}\right) \tilde{\phi} \left(t^{\prime\prime\prime} \right) \bigr\rangle \bigl\langle \tilde{\phi}^2 \left(t^{\prime\prime}\right) \bigr\rangle
\bigl\langle \tilde{\phi}^2 \left(t^{\prime\prime\prime} \right) \bigr\rangle \\ \nonumber 
& \qquad \qquad \qquad \qquad \quad
+ 2 \bigl\langle \tilde{\phi} \left(t_1\right) \tilde{\phi} \left(t^{\prime\prime} \right) \bigr\rangle \bigl\langle \tilde{\phi} \left(t^{\prime}\right) \tilde{\phi} \left(t^{\prime\prime\prime} \right) \bigr\rangle \Bigl( \bigl\langle \tilde{\phi} \left(t^{\prime\prime}\right) \tilde{\phi} \left(t^{\prime\prime\prime} \right) \bigr\rangle \Bigr)^2 \\ \nonumber
& \qquad \qquad \qquad \qquad \quad 
+ \bigl\langle \tilde{\phi} \left(t_1\right) \tilde{\phi} \left(t^{\prime\prime\prime} \right) \bigr\rangle \bigl\langle \tilde{\phi} \left(t^{\prime}\right) \tilde{\phi} \left(t^{\prime\prime} \right) \bigr\rangle
\bigl\langle \tilde{\phi}^2 \left(t^{\prime\prime}\right) \bigr\rangle
\bigl\langle \tilde{\phi}^2 \left(t^{\prime\prime\prime} \right) \bigr\rangle 
\end{align} \begin{align}  \nonumber 
& \qquad \qquad \qquad \qquad \quad 
+ 2 \bigl\langle \tilde{\phi} \left(t_1\right) \tilde{\phi} \left(t^{\prime\prime\prime} \right) \bigr\rangle \bigl\langle \tilde{\phi} \left(t^{\prime}\right) \tilde{\phi} \left(t^{\prime\prime} \right) \bigr\rangle \Bigl( \bigl\langle \tilde{\phi} \left(t^{\prime\prime}\right) \tilde{\phi} \left(t^{\prime\prime\prime} \right) \bigr\rangle \Bigr)^2 \\ \nonumber
& \qquad \qquad \qquad \qquad \quad + 2 \bigl\langle \tilde{\phi} \left(t_1\right) \tilde{\phi} \left(t^{\prime\prime\prime} \right) \bigr\rangle \bigl\langle \tilde{\phi} \left(t^{\prime}\right) \tilde{\phi} \left(t^{\prime\prime\prime} \right) \bigr\rangle \bigl\langle \tilde{\phi}^2 \left(t^{\prime\prime}\right) \bigr\rangle  \bigl\langle \tilde{\phi} \left(t^{\prime\prime}\right) \tilde{\phi} \left(t^{\prime\prime\prime} \right) \bigr\rangle \biggr) \\ \nonumber 
& -  \frac{\lambda^3}{H^3} \e^{-\tfrac{m^2}{3H} \left(t_1+t_2\right)} \int\limits_{0}^{t_1} dt^{\prime} \e^{\tfrac{m^2 t^{\prime}}{3H}} \int\limits_{0}^{t_2} dt^{\prime\prime} \int\limits_{0}^{t^{\prime\prime}} dt^{\prime\prime\prime} \e^{\tfrac{m^2 t^{\prime\prime\prime}}{3H}} \biggl( 2 \bigl\langle \tilde{\phi}^2 \left(t^{\prime}\right) \bigr\rangle \times \\ \nonumber 
& \qquad 
\times \bigl\langle \tilde{\phi} \left(t^{\prime}\right) \tilde{\phi} \left(t^{\prime\prime} \right) \bigr\rangle \bigl\langle \tilde{\phi} \left(t^{\prime\prime}\right) \tilde{\phi} \left(t^{\prime\prime\prime} \right) \bigr\rangle\bigl\langle \tilde{\phi}^2 \left(t^{\prime\prime\prime}\right) \bigr\rangle + 2\,\Bigl(\bigl\langle \tilde{\phi} \left(t^{\prime}\right) \tilde{\phi} \left(t^{\prime\prime} \right) \bigr\rangle \Bigr)^2 \bigl\langle \tilde{\phi} \left(t^{\prime}\right) \tilde{\phi} \left(t^{\prime\prime\prime} \right) \bigr\rangle \bigl\langle \tilde{\phi}^2 \left(t^{\prime\prime\prime} \right) \bigr\rangle \\ \nonumber
& \qquad 
+ 4 \, \bigl\langle \tilde{\phi} \left(t^{\prime}\right) \tilde{\phi} \left(t^{\prime\prime} \right) \bigr\rangle \Bigl(\bigl\langle \tilde{\phi} \left(t^{\prime}\right) \tilde{\phi} \left(t^{\prime\prime\prime} \right) \bigr\rangle \Bigr)^2 \bigl\langle \tilde{\phi} \left(t^{\prime\prime }\right) \tilde{\phi} \left(t^{\prime\prime\prime} \right) \bigr\rangle \qquad \qquad \,\, \\ \nonumber
& \qquad 
+ \bigl\langle \tilde{\phi}^2 \left(t^{\prime}\right) \bigr\rangle \bigl\langle \tilde{\phi} \left(t^{\prime}\right) \tilde{\phi} \left(t^{\prime\prime\prime} \right) \bigr\rangle \bigl\langle \tilde{\phi}^2 \left(t^{\prime\prime}\right)  \bigr\rangle\bigl\langle \tilde{\phi}^2 \left(t^{\prime\prime\prime}\right) \bigr\rangle \\ \nonumber
& \qquad 
+ 2\, \bigl\langle \tilde{\phi}^2 \left(t^{\prime}\right) \bigr\rangle \bigl\langle \tilde{\phi} \left(t^{\prime}\right) \tilde{\phi} \left(t^{\prime\prime\prime} \right) \bigr\rangle \Bigl(\bigl\langle \tilde{\phi} \left(t^{\prime\prime}\right) \tilde{\phi} \left(t^{\prime\prime\prime} \right) \bigr\rangle \Bigr)^2 
+ \frac{2}{3} \Bigl(\bigl\langle \tilde{\phi} \left(t^{\prime}\right) \tilde{\phi} \left(t^{\prime\prime\prime} \right) \bigr\rangle\Bigr)^3 \bigl\langle \tilde{\phi}^2 \left(t^{\prime\prime}\right) \bigr\rangle \biggr) \\ \nonumber 
& + \bigl( t_1 \leftrightarrow t_2 \bigr). \end{align}


In the case of the four-point correlation function, we follow the same way to calculate corrections. Even so, we already have the answer at the $\lambda^0$ order:
\begin{align}\label{four_point_tree}
\bigl\langle \phi \left(t_1\right) \phi \left(t_2\right)  \phi \left(t_3\right) \phi \left(t_4\right) \bigr\rangle_0
& = \bigl\langle \tilde{\phi}\left(t_1\right) \tilde{\phi}\left(t_2 \right) \bigr\rangle\bigl\langle \tilde{\phi}\left(t_3\right) \tilde{\phi}\left(t_4 \right) \bigr\rangle\\ \nonumber 
& + \bigl\langle \tilde{\phi}\left(t_1\right) \tilde{\phi}\left(t_3 \right) \bigr\rangle\bigl\langle \tilde{\phi}\left(t_2 \right) \tilde{\phi}\left(t_4 \right) \bigr\rangle  + \bigl\langle \tilde{\phi}\left(t_1\right) \tilde{\phi}\left(t_4 \right) \bigr\rangle\bigl\langle \tilde{\phi}\left(t_2\right) \tilde{\phi}\left(t_3 \right) \bigr\rangle,
\end{align} 
which is just the known two-point correlation functions multiplied in an appropriate way.
Afterwards, for the linear order in $\lambda$, we partially also have the answer, since in this case, the complete correlation function is split into connected and disconnected diagram types. The complete expression with appropriate permutations is the following: 
\begin{align} \label{four_point_OneLoop_perm}
& \bigl\langle \phi \left(t_1\right) \phi \left(t_2\right)  \phi \left(t_3\right) \phi \left(t_4\right) \bigr\rangle_\lambda 
= - \cfrac{\lambda}{H} \, \e^{-\tfrac{m^2 t_1}{3H}} \int\limits_{0}^{t_1} dt^{\prime} \e^{\tfrac{m^2 t^\prime}{3H}} \biggl( \bigl\langle \tilde{\phi}^2 \left(t^\prime\right) \bigr\rangle \bigl\langle \tilde{\phi}\left(t^\prime\right) \tilde{\phi}\left(t_2 \right) \bigr\rangle \bigl\langle \tilde{\phi}\left(t_3\right) \tilde{\phi}\left(t_4 \right) \bigr\rangle \nonumber \\
& 
+ \bigl\langle \tilde{\phi}^2\left(t^\prime\right)\bigr\rangle \bigl\langle \tilde{\phi}\left(t^\prime\right) \tilde{\phi}\left(t_3 \right) \bigr\rangle\bigl\langle \tilde{\phi}\left(t_2\right) \tilde{\phi}\left(t_4 \right) \bigr\rangle +  \bigl\langle \tilde{\phi}^2\left(t^\prime\right)\bigr\rangle \bigl\langle \tilde{\phi}\left(t^\prime\right) \tilde{\phi}\left(t_4 \right) \bigr\rangle\bigl\langle \tilde{\phi}\left(t_2\right) \tilde{\phi}\left(t_3 \right) \bigr\rangle \\ \nonumber 
& 
+ \underbrace{\,  2 \,  \bigl\langle \tilde{\phi}\left(t^\prime\right) \tilde{\phi}\left(t_2 \right) \bigr\rangle \bigl\langle \tilde{\phi}\left(t^\prime\right) \tilde{\phi}\left(t_3 \right) \bigr\rangle\bigl\langle \tilde{\phi}\left(t^\prime\right) \tilde{\phi}\left(t_4 \right) \bigr\rangle \,\, }_{\text{connected}} \biggr) \,\,\, + \,\,\, \bigl( t_1 \leftrightarrow t_2 \bigr) \,\,\, + \,\,\, \bigl( t_1 \leftrightarrow t_3 \bigr) \,\,\, + \,\,\, \bigl( t_1 \leftrightarrow t_4 \bigr).
\end{align} 
Someone can directly calculate the complete correlation function using expressions below, as we also did. However, the contribution of disconnected diagrams can be combinatorially gained via the already calculated corrections for the two-point correlation function. According to this expression~\eqref{four_point_OneLoop_perm} above, the contribution of the connected diagrams for the chosen time ordering $t_1 \geq t_2 \geq t_3 \geq t_4$ is
\begin{align} \label{four_point_OneLoop_connected}
& \bigl\langle \phi \left(t_1\right) \phi \left(t_2\right)  \phi \left(t_3\right) \phi \left(t_4\right) \bigr\rangle^{\text{connected}}_\lambda
= - \frac{81 \lambda H^{12}}{512 \, \pi^6 m^8} \Biggl( \biggl(4 + \frac{4m^2}{3H} \bigl(t_2-t_3 \bigr)\biggr) \, \e^{\tfrac{2m^2}{3H}\left(t_3+t_4\right)} \\ \nonumber
& \quad 
- \e^{-\tfrac{2m^2}{3H}\left(t_1-t_2-t_3-t_4\right)} - \e^{\tfrac{4m^2 t_4}{3H}} + \e^{-\tfrac{2m^2}{3H}\left(t_1 - t_2 - t_3\right)}  + \e^{-\tfrac{2m^2}{3H}\left(t_1 - t_2 - t_4 \right)} 
\end{align} \begin{align} \nonumber
& \quad 
+ \e^{-\tfrac{2m^2}{3H}\left(t_1 - t_3 - t_4 \right)} + \e^{-\tfrac{2m^2}{3H}\left(t_2 - t_3 - t_4 \right)} - \biggl(4 + \frac{4m^2}{3H} \bigl(t_2-t_3\bigr)\biggr) \, \e^{\tfrac{2m^2 t_3}{3H}} \\ \nonumber 
& \quad 
- \biggl(12 + \frac{4m^2}{3H} \bigl(t_2+2t_3-3t_4\bigr)\biggr) \, \e^{\tfrac{2m^2 t_4}{3H}} + 12 + \frac{4m^2}{3H} \bigl(t_2+2t_3+3t_4\bigr) - \e^{-\tfrac{2m^2}{3H}\left(t_1-t_2\right)} \\ \nonumber 
& \quad 
- \e^{-\tfrac{2m^2}{3H}\left(t_1-t_3\right)} -  \e^{-\tfrac{2m^2}{3H}\left(t_1-t_4\right)} - \e^{-\tfrac{2m^2}{3H}\left(t_2-t_3\right)} - \e^{-\tfrac{2m^2}{3H}\left(t_2-t_4\right)} - \e^{-\tfrac{2m^2}{3H}\left(t_3-t_4\right)} \\ \nonumber
& \quad  
+ \e^{-\tfrac{2m^2 t_1}{3H}} + \e^{-\tfrac{2m^2 t_2}{3H}} + \e^{-\tfrac{2m^2 t_3}{3H}}+ \e^{-\tfrac{2m^2 t_4}{3H}}\Biggr) \e^{-\tfrac{m^2}{3H}\left(t_1+ t_2+ t_3 + t_4\right)} \\
& \xrightarrow[]{\,\,\, m \rightarrow 0\,\,\,} \, -\frac{\lambda H^8}{32 \pi^6} \, t_1 t_2 t_3 t_4 \, ; 
\end{align} At the coinciding times, one has 
\begin{align}\label{connected_One_loop_equal_times}
\bigl\langle \phi^4 \left(t\right) & \bigr \rangle^{\text{connected}}_\lambda 
= - \frac{81 \lambda H^{12}}{256 \, \pi^6 m^8} \Biggl(1 - 6 \e^{-\tfrac{2m^2 t}{3H}} + \biggl(3 + \frac{4m^2t}{H} \biggr)  \e^{-\tfrac{4m^2 t}{3H}}  + 2 \e^{-\tfrac{2m^2 t}{H}}\Biggr) \\
& \qquad \quad \xrightarrow[]{\,\,\, m \rightarrow 0 \,\,\,} \, -\frac{\lambda H^8 t^4}{32 \pi^6}.
\end{align} 
While the expression for the contribution of disconnected ones at the linear $\lambda$ order is the mixed products of the zero and the linear $\lambda$ orders of already known two-point correlation functions
\begin{align} \label{4point_OneLoop_disconnected}
& \bigl\langle \phi \left(t_1\right) \phi \left(t_2\right)  \phi \left(t_3\right) \phi \left(t_4\right) \bigr\rangle^{\text{disconnected}}_\lambda
=  
\bigl\langle \tilde{\phi} \left(t_1\right) \tilde{\phi}  \left(t_2\right) \bigr\rangle \bigl\langle \phi \left(t_3\right) \phi \left(t_4\right) \bigr\rangle_{1-\text{loop}}  \\ \nonumber 
& \qquad \qquad \qquad \qquad  + \bigl\langle \tilde{\phi} \left(t_1\right) \tilde{\phi} \left(t_3\right) \bigr\rangle \bigl\langle \phi \left(t_2\right) \phi \left(t_4\right) \bigr\rangle_{1-\text{loop}} \, + \bigl\langle \tilde{\phi} \left(t_1\right) \tilde{\phi} \left(t_4\right) \bigr\rangle \bigl\langle \phi \left(t_2\right) \phi \left(t_3\right) \bigr\rangle_{1-\text{loop}} \\ \nonumber 
& \qquad \qquad \qquad \qquad + \, \bigl\langle \tilde{\phi} \left(t_2\right) \tilde{\phi} \left(t_3\right) \bigr\rangle \bigl\langle \phi \left(t_1\right) \phi \left(t_4\right) \bigr\rangle_{1-\text{loop}} + \bigl\langle \tilde{\phi} \left(t_2\right) \tilde{\phi} \left(t_4\right)   \bigr\rangle \bigl\langle \phi \left(t_1\right) \phi \left(t_3\right) \bigr\rangle_{1-\text{loop}} \\ \nonumber 
& \qquad \qquad  \qquad \qquad \qquad \qquad \qquad \qquad  + \bigl\langle \tilde{\phi} \left(t_3\right) \tilde{\phi} \left(t_4\right) \bigr\rangle \bigl\langle \phi \left(t_1\right) \phi \left(t_2\right) \bigr\rangle_{1-\text{loop}} 
\end{align} it results in the final answer \begin{align}
& \bigl\langle \phi \left(t_1\right) \phi \left(t_2\right)  \phi \left(t_3\right) \phi \left(t_4\right) \bigr\rangle^{\text{disconnected}}_\lambda
= - \frac{81 \lambda H^{12}}{1024 \, \pi^6 m^8} \Biggl( \Bigl(4 + \frac{2m^2}{3H} \bigl(t_1-t_2+t_3-t_4 \bigr)\Bigr) \, \e^{\tfrac{2m^2}{3H}\left(t_2+t_4\right)} \nonumber \\
& 
+ \Bigl(8 + \frac{4m^2}{3H} \bigl(t_1+t_2-t_3-t_4 \bigr)\Bigr) \e^{\tfrac{2m^2}{3H}\left(t_3+t_4\right)} + \e^{-\tfrac{2m^2}{3H}\left(t_1-t_2-t_4\right)} + 2 \e^{-\tfrac{2m^2}{3H}\left(t_1-t_3-t_4\right)} \\ \nonumber 
& 
+ 2 \, \e^{-\tfrac{2m^2}{3H}\left(t_2-t_3-t_4\right)} + \e^{\tfrac{2m^2}{3H}\left(t_2-t_3+t_4\right)} - \Bigl(3 + \frac{2m^2}{3H} \bigl(t_1-t_2+t_3+3t_4 \bigr)\Bigr) \, \e^{\tfrac{2m^2 t_2}{3H}} \\ \nonumber 
& 
- \Bigl(6 + \frac{4m^2}{3H} \bigl(t_1+t_2-t_3+3t_4 \bigr)\Bigr) \, \e^{\tfrac{2m^2 t_3}{3H}} - \Bigl(9 + \frac{2m^2}{3H} \bigl(3t_1+5t_2+7t_3-3t_4 \bigr)\Bigr) \, \e^{\tfrac{2m^2 t_4}{3H}} \\ \nonumber 
& 
- \e^{-\tfrac{2m^2}{3H}\left(t_1-t_2\right)} - 2 \, \e^{-\tfrac{2m^2}{3H}\left(t_1-t_3\right)} - 3\, \e^{-\tfrac{2m^2}{3H}\left(t_1-t_4\right)} - 2\, \e^{-\tfrac{2m^2}{3H}\left(t_2-t_3\right)} - 3\, \e^{-\tfrac{2m^2}{3H}\left(t_2-t_4\right)} \\ \nonumber 
& 
- 3\, \e^{-\tfrac{2m^2}{3H}\left(t_3-t_4\right)} - \e^{\tfrac{2m^2}{3H}\left(t_2-t_3\right)} -  \e^{\tfrac{2m^2}{3H}\left(t_2-t_4\right)} - 2 \, \e^{\tfrac{2m^2}{3H}\left(t_3-t_4\right)} + 6 + \frac{2m^2}{3H} \bigl(3t_1+5t_2+7t_3+9t_4 \bigr) \\ \nonumber 
& 
+ 3 \,  \e^{-\tfrac{2m^2 t_1}{3H}} + 3\, \e^{-\tfrac{2m^2 t_2}{3H}} + 3\,\e^{-\tfrac{2m^2 t_3}{3H}} + 3\,  \e^{-\tfrac{2m^2 t_4}{3H}} \Biggr) \e^{-\tfrac{m^2}{3H}\left(t_1+ t_2+ t_3 + t_4\right)} \,\,  \xrightarrow[]{\,\,\, m \rightarrow 0 \,\,\,} \\ 
& 
-\frac{\lambda H^8}{ 384 \, \pi^6} \, \biggl( 3 \, t_1^2 t_2 t_4 + 6 \, t_1^2t_3t_4 + t_2^3 t_4 + 6 \, t_2^2 t_3 t_4 + 3 \, t_2 t_3^2 t_4 
+ t_2 t_4^3 + 2 \, t_3^3 t_4 + 2 \, t_3 t_4^3 \biggr) ; 
\end{align} 
and, once again, at the coinciding times, one has 
\begin{align}\label{disconnected_One_loop_equal_times}
\bigl\langle \phi^4 \left(t\right) & \bigr \rangle^{\text{disconnected}}_\lambda 
= - \frac{243 \lambda H^{12}}{256 \, \pi^6 m^8} \Biggl(1 - \biggl( 1 + \frac{4m^2 t}{3H} \biggr) \e^{-\tfrac{2m^2 t}{3H}} - \biggl(1 - \frac{4m^2t}{3H} \biggr) \e^{-\tfrac{4m^2 t}{3H}}  + \e^{-\tfrac{2m^2 t}{H}}\Biggr) \nonumber \\ 
& \qquad \qquad \xrightarrow[]{\,\,\, m \rightarrow 0 \,\,\,}   -\frac{\lambda H^8 t^4}{16 \pi^6}.
\end{align} We provide the full answer, representing the sum of these contributions, in the main part of this article; namely, see~\eqref{four_point_OneLoop}, the massless limit~\eqref{OneLoop_FourPoint_massless}, and at equal times~\eqref{four_point_tree_equal_times}.

Finally, regarding the $\lambda^2$ order's corrections, we repeat the same reasoning. So, the complete expression for the correlation function at the $\lambda^2$ level with the necessary permutations is
\begin{align}\label{four_point_TwoLoop}
& \bigl\langle \phi \left(t_1\right) \phi \left(t_2\right) \phi \left(t_3\right) \phi \left(t_4\right) \bigr\rangle_{\lambda^2}
= \cfrac{2\lambda^2}{H^2} \, \e^{-\tfrac{m^2 t_1}{3H}} \int\limits_{0}^{t_1} dt^{\prime} \int\limits_{0}^{t^\prime} dt^{\prime\prime} \e^{\tfrac{m^2 t^{\prime\prime}}{3H}} \biggl( \Bigl( \bigl\langle \tilde{\phi}\left(t^\prime\right) \tilde{\phi}\left(t^{\prime\prime}\right) \bigr\rangle\Bigr)^2 \times \\ \nonumber
& \qquad \times \bigl\langle \tilde{\phi}\left(t^{\prime\prime}\right) \tilde{\phi}\left(t_2\right) \bigr\rangle \bigl\langle \tilde{\phi}\left(t_3\right) \tilde{\phi}\left(t_4\right) \bigr\rangle + \Bigl( \bigl\langle \tilde{\phi}\left(t^\prime\right) \tilde{\phi}\left(t^{\prime\prime}\right) \bigr\rangle\Bigr)^2 \bigl\langle \tilde{\phi}\left(t^{\prime\prime}\right) \tilde{\phi}\left(t_3\right) \bigr\rangle \bigl\langle \tilde{\phi}\left(t_2\right) \tilde{\phi}\left(t_4\right) \bigr\rangle \\  \nonumber
& \qquad \qquad
+ \Bigl( \bigl\langle \tilde{\phi}\left(t^\prime\right) \tilde{\phi}\left(t^{\prime\prime}\right) \bigr\rangle\Bigr)^2 \bigl\langle \tilde{\phi}\left(t^{\prime\prime}\right) \tilde{\phi}\left(t_4\right) \bigr\rangle \bigl\langle \tilde{\phi}\left(t_2\right) \tilde{\phi}\left(t_3\right) \bigr\rangle \\ \nonumber 
& \qquad \qquad
+ \bigl\langle \tilde{\phi}\left(t^\prime\right) \tilde{\phi}\left(t^{\prime\prime}\right) \bigr\rangle  \bigl\langle \tilde{\phi}\left(t^\prime\right) \tilde{\phi}\left(t_2\right) \bigr\rangle \bigl\langle \tilde{\phi}^2\left(t^{\prime\prime}\right) \bigr\rangle \bigl\langle \tilde{\phi}\left(t_3\right) \tilde{\phi}\left(t_4\right) \bigr\rangle \\ \nonumber & \qquad \qquad 
+ \underbrace{\, 2 \, \bigl\langle \tilde{\phi}\left(t^\prime\right) \tilde{\phi}\left(t^{\prime\prime}\right) \bigr\rangle \bigl\langle \tilde{\phi}\left(t^\prime\right) \tilde{\phi}\left(t_2\right) \bigr\rangle \bigl\langle \tilde{\phi}\left(t^{\prime\prime}\right) \tilde{\phi}\left(t_3\right) \bigr\rangle \bigl\langle \tilde{\phi}\left(t^{\prime\prime}\right) \tilde{\phi}\left(t_4\right) \bigr\rangle}_{\text{connected}} \\ \nonumber & \qquad \qquad
+ \bigl\langle \tilde{\phi}\left(t^\prime\right) \tilde{\phi}\left(t^{\prime\prime}\right) \bigr\rangle  \bigl\langle \tilde{\phi}\left(t^\prime\right) \tilde{\phi}\left(t_3\right) \bigr\rangle \bigl\langle \tilde{\phi}^2\left(t^{\prime\prime}\right) \bigr\rangle \bigl\langle \tilde{\phi}\left(t_2\right) \tilde{\phi}\left(t_4\right) \bigr\rangle \\ \nonumber
& \qquad \qquad
+ \underbrace{\, 2 \, \bigl\langle \tilde{\phi}\left(t^\prime\right) \tilde{\phi}\left(t^{\prime\prime}\right) \bigr\rangle \bigl\langle \tilde{\phi}\left(t^\prime\right) \tilde{\phi}\left(t_3\right) \bigr\rangle \bigl\langle \tilde{\phi}\left(t^{\prime\prime}\right) \tilde{\phi}\left(t_2\right) \bigr\rangle \bigl\langle \tilde{\phi}\left(t^{\prime\prime}\right) \tilde{\phi}\left(t_4\right) \bigr\rangle}_{\text{connected}} \\ \nonumber & \qquad \qquad
+ \bigl\langle \tilde{\phi}\left(t^\prime\right) \tilde{\phi}\left(t^{\prime\prime}\right) \bigr\rangle  \bigl\langle \tilde{\phi}\left(t^\prime\right) \tilde{\phi}\left(t_4\right) \bigr\rangle \bigl\langle \tilde{\phi}^2\left(t^{\prime\prime}\right) \bigr\rangle \bigl\langle \tilde{\phi}\left(t_2\right) \tilde{\phi}\left(t_3\right) \bigr\rangle \\ \nonumber
& \qquad \qquad 
+ \underbrace{\, 2 \, \bigl\langle \tilde{\phi}\left(t^\prime\right) \tilde{\phi}\left(t^{\prime\prime}\right) \bigr\rangle \bigl\langle \tilde{\phi}\left(t^\prime\right) \tilde{\phi}\left(t_4\right) \bigr\rangle \bigl\langle \tilde{\phi}\left(t^{\prime\prime}\right) \tilde{\phi}\left(t_2\right) \bigr\rangle \bigl\langle \tilde{\phi}\left(t^{\prime\prime}\right) \tilde{\phi}\left(t_3\right) \bigr\rangle}_{\text{connected}} \\ \nonumber & \qquad \qquad
+ \underbrace{ \bigl\langle \tilde{\phi}\left(t^\prime\right) \tilde{\phi}\left(t_2\right) \bigr\rangle \bigl\langle \tilde{\phi}\left(t^\prime\right) \tilde{\phi}\left(t_3\right) \bigr\rangle \bigl\langle \tilde{\phi}^2\left(t^{\prime\prime}\right) \bigr\rangle \bigl\langle \tilde{\phi}\left(t^{\prime\prime}\right) \tilde{\phi}\left(t_4 \right) \bigr\rangle}_{\text{connected}} \\ \nonumber
& \qquad \qquad
+ \underbrace{\bigl\langle \tilde{\phi}\left(t^\prime\right) \tilde{\phi}\left(t_2\right) \bigr\rangle \bigl\langle \tilde{\phi}\left(t^\prime\right) \tilde{\phi}\left(t_4\right) \bigr\rangle \bigl\langle \tilde{\phi}^2\left(t^{\prime\prime}\right) \bigr\rangle \bigl\langle \tilde{\phi}\left(t^{\prime\prime}\right) \tilde{\phi}\left(t_3 \right) \bigr\rangle}_{\text{connected}} \\ 
 \nonumber & \qquad \qquad
+ \underbrace{\bigl\langle \tilde{\phi}\left(t^\prime\right) \tilde{\phi}\left(t_3\right) \bigr\rangle \bigl\langle \tilde{\phi}\left(t^\prime\right) \tilde{\phi}\left(t_4\right) \bigr\rangle \bigl\langle \tilde{\phi}^2\left(t^{\prime\prime}\right) \bigr\rangle \bigl\langle \tilde{\phi}\left(t^{\prime\prime}\right) \tilde{\phi}\left(t_2 \right) \bigr\rangle}_{\text{connected}} \\ \nonumber
& \qquad \qquad
+ \, \frac{1}{2} \, \bigl\langle \tilde{\phi}^2\left(t^\prime\right) \bigr\rangle \bigl\langle \tilde{\phi}^2\left(t^{\prime\prime}\right) \bigr\rangle \bigl\langle \tilde{\phi} \left(t^{\prime\prime}\right) \tilde{\phi}\left(t_2 \right) \bigr\rangle \bigl\langle \tilde{\phi}\left(t_3\right) \tilde{\phi}\left(t_4 \right) \bigr\rangle \\ \nonumber  & \qquad \qquad
+ \, \frac{1}{2} \, \bigl\langle \tilde{\phi}^2\left(t^\prime\right) \bigr\rangle \bigl\langle \tilde{\phi}^2\left(t^{\prime\prime}\right) \bigr\rangle \bigl\langle \tilde{\phi} \left(t^{\prime\prime}\right) \tilde{\phi}\left(t_3 \right) \bigr\rangle \bigl\langle \tilde{\phi}\left(t_2\right) \tilde{\phi}\left(t_4 \right) \bigr\rangle \\ \nonumber
& \qquad \qquad
+ \, \frac{1}{2} \, \bigl\langle \tilde{\phi}^2\left(t^\prime\right) \bigr\rangle \bigl\langle \tilde{\phi}^2\left(t^{\prime\prime}\right) \bigr\rangle 
\bigl\langle \tilde{\phi} \left(t^{\prime\prime}\right) \tilde{\phi}\left(t_4 \right) \bigr\rangle \bigl\langle \tilde{\phi}\left(t_2 \right) \tilde{\phi}\left(t_3 \right) \bigr\rangle 
\end{align} \begin{align} \nonumber 
& \qquad \qquad 
+ \underbrace{\bigl\langle \tilde{\phi}^2\left(t^\prime\right) \bigr\rangle \bigl\langle \tilde{\phi} \left(t^{\prime\prime}\right) \tilde{\phi}\left(t_2 \right) \bigr\rangle  \bigl\langle \tilde{\phi} \left(t^{\prime\prime}\right) \tilde{\phi}\left(t_3 \right) \bigr\rangle \bigl\langle \tilde{\phi} \left(t^{\prime\prime}\right) \tilde{\phi}\left(t_4 \right) \bigr\rangle}_{\text{connected}} \\ \nonumber
& \qquad \qquad \,\,\, + \,\,\, \bigl( t_1 \leftrightarrow t_2 \bigl) \,\,\, + \,\,\, \bigl( t_1 \leftrightarrow t_3 \bigl) \,\,\, + \,\,\, \bigl( t_1 \leftrightarrow t_4 \bigl) \\ \nonumber 
& 
+ \frac{\lambda^2}{H^2} \, \e^{-\tfrac{m^2}{3H}\left(t_1+t_2\right)}  \int\limits_{0}^{t_1} dt^{\prime} \, \e^{\tfrac{m^2 t^{\prime}}{3H}} \int\limits_{0}^{t_2} dt^{\prime\prime} \e^{\tfrac{m^2 t^{\prime\prime}}{3H}} \biggl( \bigl\langle \tilde{\phi}^2 \left(t^{\prime} \right) \bigr\rangle \bigl\langle \tilde{\phi} \left(t^{\prime}\right) \tilde{\phi} \left(t^{\prime\prime} \right) \bigr\rangle \bigl\langle \tilde{\phi}^2 \left(t^{\prime\prime} \right) \bigr\rangle \bigl\langle \tilde{\phi} \left(t_3\right) \tilde{\phi} \left(t_4\right) \bigr\rangle  \\ \nonumber 
& \qquad \qquad 
+ \, \frac{2}{3} \, \Bigl(\bigl\langle \tilde{\phi} \left(t^{\prime}\right) \tilde{\phi} \left(t^{\prime\prime} \right) \bigr\rangle \Bigr)^3 \bigl\langle \tilde{\phi} \left(t_3\right) \tilde{\phi} \left(t_4 \right) \bigr\rangle  \\ \nonumber 
& \qquad \qquad 
+ \underbrace{2 \, \bigl\langle \tilde{\phi} \left(t^{\prime} \right) \tilde{\phi} \left(t_3\right) \bigr\rangle \bigl\langle \tilde{\phi} \left(t^{\prime} \right) \tilde{\phi} \left(t_4\right) \bigr\rangle \bigl\langle \tilde{\phi} \left(t^{\prime}\right) \tilde{\phi} \left(t^{\prime\prime} \right) \bigr\rangle \bigl\langle \tilde{\phi}^2 \left(t^{\prime\prime} \right) \bigr\rangle}_{\text{connected}} \\ \nonumber 
& \qquad \qquad 
+ \bigl\langle \tilde{\phi}^2 \left(t^{\prime}\right) \bigr\rangle \bigl\langle \tilde{\phi} \left(t^{\prime} \right) \tilde{\phi} \left(t_3\right) \bigr\rangle \bigl\langle  \tilde{\phi} \left(t^{\prime\prime} \right) \tilde{\phi} \left(t_4\right) \bigr\rangle 
\bigl\langle \tilde{\phi}^2 \left(t^{\prime\prime} \right) \bigr\rangle \\ \nonumber 
& \qquad \qquad  + \underbrace{ 2 \, \bigl\langle \tilde{\phi} \left(t^{\prime} \right) \tilde{\phi} \left(t_3\right) \bigr\rangle  \Bigl( \bigl\langle \tilde{\phi} \left(t^{\prime}\right) \tilde{\phi} \left(t^{\prime\prime} \right) \bigr\rangle \Bigr)^2 \bigl\langle  \tilde{\phi} \left(t^{\prime\prime} \right) \tilde{\phi} \left(t_4\right) \bigr\rangle}_{\text{connected}} \\
\nonumber 
& \qquad \qquad 
+ \bigl\langle \tilde{\phi}^2 \left(t^{\prime}\right) \bigr\rangle \bigl\langle \tilde{\phi} \left(t^{\prime}  \right) \tilde{\phi} \left(t_4\right) \bigr\rangle \bigl\langle \tilde{\phi}^2 \left(t^{\prime\prime} \right) \bigr\rangle \bigl\langle\tilde{\phi} \left(t^{\prime\prime} \right)  \tilde{\phi} \left(t_3\right) \bigr\rangle  \\ \nonumber 
& \qquad \qquad 
+ \underbrace{ 2 \, \bigl\langle  \tilde{\phi} \left(t^{\prime} \right) \tilde{\phi} \left(t_4\right) \bigr\rangle \Bigl( \bigl\langle \tilde{\phi} \left(t^{\prime}\right) \tilde{\phi} \left(t^{\prime\prime} \right) \bigr\rangle \Bigr)^2 \bigl\langle \tilde{\phi} \left(t^{\prime\prime} \right) \tilde{\phi} \left(t_3\right) \bigr\rangle}_{\text{connected}} \qquad \qquad \,\, \\
\nonumber 
& \qquad \qquad 
+ \underbrace{2 \, \bigl\langle \tilde{\phi}^2 \left(t^{\prime}\right) \bigr\rangle \bigl\langle \tilde{\phi} \left(t^{\prime}\right) \tilde{\phi} \left(t^{\prime\prime} \right) \bigr\rangle \bigl\langle \tilde{\phi} \left(t^{\prime\prime} \right) \tilde{\phi} \left(t_3 \right) \bigr\rangle \bigl\langle  \tilde{\phi} \left(t^{\prime\prime} \right) \tilde{\phi} \left(t_4 \right) \bigr\rangle}_{\text{connected}} \biggr) \qquad \qquad \qquad \qquad \qquad \\ \nonumber 
& \qquad + \,\,\, \bigl( t_2 \leftrightarrow t_3 \bigr) \,\,\, + \,\,\, \bigl( t_2 \leftrightarrow t_4 \bigr) + \,\,\, \bigl( t_1 \leftrightarrow t_2 \, ; \,  t_2 \leftrightarrow t_3 \, ; \, t_3 \leftrightarrow t_1  \bigr) \\ \nonumber 
& \qquad  + \,\,\, \bigl( t_1 \leftrightarrow t_2 \, ; \,  t_2 \leftrightarrow t_4 \, ; \, t_3 \leftrightarrow t_1 \bigr) \,\,\,  + \,\,\, \bigl( t_1 \leftrightarrow t_3 \, ;  t_2 \leftrightarrow t_4 \bigr).\end{align}
From here, the calculated connected part's contribution is appearing to be
\begin{align}  \label{four_point_TwoLoop_connected}
& \bigl\langle \phi \left(t_1\right) \phi \left(t_2\right)  \phi \left(t_3\right) \phi \left(t_4\right) \bigr\rangle^{\text{connected}}_{\lambda^2}
= \frac{729 \, \lambda^2 H^{16}}{4096 \, \pi^8 m^{12}} \, \e^{-\tfrac{m^2}{3H}\left(t_1+ t_2+ t_3 + t_4\right)}  \Biggl( \, \e^{ \, \tfrac{m^2}{3H}\left(t_2+ t_3 + 2 \, t_4 \right)} \\ \nonumber 
& \qquad 
+ \biggl( \frac{67}{2} + \frac{4m^2}{3H} \bigl(t_1+ 9t_2 -9 t_3 - t_4 \bigr) + \frac{4m^4}{9H^2} \bigl(t_2 - t_3\bigr) \bigl(t_1+ 2t_2 - 2 t_3 - t_4 \bigr)  \biggr) \e^{\, \tfrac{2m^2}{3H}\left(t_3 + t_4\right)} \\ \nonumber 
& \qquad
- \biggl( 1 + \frac{2m^2}{3H}\bigl( t_3 - t_4 \bigr) \biggr) \, \e^{ \, \tfrac{m^2}{3H}\left(t_2 - t_3 + 4 \,  t_4\right)}  - \biggl( \frac{21}{2} + \frac{m^2}{3H} \bigl(t_1+ 3t_2 + 3 t_3 - 7 t_4 \bigr) \biggr) \e^{\, \tfrac{4m^2 t_4}{3H}} \\ \nonumber 
& \qquad
+ \frac{1}{2}  \e^{-\tfrac{2m^2}{3H}\left(t_1 - t_3 - 2 \, t_4\right)} - \biggl( \frac{23}{2} + \frac{m^2}{3H} \bigl(9 t_1 - 5t_2 -3 t_3 - t_4 \bigr) \biggr) \e^{- \tfrac{2m^2}{3H}\left(t_1 -t_2 - t_3 - t_4\right)} \\ 
\nonumber 
& \qquad
+ \frac{1}{2} \, \e^{-\tfrac{2m^2}{3H}\left(t_1 - t_2 - 2 \, t_4\right)}  + \frac{1}{2} \, \e^{-\tfrac{2m^2}{3H}\left(t_1 - 2 \, t_3 - t_4 \right)}  + \frac{1}{2}  \, \e^{-\tfrac{2m^2}{3H}\left(t_2 - 2\, t_3 - t_4\right)}  + \frac{1}{2}  \, \e^{-\tfrac{2m^2}{3H}\left(t_2 -t_3 - 2 \, t_4\right)} \\ \nonumber 
& \qquad
- \, \e^{ \, \tfrac{m^2}{3H}\left( t_2 +t_3 \right)} + \biggl( 2 - \frac{4m^4}{9H^2} \bigl( t_3 - t_4 \bigr)^2 \biggr) \e^{\, \tfrac{m^2}{3H}\left(t_2 - t_3 + 2 \, t_4\right)} - \, \e^{\, \tfrac{m^2}{3H}\left( t_2 -3 \, t_3 + 4\, t_4 \right)} \\ \nonumber 
& \qquad
- \biggl( \frac{63}{2} + \frac{2m^2}{3H} \bigl(2 t_1 + 17 t_2 - 17 t_3 + 6  t_4 \bigr) + \frac{4m^4}{9H^2}\bigl(t_2 - t_3 \bigr) \bigl(t_1 + 2t_2 - 2 t_3 + 3 t_4 \bigr) \biggr) \e^{\, \tfrac{2m^2 t_3}{3H}} \\ \nonumber 
& \qquad - \biggl( 146 + \frac{2m^2}{3H} \bigl(6 t_1 + 29 t_2 + 64 t_3 - 75  t_4 \bigr) + \frac{4m^4}{9H^2}\Bigl( \bigl(t_2 +2 t_3 -3t_4 \bigr) t_1 + \bigl( 2t_2 +9 t_3 -10 t_4 \bigr) t_2 
\end{align} 
\begin{align}
\nonumber 
& \qquad + \bigl( 6t_3 - 15 t_4 \bigr) t_3 + 8t_4^2 \Bigr) \biggr) \e^{\, \tfrac{2m^2 t_4}{3H}} + \biggl( 11 + \frac{m^2}{3H} \bigl(9 t_1 - 5t_2 -3 t_3 + 3 t_4 \bigr) \biggr) \e^{- \tfrac{2m^2}{3H}\left(t_1 -t_2 - t_3\right)} \\ \nonumber 
& \qquad 
- \frac{1}{2} \, \e^{- \tfrac{2m^2}{3H}\left(t_1 - 2\, t_3\right)} + \biggl( 15 + \frac{m^2}{3H} \bigl(9 t_1 - 5t_2 +7 t_3 - 7 t_4 \bigr) \biggr) \e^{- \tfrac{2m^2}{3H}\left(t_1 -t_2 - t_4\right)} \\ \nonumber 
& \qquad 
+ \biggl( 21 + \frac{m^2}{3H} \bigl(9 t_1 +13 t_2 - 11 t_3 - 7 t_4 \bigr) \biggr) \e^{- \tfrac{2m^2}{3H}\left(t_1 -t_3 - t_4\right)} - \frac{3}{2} \, \e^{- \tfrac{2m^2}{3H}\left(t_1 - 2 \, t_4\right)} \\ \nonumber 
& \qquad
- \frac{5}{2} \,  \e^{- \tfrac{2m^2}{3H}\left(2t_1 - t_2 - t_3 - t_4\right)}  + \biggl( \frac{51}{2} + \frac{m^2}{3H} \bigl(t_1 + 21 t_2 - 11 t_3 - 7 t_4 \bigr) \biggr) \e^{- \tfrac{2m^2}{3H}\left(t_2 -t_3 - t_4\right)} \\ \nonumber 
& \qquad  - \frac{1}{2} \, \e^{- \tfrac{2m^2}{3H}\left(t_2 - 2 \, t_3\right)}  - \frac{3}{2} \, \e^{- \tfrac{2m^2}{3H}\left(t_2 - 2 \, t_4\right)} - \frac{1}{2} \, \e^{- \tfrac{2m^2}{3H}\left(t_3 - 2 \, t_4\right)} - \frac{4m^2}{3H} \bigl(t_3 - t_4 \bigr)  \e^{\,\, \tfrac{m^2}{3H}\left(t_2 - 3 \, t_3 + 2 \, t_4\right)} \\ \nonumber 
& \qquad +  \frac{4m^2}{3H} \bigl(t_3 - t_4 \bigr) \biggl( 1 + \frac{m^2}{3H}\bigl(t_3 - t_4 \bigr)\biggr)  \e^{\,\, \tfrac{m^2}{3H}\left(t_2 - t_3 \right)} - \biggl(2 + \frac{2m^2}{3H}\bigl(t_2 - t_3 \bigr)\biggr)  \e^{\,\, \tfrac{2m^2}{3H}\left(t_3 - t_4 \right)} \\ \nonumber 
& \qquad 
+ \frac{223}{2} + \frac{4m^2}{3H} \Bigl( 3 t_1 + 14 t_2 + 30 t_3 + 55 t_4\Bigr) + \frac{4m^4}{9H^2} \Bigl( \bigl( t_2 + 2 t_3 + 3 t_4\bigr) t_1 + \bigl( 2 t_2 + 9 t_3 + 12 t_4\bigr) t_2 \\  \nonumber 
& \qquad
+ \bigl( 6 t_3 + 23 t_4\bigr) t_3 + 14 t_4^4 \Bigr) - \biggl( \frac{29}{2} + \frac{m^2}{3H} \bigl( 9 t_1 - 5 t_2 + 7 t_3 +9 t_4 \bigr) \biggr) \e^{- \tfrac{2m^2}{3H}\left(t_1 -t_2\right)} \\ \nonumber 
& \qquad
+ 2 \, \e^{- \tfrac{2m^2}{3H}\left( t_1 + t_2 - t_3 - t_4\right)} - \biggl( \frac{41}{2} + \frac{m^2}{3H} \bigl( 9 t_1 + 13 t_2 - 11 t_3 +9 t_4 \bigr) \biggr) \e^{- \tfrac{2m^2}{3H}\left(t_1 -t_3\right)} \\ \nonumber 
& \qquad - \biggl( \frac{57}{2} + \frac{m^2}{3H} \bigl( 9 t_1 +13 t_2 + 17 t_3 - 19 t_4 \bigr) \biggr) \e^{- \tfrac{2m^2}{3H}\left(t_1 -t_4 \right)} + \, \e^{- \tfrac{2m^2}{3H}\left( t_1 - t_2 + t_3 - t_4\right)} \\ \nonumber   
& \qquad  + \, \frac{1}{2} \, \e^{- \tfrac{2m^2}{3H}\left( t_1 - t_2 - t_3 + t_4\right)} + \frac{5}{2} \, \e^{- \tfrac{2m^2}{3H}\left( 2 \, t_1 - t_2 - t_3 \right)} + \frac{5}{2} \, \e^{- \tfrac{2m^2}{3H}\left( 2 \, t_1 - t_2 - t_4\right)} + \frac{5}{2} \, \e^{- \tfrac{2m^2}{3H}\left( 2 \, t_1 - t_3 - t_4\right)} \\ \nonumber 
& \qquad  + \, \frac{5}{2} \, \e^{- \tfrac{2m^2}{3H}\left( 2 \, t_2 - t_3 - t_4\right)} - \biggl( 25 + \frac{m^2}{3H} \bigl( t_1 + 21 t_2 - 11 t_3 +9 t_4 \bigr) \biggr) \e^{- \tfrac{2m^2}{3H}\left(t_2 -t_3\right)} \\ 
\nonumber 
& \qquad - \biggl( 33 + \frac{m^2}{3H} \bigl( t_1 + 21 t_2 + 17 t_3 - 19 t_4 \bigr) \biggr) \e^{- \tfrac{2m^2}{3H}\left(t_2 -t_4\right)} + \biggl( 2 + \frac{4m^2}{3H} \bigl( t_3 -  t_4 \bigr) \biggr) \e^{\, \tfrac{m^2}{3H}\left(t_2 - 3\,  t_3 \right)} \\ \nonumber 
& \qquad - \biggl( \frac{75}{2} + \frac{m^2}{3H} \bigl( t_1 + 5 t_2 + 29 t_3 - 15 t_4 \bigr) \biggr) \e^{- \tfrac{2m^2}{3H}\left(t_3 -t_4\right)} 
- \biggl( \frac{3}{2} + \frac{2m^2}{3H} \bigl( t_3 -  t_4 \bigr) \biggr) \e^{\, \tfrac{m^2}{3H}\left(t_2 - t_3 -2 t_4\right)} \\ \nonumber 
& \qquad - \frac{1}{2} \,  \e^{\, \tfrac{m^2}{3H}\left(t_2 - 5 \, t_3 + 2 \, t_4\right)}  - \frac{5}{2} \,  \e^{-\tfrac{2m^2}{3H}\left(2 \, t_1 - t_2 \right)} - \frac{5}{2} \,  \e^{-\tfrac{2m^2}{3H}\left(2 \, t_1 - t_3 \right)} - \frac{5}{2} \,  \e^{-\tfrac{2m^2}{3H}\left(2 \, t_1 - t_4 \right)} \\ \nonumber 
& \qquad
- 2\,  \e^{-\tfrac{2m^2}{3H}\left( t_1 + t_2 - t_3 \right)} - 2\, \e^{-\tfrac{2m^2}{3H}\left( t_1 + t_2 - t_4 \right)} - 2\, \e^{-\tfrac{2m^2}{3H}\left( t_1 + t_3 - t_4 \right)} - \e^{-\tfrac{2m^2}{3H}\left(t_1 - t_2 + t_3 \right)} \\ \nonumber 
& \qquad 
- \e^{-\tfrac{2m^2}{3H}\left( t_1 - t_2 + t_4 \right)} - \e^{-\tfrac{2m^2}{3H}\left( t_1 - t_3 + t_4 \right)} + \biggl( 28 + \frac{m^2}{3H} \bigl( 9t_1 + 13 t_2 + 17 t_3 + 21 t_4  \bigr) \biggr) \e^{- \tfrac{2m^2 t_1 }{3H}} \\ \nonumber 
& \qquad 
- 2 \,  \e^{-\tfrac{2m^2}{3H}\left( t_2 + t_3 - t_4 \right)} - \e^{-\tfrac{2m^2}{3H}\left( t_2 - t_3 + t_4 \right)} - \frac{5}{2} \, \e^{-\tfrac{2m^2}{3H}\left( 2 \, t_2 - t_3  \right)} - \frac{5}{2} \, \e^{-\tfrac{2m^2}{3H}\left( 2 \, t_2 - t_4  \right)} \\ \nonumber 
& \qquad 
+ \biggl( \frac{65}{2} + \frac{m^2}{3H} \bigl( t_1 + 21 t_2 + 17 t_3 + 21 t_4  \bigr) \biggr) \e^{- \tfrac{2m^2 t_2}{3H}} - 2 \, \e^{-\tfrac{2m^2}{3H}\left( 2 \, t_3 - t_4  \right)} 
\\ \nonumber 
& \qquad
+ \biggl( 35 + \frac{m^2}{3H} \bigl( t_1 + 5 t_2 + 29 t_3 + 25 t_4  \bigr) \biggr) \e^{- \tfrac{2m^2 t_3}{3H}} + \, \frac{1}{2}\, \e^{\, \tfrac{m^2}{3H}\left( t_2 - t_3 - 4 \, t_4  \right)} \\ \nonumber 
& \qquad
+ \biggl( 43 + \frac{m^2}{3H} \bigl( t_1 + 5 t_2 + 11 t_3 + 43 t_4  \bigr) \biggr) \e^{- \tfrac{2m^2 t_4}{3H}}  +  \, \frac{1}{2}\,  \e^{\, \tfrac{m^2}{3H}\left( t_2 - 5 \, t_3  \right)} - \, \e^{\, \tfrac{m^2}{3H}\left( t_2 - 3\, t_3 - 2 \, t_4  \right)} \\ \nonumber 
& \qquad 
+  2 \, \e^{-\tfrac{2m^2}{3H}\left(t_1 + t_2 \right)} +  2 \, \e^{-\tfrac{2m^2}{3H}\left(t_1 + t_3 \right)} +  2 \, \e^{-\tfrac{2m^2}{3H}\left(t_1 + t_4 \right)} +  2 \, \e^{-\tfrac{2m^2}{3H}\left(t_2 + t_3 \right)} +  2 \, \e^{-\tfrac{2m^2}{3H}\left(t_2 + t_4 \right)} \qquad \qquad \qquad \qquad \qquad \qquad \qquad \qquad \qquad \\ \nonumber 
& \qquad   +  3 \, \e^{-\tfrac{2m^2}{3H}\left(t_3 + t_4 \right)} + \frac{5}{2} \, \e^{-\tfrac{4m^2 t_1}{3H}}  + \frac{5}{2} \, \e^{-\tfrac{4m^2 t_2}{3H}}  + 2\, \e^{-\tfrac{4m^2 t_3}{3H}} + 2\, \e^{-\tfrac{4m^2 t_4}{3H}} \Biggr)  \xrightarrow[]{\,\,\, m \rightarrow 0 \,\,\,} 
\end{align} \begin{align} 
&  \xrightarrow[]{\, m \rightarrow 0 \,\,} \frac{\lambda^2 H^{10}}{768 \pi^8} \biggl( 11 t_1^3t_2t_3t_4 + t_1t_2^3 t_3 t_4 + t_1 t_2 t_3^3 t_4 + t_1 t_2 t_3 t_4^3 + 5 t_2^4 t_3 t_4 + 3 \, t_3^5 t_4 + 2 \, t_4^6 \biggr),
\end{align} \vspace{-0.25cm}
and the equal time result is 
\begin{align}  \label{four_point_TwoLoop_connected_equal_times}
& \bigl\langle \phi^4 \left(t\right) \bigr\rangle^{\text{connected}}_{\lambda^2}
\,
= \,\, \frac{5103 \lambda^2 H^{16}}{2048 \, \pi^8 m^{12}} \Biggl(1 - \biggl( 8 + \frac{8m^2 t}{7H}\biggr) \e^{-\tfrac{2m^2t}{3H}} \\ \nonumber 
& \qquad \qquad \qquad \quad - \biggl( \frac{18}{7} - \frac{48m^2 t}{7H}  - \frac{16m^4 t^2}{7H^2}\biggr) \e^{-\tfrac{4m^2t}{3H}} + \biggl( 8 + \frac{40m^2 t}{7H} \biggr) \e^{-\tfrac{2m^2t}{H}} + \frac{11}{7} \, \e^{-\tfrac{8m^2t}{3H}}\Biggr) \\ 
& \qquad \xrightarrow[]{\,\,\, m \rightarrow 0 \,\,\,}  \, \frac{\lambda^2 H^{10} \, t^6}{32\pi^8} . \end{align}
Repeating ourselves once again, the contribution of disconnected ones can be calculated directly or can be gained by combining the tree-, the linear $\lambda$, and the quadratic $\lambda^2$ orders obtained for the two-point correlation function before. We calculated it straightforwardly; nonetheless, the expression for the result from what is already known is
\begin{align} \label{4point_TwoLoop_disconnected}
& \bigl\langle \phi \left(t_1\right) \phi \left(t_2\right)  \phi \left(t_3\right) \phi \left(t_4\right) \bigr\rangle^{\text{disconnected}}_{\lambda^2}
= \bigl\langle \tilde{\phi} \left(t_1\right) \tilde{\phi}  \left(t_2\right) \bigr\rangle \bigl\langle \phi \left(t_3\right) \phi \left(t_4\right) \bigr\rangle_{2-\text{loop}} \\ \nonumber 
& \quad 
+ \bigl\langle \tilde{\phi} \left(t_1\right) \tilde{\phi} \left(t_3\right) \bigr\rangle \bigl\langle \phi \left(t_2\right) \phi \left(t_4\right) \bigr\rangle_{2-\text{loop}}  + \bigl\langle \tilde{\phi} \left(t_1\right) \tilde{\phi} \left(t_4\right) \bigr\rangle \bigl\langle \phi \left(t_2\right) \phi \left(t_3\right) \bigr\rangle_{2-\text{loop}} \\ \nonumber 
& \quad 
+ \bigl\langle \tilde{\phi} \left(t_2\right) \tilde{\phi} \left(t_3\right) \bigr\rangle \bigl\langle \phi \left(t_1\right) \phi \left(t_4\right) \bigr\rangle_{2-\text{loop}} + \bigl\langle \tilde{\phi} \left(t_2\right) \tilde{\phi} \left(t_4\right) \bigr\rangle \bigl\langle \phi \left(t_1\right) \phi \left(t_3\right) \bigr\rangle_{2-\text{loop}} \\ \nonumber 
& \quad  
+ \bigl\langle \tilde{\phi} \left(t_3\right) \tilde{\phi} \left(t_4\right) \bigr\rangle \bigl\langle \phi \left(t_1\right) \phi \left(t_2\right) \bigr\rangle_{2-\text{loop}} + \bigl\langle \phi \left(t_1\right) \phi \left(t_2\right) \bigr\rangle_{1-\text{loop}} \bigl\langle \phi \left(t_3\right) \phi \left(t_4\right) \bigr\rangle_{1-\text{loop}}
\\ \nonumber 
& \quad + \bigl\langle \phi \left(t_1\right) \phi \left(t_3\right) \bigr\rangle_{1-\text{loop}} \bigl\langle \phi \left(t_2\right) \phi \left(t_4\right) \bigr\rangle_{1-\text{loop}}  + \bigl\langle \phi \left(t_1\right) \phi \left(t_4\right) \bigr\rangle_{1-\text{loop}} \bigl\langle \phi \left(t_2\right) \phi \left(t_3\right) \bigr\rangle_{1-\text{loop}} \quad \Rightarrow 
\end{align} 
the final explicit expression is the following:
\begin{align} 
& \bigl\langle \phi \left(t_1\right) \phi \left(t_2\right)  \phi \left(t_3\right) \phi \left(t_4\right) \bigr\rangle^{\text{disconnected}}_{\lambda^2}
= \frac{243 \lambda^2 H^{16}}{8192 \, \pi^8 m^{12}} \Biggl(\, \e^{-\tfrac{2m^2}{3H}\left(t_1- 2\, t_2 - t_4\right)}  \\ \nonumber 
&  \qquad + \biggl(6 + \frac{m^2}{3H} \bigl(t_1-t_2+t_3-t_4 \bigr)\biggr)\biggl(6 + \frac{m^2}{H} \bigl(t_1-t_2+t_3-t_4 \bigr)\biggr) \, \e^{\tfrac{2m^2}{3H}\left(t_2+t_4\right)}  \\ \nonumber 
& \qquad 
+ \, \e^{-\tfrac{2m^2}{3H}\left(t_1 -t_3 - 2\, t_4\right)} + \, \e^{-\tfrac{2m^2}{3H}\left(t_1 - 2\, t_3 - t_4\right)} + \,  \e^{\tfrac{2m^2}{3H}\left(t_2-t_3 + 2\, t_4\right)}  + \,  \e^{- \tfrac{2m^2}{3H}\left(t_2-t_3 - 2\, t_4\right)} \\ \nonumber 
& \qquad  
+ \biggl(4 + \frac{2m^2}{3H} \bigl(t_1+t_2-t_3-t_4 \bigr)\biggr)\biggl(18 + \frac{m^2}{H} \bigl(t_1+t_2-t_3-t_4 \bigr)\biggr) \, \e^{\tfrac{2m^2}{3H}\left(t_3+t_4\right)} \\ 
\nonumber 
& \qquad 
+ \, \e^{-\tfrac{2m^2}{3H}\left(t_2- 2 \, t_3 - t_4\right)} + \biggl( 27 + \frac{m^2}{H}\bigl(9 t_1-5t_2 + t_3 - t_4 \Bigr)\biggr) \, \e^{-\tfrac{2m^2}{3H}\left(t_1- t_2 - t_4\right)} \\  \nonumber 
& \qquad 
+ \biggl( 54 + \frac{2m^2}{H}\bigl(9 t_1 + t_2 - 3 t_3 - 3t_4 \bigr)\biggr) \, \e^{-\tfrac{2m^2}{3H}\left(t_1- t_3 - t_4\right)} - \, \e^{-\tfrac{2m^2}{3H}\left(t_1- 2 \, t_3\right)} \\ \nonumber 
& \qquad  
- \, \e^{-\tfrac{2m^2}{3H}\left(t_1- 2 \, t_4 \right)} + \biggl(27 + \frac{m^2}{H}\bigl(t_1 - t_2 +9 t_3 - 5t_4 \bigr)\biggr) \, \e^{\tfrac{2m^2}{3H}\left(t_2- t_3 + t_4\right)} \\ \nonumber 
& \qquad  
+ \biggl( 54 + \frac{2m^2}{H}\bigl(t_1 + 9 t_2 - 3 t_3 - 3t_4 \bigr)\biggr) \, \e^{-\tfrac{2m^2}{3H}\left(t_2 - t_3 - t_4\right)} - \, \e^{-\tfrac{2m^2}{3H}\left(t_2- 2 \, t_3\right)} \\ \nonumber 
& \qquad 
- \, \e^{-\tfrac{2m^2}{3H}\left(t_2- 2 \, t_4 \right)} - \biggl( \frac{m^2}{H}\bigl( 7 t_1 - 7 t_2 + 7 t_3 + 29 t_4 \bigr) + \frac{m^4}{3H^2}\bigl( t_1 - t_2 + t_3 + 3t_4 \bigr)^2\biggr) \, \e^{\tfrac{2m^2t_2}{3H}} 
\end{align} 
\begin{align} \nonumber 
& \qquad - \, \e^{-\tfrac{2m^2}{3H}\left(t_3- 2 \, t_4 \right)} - \biggl( \frac{2m^2}{H}\bigl( 7 t_1 + 7 t_2 - 7 t_3 + 29 t_4 \bigr) + \frac{2m^4}{3H^2}\bigl( t_1 + t_2 - t_3 + 3t_4 \bigr)^2\biggr) \, \e^{\tfrac{2m^2t_3}{3H}} \\ \nonumber 
& \qquad - \biggl( \frac{m^2}{H}\bigl( 21 t_1 + 43 t_2 + 65 t_3 - 21 t_4 \bigr) + \frac{m^4}{3H^2}\Bigl( \bigl(3 t_1 + 10 t_2 + 14 t_3 - 6t_4\bigr) t_1 \\ \nonumber 
& \qquad 
+ \bigl( 11 t_2 + 18 t_3 - 10 t_4 \bigr) t_2 + \bigl( 19 t_3 - 14 t_4 \bigr) t_3 + 3 t_4^2 \Bigr) \biggr) \, \e^{\tfrac{2m^2t_4}{3H}} + \frac{15}{4} \,  \e^{-\tfrac{2m^2}{3H}\left( 2 t_1 - t_2 - t_4\right)} \\ \nonumber 
& \qquad 
+ \frac{15}{2} \,  \e^{-\tfrac{2m^2}{3H}\left( 2 t_1 - t_3 - t_4\right)} - \biggl( \frac{51}{2} + \frac{m^2}{H}\bigl( 9t_1 -5 t_2 + t_3 + 3t_4 \bigr)\biggr) \, \e^{-\tfrac{2m^2}{3H}\left(t_1 - t_2 \right)} \\ \nonumber 
& \qquad 
- \biggl( 51 + \frac{2m^2}{H}\bigl( 9t_1 + t_2 -3 t_3 + 5t_4 \bigr)\biggr) \, \e^{-\tfrac{2m^2}{3H}\left(t_1 - t_3 \right)} + \frac{3}{2} \,  \e^{-\tfrac{2m^2}{3H}\left(t_1 - t_2 + t_3 - t_4\right)}  \\ \nonumber 
& \qquad 
- \biggl( \frac{153}{2} + \frac{m^2}{H}\bigl(27 t_1 + 9 t_2 + 11 t_3 - 7 t_4 \bigr)\biggr) \, \e^{-\tfrac{2m^2}{3H}\left(t_1 - t_4 \right)} + \frac{15}{2} \,  \e^{-\tfrac{2m^2}{3H}\left(2t_2 - t_3 - t_4\right)} \\ \nonumber 
& \qquad 
- \biggl( \frac{51}{2} + \frac{m^2}{H}\bigl( t_1 - t_2 + 9 t_3 + 7 t_4 \bigr)\biggr) \, \e^{\tfrac{2m^2}{3H}\left(t_2 - t_3 \right)} - \biggl( \frac{129}{4} + \frac{m^2}{H}\bigl( t_1 - t_2 + t_3 + 15 t_4 \bigr)\biggr) \times \\ \nonumber 
& \qquad \times \e^{\tfrac{2m^2}{3H}\left(t_2 - t_4 \right)} - \biggl( 51 + \frac{2m^2}{H}\bigl( t_1 +9 t_2 -3 t_3 +5 t_4 \bigr)\biggr) \, \e^{-\tfrac{2m^2}{3H}\left(t_2 - t_3 \right)} + 3 \, \e^{-\tfrac{2m^2}{3H}\left( t_1 + t_2 - t_3 - t_4\right)} \\ \nonumber 
& \qquad 
- \biggl( \frac{333}{4} + \frac{m^2}{H}\bigl(3 t_1 + 33 t_2 + 11 t_3 -7  t_4 \bigr)\biggr) \, \e^{-\tfrac{2m^2}{3H}\left(t_2 - t_4 \right)} - \, \e^{-\tfrac{2m^2}{3H}\left(t_1- 2 \, t_2 \right)} \\  \nonumber 
& \qquad 
- \biggl( 90 + \frac{m^2}{H}\bigl( 3 t_1 + 5 t_2 + 39 t_3 - 7 t_4 \bigr)\biggr) \, \e^{- \tfrac{2m^2}{3H}\left(t_3 - t_4 \right)} + \frac{15}{4} \,  \e^{\tfrac{2m^2}{3H}\left(t_2 - 2\, t_3 + t_4\right)} \\ \nonumber 
& \qquad 
- \biggl( \frac{129}{2} + \frac{2m^2}{H}\bigl( t_1 + t_2 - t_3 + 15 t_4 \bigr)\biggr) \, \e^{\tfrac{2m^2}{3H}\left(t_3 - t_4 \right)} - \frac{207}{2} + \frac{2m^2}{H} \bigl(9t_1 + 19 t_2 + 29 t_3 + 39 t_4 \bigr) \\ \nonumber 
& \qquad 
+ \frac{m^4}{3H^2} \biggl( \bigl(3 t_1 + 10 t_2 +14 t_3 + 18 t_4 \bigr)t_1 + \bigl( 11 t_2 + 18 t_3 + 30 t_4 \bigr) t_2 + \bigl( 19 t_3 + 42 t_4 \bigr)t_3 + 27  t_4^2 \biggr) \\ \nonumber  
& \qquad  
- \frac{15}{4} \, \e^{\tfrac{2m^2}{3H}\left(t_2 - 2\, t_3 \right)} - \frac{15}{4} \, \e^{\tfrac{2m^2}{3H}\left(t_2 - 2\, t_4 \right)} + \biggl( 72 + \frac{m^2}{H}\bigl( 27 t_1 + 9 t_2 + 11 t_3 + 13 t_4 \bigr)\biggr) \, \e^{-\tfrac{2m^2 t_1}{3H}} \\ \nonumber 
& \qquad  - \frac{5}{2} \, \e^{\tfrac{2m^2}{3H}\left(t_2 - t_3 - t_4 \right)} - \frac{15}{2} \, \e^{\tfrac{2m^2}{3H}\left(t_3 - 2\, t_4 \right)}   - \frac{15}{4} \, \e^{-\tfrac{2m^2}{3H}\left(2t_1 - t_2 \right)} - \frac{15}{2} \, \e^{-\tfrac{2m^2}{3H}\left(2t_1 - t_3 \right)}  \\ \nonumber 
& \qquad  
- \frac{45}{4} \, \e^{-\tfrac{2m^2}{3H}\left(2t_1 - t_4 \right)} - 3 \, \e^{-\tfrac{2m^2}{3H}\left(t_1 + t_2 - t_3 \right)} - \frac{11}{2} \, \e^{-\tfrac{2m^2}{3H}\left(t_1 + t_2 - t_4 \right)} - \frac{3}{2} \, \e^{-\tfrac{2m^2}{3H}\left(t_1 - t_2 + t_3 \right)}  \\ \nonumber 
& \qquad 
- \frac{3}{2} \, \e^{-\tfrac{2m^2}{3H}\left(t_1 - t_2 + t_4 \right)} - \frac{11}{2} \, \e^{-\tfrac{2m^2}{3H}\left(t_1 + t_3 - t_4 \right)} - 4 \, \e^{-\tfrac{2m^2}{3H}\left(t_1 - t_3 + t_4 \right)} - \frac{15}{2} \, \e^{-\tfrac{2m^2}{3H}\left(2t_2 - t_3\right)} \\ \nonumber 
& \qquad  
+ \biggl( \frac{315}{4} + \frac{m^2}{H}\bigl( 3 t_1 + 33 t_2 + 11 t_3 + 13 t_4 \bigr)\biggr) \, \e^{- \tfrac{2m^2 t_2}{3H}} - \frac{45}{4} \, \e^{-\tfrac{2m^2}{3H}\left(2t_2 - t_4\right)}  \\ \nonumber 
& \qquad 
- \frac{11}{2} \, \e^{-\tfrac{2m^2}{3H}\left(t_2 + t_3 - t_4\right)} - 4 \, \e^{-\tfrac{2m^2}{3H}\left(t_2 - t_3 + t_4\right)} - \frac{45}{4} \, \e^{-\tfrac{2m^2}{3H}\left(2t_3 - t_4\right)} \\ \nonumber 
& \qquad
+ \biggl( \frac{171}{2} + \frac{m^2}{H}\bigl( 3 t_1 + 5 t_2 + 39 t_3 + 13 t_4 \bigr)\biggr) \, \e^{-\tfrac{2m^2 t_3}{3H}} + \frac{11}{2} \, \e^{-\tfrac{2m^2}{3H}\left(t_1 + t_2\right)}  \\ \nonumber 
& \qquad
+ \frac{11}{2} \, \e^{-\tfrac{2m^2}{3H}\left(t_1 + t_3\right)} + \biggl( \frac{369}{4} + \frac{m^2}{H}\bigl( 3 t_1 + 5 t_2 + 7 t_3 + 45 t_4 \bigr)\biggr) \, \e^{-\tfrac{2m^2 t_4}{3H}} \\ \nonumber 
& \qquad
+ \frac{11}{2} \, \e^{-\tfrac{2m^2}{3H}\left(t_1 + t_4\right)} + \frac{11}{2} \, \e^{-\tfrac{2m^2}{3H}\left(t_2 + t_3\right)} + \frac{11}{2} \, \e^{-\tfrac{2m^2}{3H}\left(t_2 + t_4 \right)} + \frac{11}{2} \, \e^{-\tfrac{2m^2}{3H}\left(t_3 + t_4\right)} \\ \nonumber 
& \qquad 
+ \frac{45}{4} \, \, \e^{-\tfrac{4m^2 t_1}{3H}} + \frac{45}{4} \, \, \e^{-\tfrac{4m^2 t_2}{3H}} + \frac{45}{4} \, \, \e^{-\tfrac{4m^2 t_3}{3H}} + \frac{45}{4} \, \, \e^{-\tfrac{4m^2 t_4}{3H}} \Biggr) \e^{-\tfrac{m^2}{3H}\left(t_1+ t_2+ t_3 + t_4\right)}  \xrightarrow[]{\,\,\, m \rightarrow 0 \,\,\,} 
\end{align} 
\begin{align}
& \xrightarrow[]{\,\,\, m \rightarrow 0 \,\,\,} \, \frac{\lambda^2 H^{10}}{6144 \, \pi^8} \, \biggl( 11 \, t_1^4 \, t_2 t_4 + 22 \, t_1^4 t_3t_4 + 2\, t_1^2 \, t_2^3 \, t_4 + 12 \, t_1^2 \, t_2^2 \, t_3 t_4 + 6 \, t_1^2 \, t_2 t_3^2 \, t_4 \\ \nonumber 
& \qquad \qquad  \qquad \qquad \,\, + 2 \, t_1^2 \, t_2 t_4^3 + 4 \, t_1^2 \, t_3^3 \, t_4 + 4 \, t_1^2 \, t_3 t_4^3 + \frac{31}{5} \, t_2^5 \, t_4 + 22 \, t_2^4 \, t_3 t_4 + 2 \, t_2^3 \, t_3^2 t_4 \\ \nonumber 
& \qquad \qquad \qquad \qquad \quad + \frac{2}{3} \, t_2^3 \, t_4^3 + 4 \, t_2^2 \, t_3^3 \, t_4 + 4 \, t_2^2 \, t_3 t_4^3 + 11 \, t_2 t_3^4 \, t_4 + 2\, t_2 t_3^2 \, t_4^3 +  \frac{31}{5} \, t_2 t_4^5 \\ \nonumber 
& \qquad \qquad \qquad \qquad \qquad + \frac{62}{5} \, t_3^5 \, t_4 + \frac{4}{3} \, t_3^3 \, t_4^3  + \frac{62}{5} \, t_3 t_4^5 \biggr) \, ; 
\end{align} and at the equal times is
\begin{align} \label{four_point_TwoLoop_disconnected_equal_times}
& \bigl\langle \phi^4 \left(t\right) \bigr\rangle^{\text{disconnected}}_{\lambda^2}
= \,\, \frac{729 \, \lambda^2 H^{16}}{4096 \, \pi^8 m^{12}} \Biggl( 19 + \biggl( 26 - \frac{32m^2 t}{H} - \frac{16 m^4 t^2}{3H^2}\biggr) \e^{-\tfrac{2m^2t}{3H}} \\ \nonumber 
& \qquad \qquad \qquad \quad \,\,\,\, - \biggl( 96 + \frac{8m^2 t}{H}  - \frac{32m^4 t^2}{3H^2}\biggr) \e^{-\tfrac{4m^2t}{3H}}  + \biggl( 38 + \frac{40m^2 t}{H} \biggr) \e^{-\tfrac{2m^2t}{H}} + 13 \, \e^{-\tfrac{8m^2t}{3H}}\Biggr) \\
& \qquad \qquad \quad \,\,  \xrightarrow[]{\,\,\, m \rightarrow 0 \,\,\,}  \, \frac{23 \, \lambda^2 H^{10} \, t^6}{960 \, \pi^8} . \end{align} As before, the full $\lambda^2$ order's four-point correlation function is placed in the main part of this paper; see~\eqref{four_point_TwoLoop_full}.

\section{
Some expressions for section~\ref{Comparison_chapter}}\label{Chapter4Details}
\subsection{Starobinsky{--}Yokoyama stochastic approach}
Let us provide a few details on the content of section~\ref{Comparison_chapter}.
To calculate series for expectation values~\eqref{rho_lambda_small}, we used 
\begin{align} \label{rho_approx} 
\rho_{\text{st}} \left[\varphi\right] & = \, \frac{1}{\mathcal{N}} \,\, \e^{-\tfrac{8\pi^2}{3H^4}V\left(\varphi\right)} \approx \frac{1}{\mathcal{N}} \, \biggl(1- \frac{2\lambda \pi^2}{3H^4} \, \varphi^4 + \frac{2\lambda^2 \pi^4}{9H^8} \, \varphi^8 - \frac{4\lambda^3 \pi^6}{81H^{12}} \, \varphi^{12} \biggr) \e^{ \, -\tfrac{4\pi^2m^2 \varphi^2}{3H^4}}. \\ 
\label{phi2_approx}
\bigl\langle \varphi^2 \bigr\rangle & = \cfrac{\int\limits_{-\infty}^{+\infty} d \varphi \, \varphi^2 \rho_{\text{st}} \left[\varphi\right]}{\int\limits_{-\infty}^{+\infty} d \varphi \,  \rho_{\text{st}} \left[\varphi\right]} \approx \,\,\, \frac{ \dfrac{3H^4}{8\pi^2m^2}-\dfrac{135\lambda H^8}{256 \, \pi^4 m^6}+\dfrac{25515\lambda^2 H^{12}}{16384\, \pi^6 m^{10}}-\dfrac{3648645\lambda^3 H^{16}}{524288\, \pi^8 m^{14}}}{1-\dfrac{9\lambda H^4}{32 \, \pi^2 m^4}+\dfrac{945\lambda^2 H^{8}}{2048\, \pi^4 m^{8}}-\dfrac{93555\lambda^3 H^{12}}{65536\, \pi^6 m^{12}}} \\ \nonumber 
& \qquad \qquad \qquad \qquad  \xrightarrow[\quad]{} \,\,\, \dfrac{3H^4}{8\pi^2m^2}-\dfrac{27\lambda H^8}{64 \, \pi^4 m^6}+\dfrac{81\lambda^2 H^{12}}{64\, \pi^6 m^{10}}-\dfrac{24057\lambda^3 H^{16}}{4096\, \pi^8 m^{14}} + O\left( \lambda^4 \right); \\ 
\label{phi4_approx}
\bigl\langle \varphi^4 \bigr\rangle & = \cfrac{\int\limits_{-\infty}^{+\infty} d \varphi \, \varphi^4 \rho_{\text{st}} \left[\varphi\right]}{\int\limits_{-\infty}^{+\infty} d \varphi \,  \rho_{\text{st}} \left[\varphi\right]} \approx  \,\,\, \frac{ \dfrac{27H^8}{64 \, \pi^4m^4}-\dfrac{2835\lambda H^{12}}{2048 \, \pi^6 m^8}+\dfrac{841995\lambda^2 H^{16}}{131072\, \pi^8 m^{12}}}{1-\dfrac{9\lambda H^4}{32 \, \pi^2 m^4}+\dfrac{945\lambda^2 H^{8}}{2048\, \pi^4 m^{8}}} \\ \nonumber 
& \qquad \qquad \qquad \qquad 
\xrightarrow[\quad]{} \,\,\, \dfrac{27H^8}{64 \, \pi^4 m^4}-\dfrac{81\lambda H^{12}}{64 \, \pi^6 m^8}+\dfrac{24057\lambda^2 H^{16}}{4096\, \pi^8 m^{12}} + O\left( \lambda^3 \right). \end{align}

\vspace{0.15cm}
\subsection{Relation of the Starobinsky stochastic approach to perturbative QFT series within the leading-logarithm approximation}\label{Woodard}
To expand~\eqref{massles_stochastic_Woodard} up to the next $\lambda^3$ order, we obtain the same expressions in the spirit of~\cite{2005NuPhB.724..295T} in this part of an appendix, as well as consider the free massive case.

In the case of massive theory, $V(\varphi) = \cfrac{m^2 \varphi^2}{2}$, one can multiply both sides of the Fokker{--}Planck equation~\eqref{Fokker-Planck} by $
\varphi^{2\n}$ and integrate with the probability density $\rho \, \bigl[
\varphi(t,\x)\bigr]$, resulting in
\begin{equation}\label{Fokker-Planck_massive_almost_final}
\frac{\partial}{\partial t} \Bigl\langle \varphi^{2\n} (t,\x) \Bigr \rangle = - \frac{2 \n m^2}{3H} \, \Bigl\langle \varphi^{2\n} (t,\x) \Bigr \rangle + \n \, \bigl( 2\n - 1 \bigr) \, \frac{H^3}{4\pi^2} \, \Bigl\langle \varphi^{2\n-2} (t,\x) \Bigr \rangle. \,\,\,
\end{equation}
After some redefinitions, namely,  \begin{equation}\label{redefenition_1}
\alpha \equiv \frac{Ht}{4 \pi^2} \qquad \text{and} \qquad \bar{m}^2 \equiv \frac{8\pi^2 m^2}{3 H^2},
\end{equation}
one obtains a bit more appropriate form of~\eqref{Fokker-Planck_massive_almost_final} to deal with: \begin{equation}\label{final_form}
\frac{\partial}{\partial \alpha} \Biggl\langle \biggl(\frac{\varphi}{H}\biggr)^{2\n} \Biggr \rangle = - \n \,\bar{m}^2 \, \Biggl\langle \biggl(\frac{\varphi}{H}\biggr)^{2\n} \Biggr \rangle + \n \, \bigl( 2\n - 1 \bigr) \, \Biggl\langle \biggl(\frac{\varphi}{H}\biggr)^{2\n-2} \, \Biggr \rangle.
\end{equation}  
The solution to equation~\eqref{final_form} above can be found in the following way:
\begin{equation}\label{solution_with_coeff_massive_series}
\Biggl\langle \biggl(\frac{\varphi}{H}\biggr)^{2\n} \Biggr \rangle = \bigl( 2\n -1\bigr)!! \, \alpha^\n \, F_\n \bigl(\alpha \bar{m}^2\bigr) , \quad F_\n (z) = 1 + \alpha^1_\n z + \alpha^2_\n \, z^2 + \alpha^3_\n z^3 + \alpha^4_\n z^4 
+  \, ... \,,
\end{equation} 
where $F_\n (z)$ obeys 
\begin{equation}
\n F_\n (z) + z F^\prime_\n (z) = -z \n F_\n (z) + \n F_{\n-1} (z).
\end{equation} The recurrence relations on the unknown coefficients $\alpha^i_\n$ are $\bigl(\n+i\bigr)\alpha^i_\n = - \n \, \alpha^{i-1}_\n + \n \, \alpha ^i_{\n-1}$ and  $i= \overline{1, \infty}$.
Therefore, up to $O(z^6)$ in~\eqref{solution_with_coeff_massive_series}, one obtains the perturbative expansion
\begin{align}\label{Woodard_pert_series}
\Bigl\langle & \varphi^{2\n} (t) \Bigr \rangle = \bigl( 2\n -1 \bigr) !! \, \biggl( \frac{H^3 t}{4 \pi^2} \biggr)^\n  \Biggl( 1 - \n \, \frac{m^2 t}{3H} + \frac{\n \bigl(3 \n + 1 \bigr)}{54} \frac{m^4t^2}{H^2} - \frac{\n^2 \bigl(\n + 1 \bigr)}{162} \frac{m^6t^3}{H^3} \\ \nonumber 
& \qquad \qquad + \frac{\n}{29160} \Bigl(15 \n^3 + 30\n^2 + 5\n -2\Bigr) \frac{m^8 t^4}{H^4} - \frac{\n^2 \bigl(\n +1 \bigr)}{87480} \Bigl( 3\n^2 + 7\n -2\Bigr) \frac{m^{10} t^5}{H^5} + \, ...\Biggr). 
\end{align} And for specific $\n=1$ and $\n=2$, we have 
\begin{equation}\label{massive_pert_series_Woodard_2pt}
\Bigl\langle \varphi^{2} (t) \Bigr \rangle  = \, \frac{H^3 t}{4 \pi^2}  -  \frac{ m^2H^2 t^2}{12 \, \pi^2} + \frac{ m^4 H t^3}{54 \, \pi^2} - \frac{ m^6 t^4}{324 \, \pi^2} + \frac{ m^8 t^5}{2430 \, H \, \pi^2} - \frac{ m^{10} t^6}{21870 \, H^2 \, \pi^2}  +  ... \, ;  \end{equation} \begin{equation}\label{massive_pert_series_Woodard_4pt}
\Bigl\langle \varphi^{4} (t) \Bigr \rangle =  \frac{3 \, H^6 t^2}{16 \,\pi^4} - \frac{ m^2 H^5 t^3}{8 \pi^4} + \frac{7 m^4 H^4 t^4}{144 \, \pi^4} - \frac{m^6 H^3 t^5}{72 \, \pi^4} + \frac{31 m^8 H^2 t^6}{9720 \,  \pi^4} - \frac{m^{10} H t^7}{1620 \,  \pi^4} + ... . \end{equation} 
These series are precisely the same as if someone expands in the small regime $m^2/H^2$ our $\bigl\langle \phi^{2} (t) \bigr \rangle_0$ and $\bigl\langle \phi^{4} (t) \bigr \rangle_0$ in~\eqref{TwoPoint_equal_times_full_series} and~\eqref{four_point_tree_equal_times}.

In $V(\varphi) = \cfrac{\lambda \varphi^4}{4}$ case, we again multiply both sides of the Fokker{--}Planck equation~\eqref{Fokker-Planck} by $
\varphi^{2\n}$ and integrate over with the probability density $\rho \bigl[
\varphi(t,\x)\bigr]$. It leads to
\begin{equation}
\frac{\partial}{\partial t} \Bigl\langle \varphi^{2\n} (t,\x) \Bigr \rangle = - \frac{2 \n\lambda}{3H} \, \Bigl\langle \varphi^{2\n+2} (t,\x) \Bigr \rangle + \n \, \bigl( 2\n - 1 \bigr) \, \frac{H^3}{4\pi^2} \, \Bigl\langle \varphi^{2\n-2} (t,\x) \Bigr \rangle.
\end{equation}
Proceeding with analogy redefinitions~\eqref{redefenition_1}, one gets
\begin{equation}\label{redefinition2}
\frac{\partial}{\partial \alpha} \Biggl\langle \biggl(\frac{\varphi}{H}\biggr)^{2\n} \Biggr \rangle = - \n \,\bar{\lambda} \, \Biggl\langle \biggl(\frac{\varphi}{H}\biggr)^{2\n+2} \, \Biggr \rangle + \n \, \bigl( 2\n - 1 \bigr) \, \Biggl\langle \biggl(\frac{\varphi}{H}\biggr)^{2\n-2} \, \Biggr \rangle, \,\,\, 
\bar{\lambda} \equiv \frac{8\pi^2}{3} \lambda.
\end{equation}
The solution can be found in the following form 
\begin{equation}\label{solution_with_coeff_lambda_series}
\Biggl\langle \biggl(\frac{\varphi}{H}\biggr)^{2\n} \Biggr \rangle = \bigl( 2\n -1\bigr)!! \, \alpha^\n \, G_\n \bigl(\alpha^2 \bar{\lambda}\bigr) , \qquad G_\n (z) = 1 + \beta^1_\n z + \beta^2_\n \, z^2 + \beta^3_\n z^3 + \, ...,
\end{equation} 
and $G_\n (z)$ obeys 
\begin{equation}
\n \, G_\n (z) + 2z \, G^\prime_\n (z) = -\n \bigl(2\n+1\bigr) \, z \, G_{\n+1} (z) + \n \, G_{\n-1} (z).
\end{equation} Here, recurrence relations for coefficients are $\bigl(\n+2i\bigr)\beta^i_\n = - \n \, \bigl(2\n+1\bigr)\beta^{\,i-1}_{\n+1} + \n \, \beta ^i_{\n-1}$, and  
up to $O(z^4)$ in~\eqref{solution_with_coeff_lambda_series} one can obtain the perturbative expansion 
\begin{align}
\Bigl\langle \varphi^{2\n} (t) \Bigr \rangle = \bigl( 2\n -1 \bigr) !! & \biggl( \frac{H^3 t}{4 \pi^2} \biggr)^\n \Biggl(  \, 1 - \frac{\lambda \, \n \, \bigl(\n+1\bigr)}{12\pi^2} \, H^2 t^2 \\ \nonumber 
& + \frac{ \lambda^2 \, \n}{10080 \, \pi^4} \, \Bigl(35\n^3 + 170 \n^2 + 225\n+74\Bigr) H^4 t^4  \\ \nonumber 
&  \quad  - \frac{ \lambda^3 \n  \bigl(\n+1\bigr)  \bigl(\n+2\bigr)}{362880 \, \pi^6} \, \Bigl(35\n^3 + 300 \n^2 + 685\n + 252\Bigr) H^6 t^6  + \, ...  \, \Biggr),
\end{align} and for $\n=2$ and $\n=4$ that reads
\begin{align}
\Bigl\langle \varphi^{2} (t) \Bigr \rangle & = \, \, \frac{H^3 t}{4 \pi^2}  - \lambda \,  \frac{ H^5 t^3}{24 \, \pi^4} + \lambda^2 \, \frac{H^7 t^5}{80 \, \pi^6} - \lambda^3 \, \frac{ 53 H^9 t^7}{10080 \, \pi^8} + \, ... \, ; \\
\Bigl\langle \varphi^{4} (t) \Bigr \rangle & =  \, \frac{3 H^6 t^2}{16 \, \pi^4} - \lambda \, \frac{3 H^8 t^4}{32\, \pi^6} +  \lambda^2 \, \frac{53 H^{10} t^6}{960 \, \pi^8} - \lambda^3 \,  \frac{517 H^{12} t^8}{13440\, \pi^{10}}  + \, ... \, .
\end{align} One can easily compare it with~\eqref{loop_series_massless_coincide} and~\eqref{massle_four_point_full}.

For the general case with $V(\varphi) = \cfrac{m^2\varphi^2}{2} + \cfrac{\lambda \varphi^4}{4}$, we present an alternative approach for computing the perturbative series for equal-time and multi-time correlation functions in the leading logarithm approximation in~\cite{Kamenshchik:2025ses}.

\subsection{The Hartree-Fock approximation}\label{HF_appendix}
We have also found the solution to equation~\eqref{HFA_equation}; here we present some intermediate steps. Denoting $ \bigl\langle \phi^2(t) \bigr\rangle \equiv f(t)$, equation~\eqref{HFA_equation} reads
\begin{equation}\label{abc}
\frac{df}{dt} = a - b \, f - c \, f^2 \,\, ; \qquad \text{where} \quad a \equiv \frac{H^3}{4\pi^2} \, , \quad b \equiv \frac{2m^2}{3H} \, , \quad \text{and} \quad c \equiv \frac{2\lambda}{H}.
\end{equation} Then, the solution can be defined through 
\begin{equation}\label{f12_}
\frac{df}{\left(f - f_1\right)\left( f - f_2\right)} = - c \, dt\,  \quad \text{and} \quad f_{1,2} = - \frac{b}{2c} \pm \sqrt{\frac{b^2}{4c^2} + \frac{a}{c}},
\end{equation} therefore, 
\begin{equation}\label{f12}
f(t) = \frac{f_1 f_2 \Bigl(1 - \e^{-c \, \left( f_1 - f_2\right) \, t} \Bigr)}{f_2 - f_1\,\e^{-c \, \left( f_1 - f_2\right) \, t}} \, .
\end{equation}
Substituting these expressions~\eqref{f12} and assigned values~\eqref{abc}, one can get solution~\eqref{Hartree_Fock_solution} in the main part of this paper.

\section{A few details for 
an autonomous equation in section~\ref{Autonomous_chapter}}\label{Starob_Hartee_details}
Since we have supposed that the correcting 
term in our autonomous equation~\eqref{aut_second_eq} is small, our equation~\eqref{linear_correction2} is
\begin{equation}\label{linear_W}
\frac{d}{dt} \Bigl(\delta f (t)
\Bigr) = - \biggl( \frac{2m^2}{3H} +\frac{3\lambda H^3}{\pi^2 m^2} f_{\mathcal{Z}}(t)
\biggr) \, \delta f (t)
+ \mathcal{W} (t),
\end{equation} 
where we denote the correcting term with the help of new notations as follows
\begin{equation}
\mathcal{W} (t)
= - \frac{ 81\lambda^2 H^{11}}{64\, \pi^6 \, m^8} \Biggl( 2\, f_{\mathcal{Z}}  (t)
- 6\,  f^{\,2}_{\mathcal{Z}}  (t)
+ \frac{8}{3} \, f^{\,3}_{\mathcal{Z}} (t)
+ \Bigl(1-2f_{\mathcal{Z}} (t)
\Bigr)^2 \text{ln}\Bigl(1-2f_{\mathcal{Z}} (t)
\Bigr) \Biggr).
\end{equation} The solution to this linearized equation~\eqref{linear_W} can be found in the form
\begin{align}\label{Eq_form}
\delta f (t)
= \text{exp}\Biggl(-\int\limits_0^t dt^\prime  \biggl(\,  \frac{2m^2}{3H} & + \frac{3\lambda H^3}{\pi^2 m^2} \, f_{\mathcal{Z}}(t^\prime)
\biggr) \Biggr) \times \\ \nonumber 
& \times \int\limits_0^t dt^{\prime\prime} \,\, \mathcal{W} (t^{\prime\prime})
\,\, \text{exp}\Biggl( \,\, \int\limits_0^{t^{\prime\prime}} dt^{\prime\prime\prime} \biggl( \frac{2m^2}{3H} +\frac{3\lambda H^3}{\pi^2 m^2} f_{\mathcal{Z}}(t^{\prime\prime\prime})
\biggr) \Biggr).
\end{align}
Here the first multiplier is 
\begin{equation}\label{first_mult}
\text{exp}\Biggl(-\int\limits_0^t dt^\prime \, \biggl( \frac{2m^2}{3H} +\frac{3\lambda H^3}{\pi^2 m^2} f_{\mathcal{Z}}(t^\prime)
\biggr) \Biggr) = 4 \mathcal{Z}^2 \biggl( \bigl( \mathcal{Z} +1 \bigr) \, \e^{\tfrac{m^2}{3H}\mathcal{Z}t} + \bigl(  \mathcal{Z} - 1 \bigl) \, \e^{-\tfrac{m^2}{3H} \mathcal{Z}t} \, \biggl)^{-2},   
\end{equation}  and after cumbersome though straightforward integration of the second multiplier in ~\eqref{Eq_form}, the full answer is~\eqref{Full_aut}.

\acknowledgments This research was partially supported by the INFN grant FLAG.
A.K. is grateful to Tereza Vardanyan for numerous fruitful discussions.

\bibliographystyle{JHEP}
\bibliography{biblio.bib}
\end{document}